\documentclass[prx,letterpaper,aps,superscriptaddress,floatfix,twocolumn,nofootinbib,table]{revtex4-2}
\pdfoutput=1
\usepackage{graphicx}
\usepackage{amsmath}
\usepackage{amsfonts}
\usepackage{amssymb}
\usepackage[normalem]{ulem}
\usepackage[small,bf]{subfigure}
\usepackage[utf8]{inputenc}
\usepackage{dsfont}
\usepackage{slashed}
\usepackage{soul}
\usepackage{comment}
\usepackage{tikz-cd}
\usepackage{hyperref}
\usepackage[raggedrightboxes]{ragged2e}
\usepackage{makecell}
\usepackage{xcolor}

\newcommand{\beq}{\begin{equation}}
\newcommand{\eeq}{\end{equation}}

\newcommand{\beqa}{\begin{eqnarray}}
\newcommand{\eeqa}{\end{eqnarray}}

\usepackage{amsthm}
\theoremstyle{plain}
\newtheorem*{theorem*}{Theorem}
\newtheorem{theorem}{Theorem}
\newtheorem{corollary}{Corollary}[theorem]

\usepackage{diagbox}
\newcolumntype{P}[1]{>{\centering\arraybackslash}p{#1}}

\begin{document}
\title{No-go theorem for single time-reversal invariant symmetry-protected Dirac fermions in 3+1d}
\author{Lei Gioia}
\affiliation{Walter Burke Institute for Theoretical Physics, Caltech, Pasadena, CA, USA}
\affiliation{Department of Physics, Caltech, Pasadena, CA, USA}
\author{Anton A. Burkov}
\affiliation{Department of Physics and Astronomy, University of Waterloo, Waterloo, Ontario 
N2L 3G1, Canada}
\affiliation{Perimeter Institute for Theoretical Physics, Waterloo, Ontario N2L 2Y5, Canada}
\author{Taylor L. Hughes}
\affiliation{Department of Physics, University of Illinois Urbana-Champaign, Urbana, IL 61801, USA}
\begin{abstract}

We employ a general method, known as anomaly-matching, to derive new no-go theorems of fermionic lattice models. For our main result, we show that time-reversal invariant 3+1d lattice systems (such as Dirac and Weyl semimetals) can never admit a lone low-energy symmetry-protected Dirac fermion (or node), i.e., it must always come in higher muliplets or be fine-tuned. This theorem holds for both non-interacting and interacting systems as long as the electromagnetic $U(1)_{V,{\rm UV}}$ symmetry is a normal subgroup of the microscopic symmetry group $G_{\mathrm{UV}}$; a condition that is ubiquitous in physical $U(1)_{V,{\rm UV}}$ preserving lattice models. To show that our theorems are tight, we also explore both well-known and new systems that are converses of the no-go theorem, obtained by forfeiting certain assumptions such as a broken time-reversal symmetry (magnetic Weyl semimetal), a non-compact non-on-site $U(1)$ (almost local Dirac node model), no-symmetry protection (fine-tuned Dirac semimetal), or multiple low-energy Dirac nodes (time-reversal invariant Weyl and Dirac semimetals). We will also explicitly demonstrate that, while this theorem strictly prohibits single time-reversal invariant symmetry-protected Dirac node, it does allow for other odd numbers of Dirac nodes under certain circumstances, such as three Dirac nodes in the Fu-Kane-Mele diamond lattice model. This is akin to the Nielsen-Ninomiya theorem for an odd number of differently-charged chiral fermions, whose lattice realizations are allowed if certain anomaly cancellation conditions are met.

\end{abstract}
\maketitle

\tableofcontents

\section{Introduction}
\label{sec:intro}

Chiral fermions regularized on a lattice famously obey the Nielsen-Ninomiya fermion doubling theorem~\cite{NIELSEN198120,NIELSEN1981173,Friedan82}. This theorem dictates that a $U(1)$ symmetric lattice spectrum contains the same number of $U(1)$-charged left and right-handed chiral fermions, assuming general conditions such as a quantized local $U(1)$ charge with all fermions having the same $U(1)$ charge, as well as the presence of crystalline translation symmetry, hermiticity, and locality. Although this statement was originally derived for free fermions, it has since been generalized to interacting systems~\cite{Fidkowski_2023}. As an example, in 3+1d systems this theorem implies that one cannot create a model of a single symmetry-protected Weyl fermion on a lattice, unless one or more of the assumptions are broken, such as charge quantization~\cite{PhysRevLett.118.207701,gioia2025exactchiralsymmetries31d,Meng12}, or locality~\cite{Meng19}.

In contrast, this theorem does not restrict the lattice realization of 3+1d Dirac fermions since they are composed of a single pair of opposite-handed chiral fermions. Thus, it is perhaps unsurprising that single Dirac fermions are fully regularizable and realizable in free-fermion lattice models~\cite{Qi08}, though they are perturbatively gappable (i.e., \textit{fine-tuned}) unless there are additional required symmetries, such as rotation symmetry, that can prevent a gap opening. Indeed, there have been many works that classify sets of internal and crystalline symmetries that \textit{symmetry-protect} Dirac fermions~\cite{Nagaosa14,RevModPhys.90.015001}. 

Interestingly, in the presence of time-reversal symmetry, there are no known examples of lattice models harboring a single symmetry-protected Dirac fermion in the infrared (IR)~\footnote{We will revisit some claimed models of single symmetry-protected Dirac nodes later in the paper.}. In this paper, we will explicitly prove that this absence is actually because of a no-go theorem that is summarized as follows:
\begin{theorem}
    Given locality, hermiticity, time-reversal symmetry $\Theta$ with $\Theta^2=(-1)^{\hat{F}}$, and a `natural' $U(1)$ symmetry, all lattice theories exhibiting a single low-energy Dirac fermion are fine-tuned.
    \label{thm:1}
\end{theorem}
\noindent Here, a `natural' $U(1)$ symmetry is defined to be one that acts \textit{on-site}, i.e., acts as a tensor-product operator on the underlying Hilbert space, and is also a normal subgroup of the full microscopic (UV) symmetry group~\footnote{This is proven up to exceptional cases detailed in Section~\ref{sec:case2}.}. This is an implicit assumption of almost all number-conserving free-fermionic systems, and it allows for symmetries that commute with the $U(1)$ charge, as well as those such as charge conjugation that flip the sign of the charge. Additionally, $\hat{F}$ is the fermion parity operator, and hence $\Theta^2=(-1)^{\hat{F}}$ is the conventional many-body generalization of the single-particle notion that $\Theta^2=-1$ for fermions.  

Theorem~\ref{thm:1} can be viewed as a generalization of the Nielsen-Ninomiya theorem for the lattice regularizability of a single Dirac fermion. However, while \emph{single} symmetry-protected Dirac fermions are strictly forbidden in the presence of time-reversal symmetry, an odd number of symmetry-protected Dirac fermions is allowed under certain circumstances~\footnote{In fact, this is similar to the Nielsen-Ninomiya theorem, where an odd number (larger than one) of microscopic $U(1)$ symmetric Weyl fermions can exist if they are charged differently under the $U(1)$ symmetry. See Section~\ref{sec:NNthm} for more details.}, e.g., the Fu-Kane-Mele model exhibits three low-energy Dirac fermions~\cite{FKM} as we will examine carefully in Section~\ref{sec:FKM}. We note that previous literature has claimed the existence of lattice models having a single symmetry-protected time-reversal invariant Dirac node which would violate our theorem~\cite{RevModPhys.90.015001,PhysRevLett.112.036403,Nagaosa14,PhysRevB.96.195105}. However, upon careful examination, we show that these systems actually contain additional low-energy Dirac fermions (see Section~\ref{sec:twoormore}).

\begin{table*}[t]
    \centering
    {\renewcommand{\arraystretch}{1.7}
    \begin{tabular}{|P{3.9cm}||P{3.2cm}|P{3.4cm}|P{2.8cm}|P{3.2cm}|}\hline
        \vspace*{0.1cm} IR theory\textbackslash UV symmetries& On-site $U(1)_{V,{\rm UV}}$ \newline On-site $\Theta_{\rm UV}$ \newline Crystalline symmetry-protected & On-site $U(1)_{V,{\rm UV}}$ \newline Broken $\Theta_{\rm UV}$ \newline Crystalline symmetry-protected & \vspace*{0.001cm} On-site $U(1)_{V,{\rm UV}}$ \newline On-site $\Theta_{\rm UV}$ & Non-on-site $U(1)_{V,{\rm UV}}$ \newline On-site $\Theta_{\rm UV}$ \newline Crystalline symmetry-protected \\\hline\hline
        Single massless Dirac theory \newline (Lorentz invariant, Eq.~\ref{eq:IRaction}) & No-go\newline Theorem~\ref{thm:1}; \newline Corollary~\ref{thm:nodirac} & Non-symmorphic Dirac semimetal (Section~\ref{sec:nonsymDirac}) & Fine-tuned Dirac \newline semimetal (Section~\ref{sec:finetuned}) & Almost-local Dirac node model (Section~\ref{sec:nononsiteU(1)}) \\\hline
        Single massless Dirac theory \newline (Lorentz broken, Eq.~\ref{eq:IRactionbroken}) & No-go \newline Theorem~\ref{thm:1}; \newline Corollary~\ref{thm:noweyl} & Time-reversal broken Weyl semimetal (Section~\ref{sec:TbrokenWeyl}) & \cellcolor{gray!25} & \cellcolor{gray!25} \\\hline
        Double massless Dirac theory \newline (Lorentz invariant) &  Type-I and II Dirac semimetals (Section~\ref{sec:spinel}) & \cellcolor{gray!25} & \cellcolor{gray!25} & \cellcolor{gray!25} \\\hline
        Double massless Dirac theory \newline (Lorentz broken) & $\Theta_{\rm UV}$-invariant Weyl semimetal (Section~\ref{sec:TinvariantWeyl}) & \cellcolor{gray!25} & \cellcolor{gray!25} & \cellcolor{gray!25} \\\hline
    \end{tabular}}
    \caption{Table depicting the possible UV theory resulting from the assumed IR theory and UV symmetries. Note that a UV system is said to be have IR Lorentz symmetry if its IR theory up to first order in momentum possesses Lorentz symmetry. We give Hamiltonian examples linked in brackets for the mentioned lattice systems. The grayed out cells represent non-interesting examples that can be derived via simple perturbations or doubling of existing examples.}
    \label{tab:results}
\end{table*}

As will be demonstrated in the method of proof below (see Section~\ref{sec:proof}), our results apply to both time-reversal invariant free-fermionic systems such as Weyl and Dirac semimetals~\cite{RevModPhys.90.015001,Xu613,Burkov11-1,Burkov_ARCMP,Felser_ARCMP}, as well as strongly-interacting lattice models~\cite{wang2024interactionrobustnesschiralanomaly,PhysRevB.107.115147}. For now, we can outline an intuitive argument for our theorem as follows. If a time-reversal invariant and $U(1)$ symmetric lattice model possesses an IR theory with a single, symmetry-protected Dirac node, there exists a time-reversal and $U(1)$ symmetric mass that must be prevented by an additional symmetry $g$. Suppose this symmetry $g$ is a lattice (i.e., crystalline) symmetry. One can break the $g$ symmetry and induce a Dirac mass $\pm m.$ The choice of sign will pick one of the two distinct 3+1d time-reversal symmetric topological insulator (TI) $\mathbb{Z}_2$ phases, i.e., either the strong TI or trivial insulator phases. Because the Dirac node was symmetry-protected by $g$, the two signs of the mass are related to each other by the symmetry action of $g$ (i.e., $g:\pm m\mapsto \mp m$ and an unbroken $g$ would require $m=0$). This would allow one to change the intrinsic quantized magneto-electric polarizability (i.e., the $\theta$ angle), or equivalently, transform from a trivial atomic insulator to a strong TI, via a simple crystalline $g$ action which seems impossible. And thus, such a situation should not arise. However, this argument is non-rigorous and relies on several assumptions, such as the commutativity between $g$ and time-reversal symmetry. In this paper, we will carefully detail the conditions under which this intuitive argument is accurate via rigorous symmetry and anomaly analysis in Section~\ref{sec:proof}.

Our article is organized as follows. We begin with a warm-up exercise in Section~\ref{sec:NNthm} where we prove the Nielsen-Ninomiya theorem in 3+1d using a renormalization group map and anomaly focused framework in order to convey the ideas we will use for the single time-reversal invariant Dirac no-go theorem. With this in hand, we will prove Theorem~\ref{thm:1} in Section~\ref{sec:proof}, which consists of a few important steps: (1) We will first analyze the IR theory of a single Dirac fermion in 3+1d, including the Dirac masses in Section~\ref{sec:IRmodel}, and its symmetry properties in Section~\ref{sec:IRsymmetriesmasses}. (2) In Section~\ref{sec:UVtoIRmap} we employ a renormalization group perspective to map from the microscopic (UV) symmetries of a lattice to the IR symmetries. (3) After carefully defining and assuming two `natural' notions of $U(1)$ symmetry in the UV in Sections~\ref{sec:case1symm} and \ref{sec:case2}, we will present the associated IR quantum anomalies in Section~\ref{sec:IRanomalies}. (4) Using this knowledge, we will match the resulting IR and UV anomalies in Sections~\ref{sec:gaugefieldmatch} and \ref{sec:nogo}, and derive the associated lattice regularizability theorems. In Section~\ref{sec:circumvent} we will explore the consequences of violating each assumption of the theorem, thus arriving at converse examples, e.g., we can break time-reversal symmetry to generate a magnetic Weyl semimetal that has a single low-energy pair of opposite handed chiral fermions (albeit with different symmetry charges under translation). These examples are summarized in Table~\ref{tab:results} and will help us show that our conclusions are tight. Finally, in Section~\ref{sec:discussion} we will discuss potential generalizations and open questions. Our statements provide general rules for the realizability of semimetal systems under lattice symmetries that may guide both theorists and experimentalists in the search for 3+1d semimetallic and gapless states in non-magnetic materials.

\section{Warm up: Nielsen-Ninomiya no-go via anomaly matching}
\label{sec:NNthm}

Some aspects of our proof are subtle, so we first present an analysis of the Nielsen-Ninomiya theorem which illustrates our general approach in a simpler example. We will show that all 3+1d theories with an on-site $G_{\rm UV} = U(1)_{\rm UV}$ symmetry, admitting a single low energy massless Weyl fermion, are fine-tuned.\footnote{In the conventional literature, Weyl points are not thought of as fine-tuned. However, if we allow superconducting masses, then indeed we require the $U(1)$ charge conservation symmetry to symmetry-protect the Weyl fermion in the IR, as otherwise the gaplessness is accidental.} Some details will be omitted as we just want to give a flavor of the argument, which will be generalized in the main section.

A single massless Weyl fermion in 3+1d possesses an action, given by
\begin{align}
    S_{\rm IR}=\int dt\,d^3 r\, i\psi^\dag\sigma^\mu \partial_\mu \psi\quad,
    \label{eq:singleweylaction}
\end{align}
where $\sigma^\mu=(1,\,\sigma^x,\sigma^y,\sigma^z)$, $\psi$ is the two-component Weyl spinor, we employ the Einstein summation notation for indices, and we use $\hbar = c= e = 1$ units. The massless Weyl theory possesses an internal $U(1)$ chiral symmetry
\begin{align}
    U(1)_{\rm IR}: \quad\psi(x^\mu)\mapsto e^{i\theta}\psi(x^\mu) \quad,
\end{align}
where $\theta\in[0,2\pi)$.

The single Weyl fermion has two independent allowed mass terms summarized by the action
\begin{align}
    S_{M}=\int dt\,d^3 r\left( |M| e^{i\alpha}\psi^T\sigma^y \psi+|M| e^{-i\alpha}\psi^\dag\sigma^y \psi^*\right),
    \label{eq:massWeyl}
\end{align}
where $\alpha\in[0,2\pi)$, and $M\in\mathbb{R}$. Typically we think of Weyl fermions as being locally stable in momentum space. However, that stability assumes $U(1)$-symmetry, i.e., assumes no superconducting masses. Indeed, the masses in Eq.~\ref{eq:massWeyl} are not invariant under the $U(1)_{\rm IR}$ symmetry and transform as
$$ U(1)_{\rm IR}: \quad \alpha\mapsto \alpha+2\theta\quad.$$
Thus, one sees that in the presence of $U(1)_{\rm IR}$ the Weyl fermion remains massless. One can check that, ignoring Lorentz symmetries, the complete IR symmetry group $G_{\rm IR}$ of the massless Weyl fermion is given solely by the $U(1)_{\rm IR}$ symmetry, i.e.,
\begin{align}
    G_{\rm IR}= U(1)_{\rm IR}\quad.
\end{align}

To connect the discussion of the IR physics to the potential realization of a single Weyl fermion on the UV lattice we can apply the renormalization group (RG) framework.  When considering symmetries, let us define the ``RG map'' $\rho$, which is a homomorphism induced by the renormalization group. The homomorphism $\rho$ maps the UV symmetry group $G_{\mathrm{UV}}$ into the IR symmetry group $G_{\rm IR}$ as the lattice becomes more and more coarse-grained:
\begin{align}
    \rho:\quad G_{\mathrm{UV}}\rightarrow G_{\mathrm{IR}}\quad.
    \label{eq:rho}
\end{align}

Beyond symmetries, the RG procedure also defines a linear and surjective map between the UV and IR operators $\mathcal{R}: \mathcal{A}_{\mathrm{UV}}\rightarrow \mathcal{A}_{\mathrm{IR}}$ where the UV/IR operator algebras are labeled $\mathcal{A}_{\mathrm{UV}/IR}.$ 
The symmetry map $\rho$ obeys the following commutative diagram with the RG operator map $\mathcal{R},$
\[
\begin{tikzcd}
\mathcal{A}_{\mathrm{UV}} \arrow[r,"\mathcal{R}"] \arrow[d,"g"'] &
\mathcal{A}_{\mathrm{IR}}  \arrow[d,"\rho(g)"'] 
\\
\mathcal{A}_{\mathrm{UV}} \arrow[r,"\mathcal{R}"] &
\mathcal{A}_{\mathrm{IR}}
\end{tikzcd}
\]
which can be written succinctly as
$$\mathcal{R}\circ g=\rho(g)\circ \mathcal{R}\quad.$$ A crucial feature of $\mathcal{R}$ is that it is surjective, i.e., every IR operator comes from the UV. As a consequence of this, and the commutative diagram, any combination of symmetries that acts trivially on UV operators (and thus is the identity in $G_{\rm UV}$), must also act trivially on IR operators. This means, as stated above, that $\rho$ is a homomorphism and preserves the group laws
\begin{align}
    \rho(g) \rho( h)\rho((g h)^{-1})=1\in G_{\mathrm{IR}}\quad,
\end{align}
where $g,h\in G_{\rm UV}.$

From this perspective we see that to create a \textit{symmetry-protected} gapless fermion in a lattice system we need the image of $G_{\mathrm{UV}}$ under $\rho$, i.e., $\rho(G_{\mathrm{UV}})\subseteq G_{\mathrm{IR}}$, to forbid all masses. To accomplish this for the single Weyl fermion in Eq.~\ref{eq:singleweylaction}, $\rho(G_{\rm UV})$ could be $U(1)_{\rm IR}$, or even a subgroup of $U(1)_{\rm IR}$ such as $\mathbb{Z}_n$ with $n>2$, as these would prevent $M\neq 0$ in Eq.~\ref{eq:massWeyl}. In contrast, we call gapless systems that are not symmetry-protected \textit{fine-tuned} instead, as one may generically add a mass without violating any symmetries.
It seems plausible that one could find cases where the homomorphism $\rho$ would map the $U(1)_{\rm UV}$ symmetry into $U(1)_{\rm IR}$ and hence allow for symmetry-protected single Weyl fermions. However, we have ignored the possibility of the symmetry being anomalous. Furthermore, the consequences of anomalies in the IR depend crucially on a subtlety that distinguishes two classes of IR symmetries that we now discuss. 

In general,  $G_{\mathrm{IR}}$ may be larger than $\rho(G_{\mathrm{UV}})$, and hence it is helpful to distinguish between two symmetry types: \textit{emanant}~\cite{10.21468/SciPostPhys.15.2.051} and \textit{emergent} symmetries. Emanant symmetries are inherited from the UV symmetry group and appear in the image of $\rho$, i.e., $\rho(G_{\mathrm{UV}})$. On the other hand, emergent symmetries in $G_{\mathrm{IR}}$ do not come directly from the UV, and are accidental, and, e.g., generically broken by irrelevant operators. Emergent symmetries live in the complement of $\rho(G_{\mathrm{UV}})$, i.e., $G_{\mathrm{IR}}\backslash\, \rho(G_{\mathrm{UV}})$. 

Both emanant and emergent symmetries may have associated IR quantum anomalies such as the well-known chiral anomaly. However, only IR anomalies of \emph{emanant} symmetries enforce non-trivial conditions on the UV theory via anomaly matching. We have just argued that symmetry-protection requires the presence of certain emanant IR symmetries in order to forbid mass terms. Thus, enforcing non-trivial anomaly matching conditions between the IR and UV anomalies may be required if we have symmetry protection. This matching may not always be possible, which is exactly how no-go theorems such as the Nielsen-Ninomiya theorem arise.
We will now introduce the anomaly matching condition for the single Weyl fermion.

Any IR anomaly resulting from an emanant symmetry must be matched in the UV. This is sometimes called anomaly matching or the ``emergeability condition''~\cite{PhysRevB.104.075132,PhysRevX.11.021005,10.21468/SciPostPhys.13.3.066}. Indeed, if the anomalies do not match, the UV and IR theories are inconsistent with each other. To write the UV-IR anomaly matching conditions more explicitly, we consider the $G_{\rm UV}$ and $G_{\rm IR}$ gauge fields $A_{\rm UV}$ and $A_{\rm IR}$. Gauge field configurations can be represented in terms of their Wilson loops $W_\gamma$ over closed loops $\gamma.$ The Wilson loops form conjugacy classes in $G_{\rm UV}$ and $G_{\rm IR}$ respectively, and define the gauge fields up to gauge transformations. Using this fact and our RG map, we can define the ``pushforward'' of the UV gauge field, $\rho_*\left(A_{\mathrm{UV}} \right)$, to be
\begin{align}
    W_\gamma(\rho_* (A_{\rm UV})):= \rho(W_\gamma(A_{\rm UV}))\quad.
    \label{eq:gaugefieldsmatch}
\end{align}
The anomaly matching (emergeablility) condition can then be expressed as
\begin{align}
    \Omega_{\mathrm{IR}}\left(\rho_*(A_{\mathrm{UV}})\right)=\Omega_{\mathrm{UV}}\left(
A_{\mathrm{UV}}\right)\quad,
\label{eq:emergeability}
\end{align}
where $\Omega_{\mathrm{UV}}(A_{\mathrm{UV}})$ and $\Omega_{\mathrm{IR}}(A_{\mathrm{IR}})$ are the UV and IR anomaly terms, respectively (see Eq. \ref{eq:weylIRanomaly} below for an example).  The anomaly matching condition comes from the necessary consistency of gauge transformations of $A_{\rm UV}$ when considered in the UV or IR. It follows that in the case of emergent (as opposed to emanant) symmetries, there is no anomaly matching condition since the IR gauge field of an emergent symmetry would not emanate from a UV gauge field and hence we need not be concerned with UV-IR consistency. Hence we consider anomalies of emanant gauge fields, and the connection between the gauge fields in the UV and IR is 
\begin{align}
    A_{\mathrm{IR}}=\rho_*\left(A_{\mathrm{UV}} \right)\quad.
    \label{eq:gaugefieldrelation}
\end{align}

In order to derive the Nielsen-Ninomiya theorem, let us now assume, as was assumed in the original proof, that the structure of $G_{\mathrm{UV}}$ is given by
\begin{align}
    G_{\mathrm{UV}}=U(1)_{\rm UV}=e^{i \theta\hat{Q}}\quad,
\end{align}
where $U(1)_{\rm UV}$ is on-site, with $\hat{Q}$ representing some total charge operator that acts in an on-site manner. The usual example in condensed-matter systems is when $U(1)_{\rm UV}$ corresponds to the electromagnetic charge conservation symmetry such that $\hat{Q}=\sum_{\mathbf{r}}\hat{n}_{\mathbf{r}}$ with number operator $\hat{n}_{\mathbf{r}}=\sum_j \hat{n}_{j,\mathbf{r}}$ where $j$ encompasses orbital, spin, and other local degrees of freedom in lattice systems of electrons. However, more generally, $U(1)_{\rm UV}$ just corresponds to a generic microscopic on-site $U(1)$ symmetry that does not have to act non-trivially on all degrees of freedom.

As discussed previously, a symmetry-protected single Weyl fermion must possess a map $\rho$ from $U(1)_{\rm UV}$ to $U(1)_{\rm IR}$ such that $\rho(U(1)_{\rm UV})$ prevents the fermion masses in Eq.~\ref{eq:massWeyl}. The only possible maps, that satisfy this condition and are group homomorphisms, take the form
\begin{align}
    \rho(e^{i\theta \hat Q}):\quad\psi \mapsto e^{iq\theta}\psi\quad,
    \label{eq:nontrivialwrapping}
\end{align}
where $q\in\mathbb{Z}\backslash 0$. Such a mapping means that the $U(1)_{\rm IR}$ is an emanant symmetry group such that the Weyl fermion is symmetry-protected.

Importantly, such an IR symmetry group for the massless Weyl fermion is associated with a well-known chiral and chiral-gravitational anomaly, which is an obstruction to gauging the chiral $U(1)_{\rm IR}$ symmetry. There are a few ways to see this anomaly. First, one may observe that minimally coupling the Weyl action in Eq.~\ref{eq:singleweylaction} to a background gauge field $A_{\rm IR}$ results in a partition function
\begin{align}
    Z=\int D\psi D\psi^\dag\,e^{iS_{\rm IR}[\psi,\psi^\dag,A_{\rm IR}]}\quad,
\end{align}
with
$$S_{\rm IR}[\psi,\psi^\dag,A_{\rm IR}]=\int dt\,d^3 r\, i\psi^\dag\sigma^\mu \left(\partial_\mu -iA_\mu\right)\psi\quad.$$
Without the proper regularization this partition function is ill-defined, i.e., zero, due to the presence of zero-modes for certain gauge field configurations and manifolds~\cite{nakahara}. Naturally, if we were to proceed with gauging $U(1)_{\rm IR}$ without regularizing the theory we would integrate over all gauge configurations and always result in an ill-defined theory. This is the aforementioned obstruction to gauging that we may attempt to sort out using some regularization scheme. However, the only regularization that allows for both the conservation of the chiral $U(1)_{\rm IR}$ symmetry, as well as preserving its on-site character, is to realize the single Weyl fermion (or other anomalous theories) on the boundary of a higher-dimensional symmetry-protected topological (SPT) phase of matter~\cite{RevModPhys.88.035001}. Such an SPT phase is described by a topological term in one higher dimension than the original anomaly, and is a convenient method to encapsulate the gauge-field dependence of the anomaly. 

In our case, the bulk SPT response action corresponding to the boundary Weyl chiral and chiral-gravitational anomaly is given by the 4+1d Chern-Simons term
\begin{align}
\label{eq:weylIRanomaly}
    &\Omega_{\mathrm{IR}} (A_{\rm IR})\\
    &=\frac{1}{2} \frac{1}{4\pi^2}\int A_{\rm IR}\wedge\left[\frac{1}{3} dA_{\rm IR}\wedge dA_{\rm IR}
    +\frac{1}{24}\mathrm{Tr}\left[R\wedge R\right]\right]\,,
    \nonumber
\end{align}
where $A_{\rm IR}$ is the gauge field of $G_{\rm IR}$, $R$ is the Riemann curvature, and the trace is over the Lorentz indices of the Riemann curvature ~\cite{10.21468/SciPostPhysLectNotes.62}.  The surface of the SPT described by this term will harbor a single Weyl fermion. The anomaly on the boundary can be observed by noticing that under a gauge transformation $A_{\rm IR}\rightarrow A_{\rm IR}+d\theta(x^\mu)$, the boundary is not gauge invariant, and transforms as
\begin{align}
    &\partial\Omega_{\mathrm{IR}} =\frac{1}{2} \frac{1}{4\pi^2}\int \theta(x^\mu)\left[\frac{1}{3} dA_{\rm IR}\wedge dA_{\rm IR}
    +\frac{1}{24}\mathrm{Tr}\left[R\wedge R\right]\right]\,,
    \nonumber
\end{align}
which is an obstruction to gauging the $U(1)_{UV}$ symmetry boundary, since the gauge non-invariance would result in the partition function being zero after integration over all the gauge configurations. The correspondence between the boundary anomaly and the bulk SPT response action gives us a convenient way to encode the anomaly in the bulk-boundary correspondence. We will employ this higher-dimensional SPT language to describe anomalies for the purposes of anomaly matching when we consider Dirac fermions below.

Another hallmark of the chiral anomaly is the non-conservation of current in the presence of certain gauge field configurations. This effect can be observed when we compute the chiral current $j_{\rm IR}^\mu$ by varying Eq.~\ref{eq:weylIRanomaly} with respect to $A_{\rm IR}$, written  
$$\frac{\delta \Omega_{\rm IR}}{\delta A_{{\rm IR}\mu}}=j_{\rm IR}^\mu\quad.$$
In the presence of a boundary perpendicular to the fourth spatial dimension $x^4\equiv u$ at $u=0$ this gives us in flat-space ($R=0$)
\begin{align}
    \partial_\mu j_{\rm IR}^\mu
    &=  \frac{1}{2} \frac{1}{4\pi^2}\delta(u)\epsilon^{\alpha\beta\gamma\delta}\partial_\alpha A_{\rm IR\beta}\partial_\gamma A_{\rm IR\delta}\quad,
\end{align}
where $\delta(u)$ is the Dirac delta function.
We see that the boundary $U(1)_{\rm IR}$ current is not conserved for non-trivial $A_{\rm IR}$ configurations, once again exemplifying the features of the chiral anomaly via a bulk SPT term.

The UV and IR gauge fields can then be related via Eqs. \ref{eq:gaugefieldrelation} and \ref{eq:nontrivialwrapping} to be
\begin{align*}
     A_{\rm IR}= q A_{\rm UV}\quad,
\end{align*}
such that the UV anomaly, via Eqs.~\ref{eq:emergeability} and \ref{eq:weylIRanomaly}, is given by
\begin{align}
    &\Omega_{\mathrm{UV}}(A_{\rm UV})=\\
    &\frac{1}{2} \frac{1}{4\pi^2}\int  A_{\rm UV}\wedge\left[ \frac{q^3}{3}dA_{\rm UV}\wedge dA_{\rm UV}
    +\frac{q}{24}\mathrm{Tr}\left[R\wedge R\right]\right]\quad,\nonumber
\end{align}
which is a non-trivial chiral and chiral-gravitational anomaly, albeit potentially with a different coefficient than Eq.~\ref{eq:weylIRanomaly}. As mentioned previously, this anomaly (now also in the UV) is an obstruction to gauging $U(1)_{\rm UV}$ without, for example, the presence of a higher-dimensional bulk. However, by definition all on-site symmetries have a well-defined gauging procedure and are thus automatically gaugeable! This leads us to a contradiction in 3+1d, which can only be resolved if $q=0$ in Eq.~\ref{eq:nontrivialwrapping}, i.e., $U(1)_{\rm UV}$ maps to the identity in the IR. If $U(1)_{\rm UV}$ maps to the identity in the IR then $U(1)_{\rm IR}$ is not an emanant symmetry, and hence the system is not symmetry-protected, and is at best fine-tuned~\footnote{A condensed-matter free-fermionic example of a single fine-tuned Weyl fermion can be constructed using bands that are uncharged under $U(1)_{\rm UV}$, alongside other degrees of freedom that can be charged under $U(1)_{\rm UV}$.}. This is in fact the Nielsen-Ninomiya no-go theorem, formulated in the framework of anomalies:
\begin{theorem}
    (Nielsen-Ninomiya theorem in 3+1d) All single Weyl fermion systems are fine-tuned if $U(1)_{\rm UV}\subseteq G_{\rm UV}$ with on-site $U(1)_{\rm UV}$.
\end{theorem}
Note that such a formulation of the theorem may appear new to some readers, as the concept of a single Weyl fermion is often unfamiliar, so let us construct an explicit tight-binding example. Consider a 3+1d cubic lattice having two flavors of fermions ($a$ and $b$), denoted by creation operators $a_\sigma^\dag$ and $b_\sigma^\dag$ where $\sigma\in\{A,B\}$ denotes an orbital degree of freedom. Let the $U(1)_{\rm UV}$ symmetry act as
\begin{align}
    U(1)_{\rm UV}:
    \begin{cases}
        \quad a_\sigma^\dag\mapsto e^{i\theta}a_\sigma^\dag\\
        \quad b_\sigma^\dag\mapsto b_\sigma^\dag
    \end{cases}\quad,
\end{align}
where we see that $a_\sigma^\dag$ has charge 1 and $b_\sigma^\dag$ has charge 0 under this symmetry. 

The following Hamiltonian possesses this symmetry as well as a single fine-tuned Weyl fermion in its spectrum:
\begin{align}
    H=\sum_{\mathbf{k}}&\bigg[\sum_\sigma a^\dag_{\sigma\mathbf{k}}a_{\sigma\mathbf{k}}+b_{\mathbf{k}}^\dag \sum_{\alpha}\sin k_\alpha \sigma^\alpha b_{\mathbf{k}}\nonumber\\
    &+ m(\mathbf{k})\left(b_{\mathbf{k}}\sigma^y b_{-\mathbf{k}}+b^\dag_{-\mathbf{k}}\sigma^y b^\dag_{\mathbf{k}}\right)\bigg]\quad,
\end{align}
where $b^\dag_{\mathbf{k}}=(b^\dag_{A\mathbf{k}},b^\dag_{B\mathbf{k}})$, and $\alpha\in\{x,y,z\}$, $m(\mathbf{k})=3-\cos k_x-\cos k_y-\cos k_z$. The first term represents a flat band of $a$ fermions that are gapped away from zero-energy (i.e., all $a$ states are unoccupied in the ground state), the second term is the usual Weyl semimetal term for the $b$ fermions which has 8 Weyl fermions at the 8 time-reversal invariant (TRIM) points in the Brillouin zone, and the last term opens a gap at all TRIM points except at $\mathbf{k}=\mathbf{0}$. We see that $U(1)_{\rm UV}$ is unbroken, since the $b$ fermions are uncharged under the symmetry. Indeed, in the UV to IR map given by Eq.~\ref{eq:nontrivialwrapping}, we have $q=0$. Such a system is fine-tuned since we may easily symmetrically perturb the system with $m(\mathbf{k})\rightarrow m(\mathbf{k})+\epsilon$ with $\epsilon>0$ to give the Weyl fermion at $\mathbf{k}=0$ a mass of $O(\epsilon)$, i.e., $U(1)_{\rm UV}$ offers no symmetry-protection.

Such a no-go theorem can also be generalized to other odd numbers of Weyl fermions under certain charge assumptions, e.g, when all Weyl fermions have the same charge under $U(1)_{\rm UV}$. However, this does not hold in general due to the possibility of anomaly cancellation, with famous examples including 5 chiral Weyl fermions which are allowed if their charges are $\{1,5,-7,-8,9\}$~\footnote{Anomaly cancellation can be checked by observing that $\sum_j q_j=0$ and $\sum_j q_j^3=0$ with $q_j$ being the $U(1)_{\rm UV}$ charge of the $j$th Weyl fermion.}, as well as the Standard model chiral gauge theory which contains 15 chiral Weyl fermions~\cite{tonggaugetheory,bilal2008lecturesanomalies}. We will now use the methods we presented in this section to derive a new and more complicated no-go theorem involving single Dirac nodes in the presence of time-reversal symmetry.

\section{No-go theorem for single Dirac fermions}
\label{sec:proof}

In the previous section, we analyzed single 3+1d Weyl fermions and showed how the Nielsen-Ninomiya no-go theorem can be deduced using the RG map and anomaly matching methods. Now, we can use these methods to study the more complicated case of 3+1d Dirac fermions in the presence of time-reversal symmetry in order to deduce a new no-go theorem.

\subsection{IR theory of the 3+1d Dirac fermion}
\label{sec:IRmodel}

We start the discussion by analyzing the possible symmetries and masses associated with a low-energy (IR) 3+1d Dirac fermion.
The IR theory of a single Dirac fermion in 3+1d is described by the action
\begin{align}
    S_{\mathrm{IR}}=\int dt\, d^3r\,\,i\Psi^\dag(\gamma^0\gamma^\mu \partial_\mu) \Psi\quad,
    \label{eq:IRaction}
\end{align}
where $\gamma^\mu$ are $4\times 4$ Dirac matrices that obey the Clifford algebra $\{\gamma^\mu,\gamma^\nu\}=2\eta^{\mu\nu}\mathds{1}_4$ with $\eta_{\mu\nu}$ being the Minkowski metric with signature $(+,-,-,-)$, and $\Psi$ is the four-component Dirac spinor.
There exist six Lorentz-invariant~\footnote{The Lorentz transformations are given by $\Psi\mapsto e^{\frac{ \theta_{\mu\nu}}{4}[\gamma^\mu,\gamma^\nu]}\Psi$ where $\theta_{\mu\nu}$ is an angle. Recall that $\gamma^{0,5}$ are Hermitian and symmetric, $\gamma^{1,3}$ are anti-Hermitian and anti-symmetric, and $\gamma^{2}$ is anti-Hermitian and symmetric.}, free-fermionic, dimensionally-relevant masses that can open a gap, given by
\begin{align}
    \int  & dt\, d^3r \,\bigg(M_1\Psi^\dag \gamma^0\Psi-M_2\Psi^\dag i \gamma^0\gamma^5\Psi\nonumber\\
    &+\Psi^T[M_3-M_4\gamma^5+iM_5-iM_6\gamma^5]\hat{C}\Psi\bigg)+h.c.\,,
    \label{eq:3dmasses}
\end{align}
where $M_{\alpha}\in\mathbb{R}$ $\forall \alpha\in\{1,..,6\}$, $\hat{C}\equiv i\gamma^0\gamma^2$ is the charge conjugation operator, and $\gamma^5\equiv i\gamma^0\gamma^1\gamma^2\gamma^3$ is the chiral operator. We note that four out of the six masses violate $U(1)$ charge conservation ($M_3, M_4, M_5, M_6$), i.e., they would be generated by a superconducting pairing potential. 
Via dimensional counting of units of mass (in the context of renormalization group with $\hbar=c=1$), these six masses are the only relevant fermion bilinear operators of the gapless Dirac fermion theory. In Appendix~\ref{app:masses} we show that these masses open an energy gap for the Dirac fermion by using the equivalent Hamiltonian formulation that may be more familiar to condensed matter readers. 

Without loss of generality, for the remainder of Section \ref{sec:proof}  we use the Weyl basis with the Dirac matrices given by
\begin{align}
    \gamma^0=\sigma^x,\,\,\gamma^{1,2,3}=i\sigma^y s^{x,y,z}, \,\,\gamma^5=-\sigma^z,\,\, \hat{C}=i\sigma^z s^y,
    \label{eq:weylbasis}
\end{align}
where $\sigma^{x,y,z}$ and $s^{x,y,z}$ are two independent sets of Pauli matrices, and tensor products with the identity matrix are left implicit.

\subsection{IR symmetries of the 3+1d Dirac fermion}
\label{sec:IRsymmetriesmasses}

To explore the symmetry protection of the 3+1d Dirac fermion we need to see how IR symmetries act on these masses (and potentially prohibit their existence). For convenience, let us rewrite $\Psi=(\psi_1,-is^y\psi_2^*)^T$ so that $\psi_1$ and $\psi_2$ are both \emph{left-handed} Weyl spinors, and the subscripts represent different flavors. In this basis, the masses in Eq.~\ref{eq:3dmasses} can be rewritten as
\begin{align}
\int dt\,d^3r \left(\,M^{ij}\psi_{i}^Tis^y\psi_{j}+h.c.\right)\quad,
\label{eq:massesweyl}
\end{align}
where $i,j\in\{1,2\}$ indices represent flavors of Weyl fermions, and we have gathered the mass terms into the (symmetric) matrix $M^{ij}$, given by
\begin{align}
    M=\begin{pmatrix}
        M_3+M_4+iM_5+iM_6 & M_1+i M_2 \\
        M_1+iM_2& M_3-M_4+iM_5-iM_6
    \end{pmatrix}.
    \label{eq:massM}
\end{align}

Conveniently, in this basis, we can immediately identify unitary symmetries that act trivially on the masses, i.e., leave them invariant, such as the Lorentz rotation group  $SU(2).$ These transformations are given by
$$\psi_j(x^\mu)\mapsto e^{-i\frac{s^{x,y,z}}{2}\theta}\psi_j(R^\mu_\nu x^\nu)\quad,$$
along with the appropriate spatial rotation $x^\mu\mapsto \left[R^{-1}\right]^\mu_\nu x^\nu$ where we note $R^\mu_\nu\in SO(3)$, leaves Eq.~\ref{eq:massesweyl} invariant. Similarly, the non-unitary Lorentz boost transformations, which act as
$$\psi_j(x^\mu)\mapsto e^{-\frac{s^{x,y,z}}{2}\theta}\psi_j(B^\mu_\nu x^\nu)\quad,$$
alongside the respective spatial boosts $x^\mu\mapsto \left[B^{-1}\right]^\mu_\nu x^\nu,$ where $B^\mu_\nu$ is a boost transformation, also leave the masses invariant. Hence, these symmetries will not protect~\footnote{The IR rotation symmetries do not symmetry-protect the Dirac node, however UV lattice rotation symmetries can protect Dirac nodes that lie on an invariant rotation axis in momentum space because such a UV symmetry will act like an internal unitary symmetry in the IR and may not leave the masses invariant.} the Dirac node and, together, the rotations and boosts form the Lorentz transformation $\mathrm{Spin}(1,3)$ group.

In contrast, there are internal unitary symmetries that commute with the gapless Dirac Hamiltonian (see Appendix \ref{app:masses})~\footnote{Internal symmetries are those that purely act on the internal flavor and spin degrees of freedom, as opposed to spatial degrees of freedom (such as for Lorentz transformations). Examples of internal symmetries include $U(1)_V$ and $U(1)_A$.} that can act non-trivially on the masses. These symmetries are represented by actions on the Weyl flavors
$$\psi_i(x^\mu)\mapsto U_{i}^j\psi_j(x^\mu)\quad,$$
where $U\in U(2)$. Under these symmetries, the mass term in Eq.~\ref{eq:massesweyl} transforms via the symmetric-squared representation of $U(2)$, given by
\begin{align}
    U(2):\quad M^{ij}\mapsto U_k^i U_l^j M^{kl}=\left[U M U^T\right]^{ij}\quad.
\end{align}

To study the possible symmetry protection mechanisms, we recall that symmetry subgroups in $U(2)$ that force $M_\alpha=0\,\forall \alpha$ are said to symmetry-protect the gapless Dirac node theory in Eq.~\ref{eq:IRaction}, i.e., small symmetric perturbations will not open a gap. To more clearly see the group structure, it is convenient to write $U(2)$ as
\begin{align}
    U(2)=\frac{U(1)_A\times SU(2)}{\mathbb{Z}_2}\quad,
    \label{eq:U(2)decomp}
\end{align}
where $U(1)_A$ is the axial (or chiral) symmetry which acts as $\psi_j\mapsto e^{i\theta}\psi_j$ (recall that our rewriting of $\Psi$ results in $\psi_{1,2}$ having the same handedness), $SU(2)$ acts as $\psi_j\mapsto g_{j}^k\psi_k$ with $g\in SU(2)$, and the $\mathbb{Z}_2$ refers to quotienting by $(e^{i\pi},-1)\in U(1)_A\times SU(2)$~\footnote{Note that for the rest of this paper we will be taking various $\mathbb{Z}_2$ quotients that identify fermion parity elements in the respective groups such that the total symmetry group acts faithfully on the IR operators.}. Generators of the $U(2)$ Lie algebra are~\footnote{We use the physics convention which has a factor of $i$ in the exponentialization process.}
\begin{align}
    \{\mathds{1},\tau^x,\tau^y,\tau^z\}\quad,
\end{align}
where $\tau$ are Pauli matrices that act on the different Weyl flavors, with $\mathds{1}$ being the $U(1)_A$ generator, and $\tau^z$ being the $U(1)_V$ generator, i.e., the vector/electric charge generator. We translate the above symmetry analysis into the Hamiltonian language in Appendix~\ref{app:symmDirac}.

Finally, let us discuss the action of time-reversal symmetry. The gapless Dirac fermion is symmetric under an anti-unitary time-reversal operator $\Theta$ that acts as
\begin{align}
    \Theta: \quad\gamma^{1,2,3}\mapsto -\gamma^{1,2,3}\,,\,\gamma^{0,5}\mapsto \gamma^{0,5}\,, \quad\Theta^2=-1\,.
\end{align}
An explicit example of the anti-unitary time-reversal symmetry is given by~\footnote{This is the operator that is commonly used in condensed-matter literature.}
\begin{align}
    \Theta:\quad \Psi(t)\mapsto -is^y\Psi(-t)\,,\quad \psi_j(t)\mapsto -is^y \psi_j(-t)\,,
    \label{eq:exampleTR}
\end{align}
alongside reversal of the time direction $t\mapsto -t$. All other time-reversal symmetries are related to this one via operations in $U(2)$ and the Lorentz group. Such a symmetry will prevent certain mass terms, e.g., for the $\Theta$ in Eq.~\ref{eq:exampleTR} $M_2,M_5,M_6$ are all forbidden. However, we must require additional elements of $U(2)$ to fully symmetry-protect the system since there are still three allowed masses. We will primarily focus on the time-reversal operators that commute with the $U(1)_V$ charge generator $\tau^z,$ as mentioned in the next subsection.

Combining the above observations, we can conclude that the unitary free fermionic~\footnote{Note that we do not consider `non-linear' symmetries such as $\Psi\mapsto \Psi^\dag\gamma^0\Psi$.} IR symmetry group $G_{\mathrm{IR}}$ of a single Dirac fermion is given by
\begin{align}
    G_{\mathrm{IR}}=\frac{U(2)\times \mathrm{Pin}(1,3)}{\mathbb{Z}_2}\quad,\label{eq:irGroup}
\end{align}
where we have used the $\mathrm{Pin}(1,3)$ group, instead of the Lorentz group $\mathrm{Spin}(1,3)$, so that parity and time-reversal symmetries are also included in the allowed IR symmetry group, and the $\mathbb{Z}_2$ quotient refers to quotienting by $(-1,-1)\in U(2)\times \mathrm{Pin}(1,3)$. We summarize which subgroups of the full set of IR symmetries prevent which Dirac fermion masses in Table~\ref{tab:symtomass}.
\begin{table}
    \centering
    \bgroup
    \def\arraystretch{1.5}
    \begin{tabular}{|c|c|}
    \hline
         IR Symmetry & Mass prevented \\ \hline
         Lorentz $\mathrm{Spin}(1,3)$ & \text{None} \\
         Internal $U(2)$ & \text{All} \\
         Internal $SU(2)$ & \text{All} \\
         Internal $U(1)_A$ & \text{All} \\
         Internal $U(1)_V$ & $M_3,M_4,M_5,M_6$ \\
         Time-reversal $\Theta$ & $M_2,M_5,M_6$ \\
         Internal $\mathbb{Z}_2^R$ & $M_1,M_2$ \\
         Internal $\mathbb{Z}_2^R\times U(1)_V$ & All\\
         \hline
    \end{tabular}
    \egroup
    \caption{Here we list some possible IR symmetries and which fermion bilinear mass terms (in Eq.~\ref{eq:massM}) they prevent. ``All'' refers to $M_1,M_2,M_3,M_4,M_5,M_6$. We define a $\mathbb{Z}_2$ internal symmetry $\mathbb{Z}_2^R:\psi_1\mapsto \psi_1$, $\psi_2\mapsto -\psi_2$ that we will use later on in the paper.}
    \label{tab:symtomass}
\end{table}

As an aside, we note that all of the above symmetries and gappability observations hold even if we break Lorentz symmetry (which is generically true in crystalline lattices) by adding terms such as $m_\mu\Psi^\dag\gamma^0\gamma^5\gamma^\mu \Psi.$ These terms preserve $U(1)_A$ and do not open a mass gap because $\gamma^0\gamma^5\gamma^\mu$ does not anticommute with all of the $\gamma^0\gamma^\mu$ matrices in the kinetic term of Eq.~\ref{eq:IRaction}. Via Fourier transformation to momentum space, one can observe that, instead of creating a mass for the Dirac fermion, these terms shift the relative momenta of the Weyl fermions that comprise the gapless Dirac fermion. An example of a Lorentz symmetry-broken action is given by
\begin{align}
    S_{\mathrm{IR},\mathrm{broken}}=\int dt\,d^3r\,\Psi^\dag\gamma^0\gamma^\mu(i\partial_\mu+\gamma^5 m_\mu)\Psi,
    \label{eq:IRactionbroken}
\end{align}
where we see that such an IR action corresponds to the low-energy theory of a time-reversal broken Weyl semimetal~\cite{Burkov_ARCMP}.

\subsection{Assumptions on the UV symmetries and the RG symmetry map}
\label{sec:UVtoIRmap}

To see whether a 3+1d time-reversal invariant Dirac fermion can be regularized on a 3+1d lattice without fine-tuning, we must now consider the relevant RG symmetry map and its consequences in the UV.

Recall that in order for a 3+1d Dirac fermion to be symmetry-protected, we require the image of the UV symmetry group under $\rho$ to forbid all masses. Since the Lorentz symmetry group $\mathrm{Spin}(1,3)$ acts trivially on the masses, we can focus, without loss of generality, on the map $\bar\rho$, defined as $\rho$ followed by a $\mathrm{Spin}(1,3)$ quotient:
\begin{align}
    \bar\rho:\quad G_{\rm UV}\rightarrow \bar{G}_{\mathrm{IR}}\quad,
\end{align}
with
\begin{align}
    \bar{G}_{\mathrm{IR}}\equiv \frac{G_{\mathrm{IR}}}{\mathrm{Spin}(1,3)}=\frac{U(2)\rtimes \mathbb{Z}_4^T}{\mathbb{Z}_2}\quad,
    \label{eq:GIRbar}
\end{align}
where the $\mathbb{Z}_2$ quotient refers group elements equivalent up to $(-1,-1)\in U(2)\rtimes \mathbb{Z}_4^T$, and $\mathbb{Z}_4^T$ is the time-reversal symmetry group that squares to fermion parity. 

In analogy with Section~\ref{sec:NNthm},
$\bar\rho(G_{\mathrm{UV}})$ is an emanant symmetry group, and for the system to be symmetry-protected, the emanant symmetries must be sufficient to forbid all masses. Examples of sufficient emanant symmetry groups that prevent fermion masses include $U(2)$, $U(1)_A$, $SU(2)$, etc.(see Table~\ref{tab:symtomass}). Just as in the Nielsen-Ninomiya proof above, such emanant symmetry groups may place non-trivial anomaly matching conditions on the UV and lead to no-go theorems, as we will see below.

Before we proceed, we make some natural, but crucial assumptions on the structure of $G_{\mathrm{UV}}$:
\begin{enumerate}
    \item We assume that $U(1)_{V,{\rm UV}}$ is a subgroup of $G_{\mathrm{UV}}$, written
$$U(1)_{V,{\rm UV}}\subset G_{\mathrm{UV}}\quad,$$
and that it is also on-site, i.e., $U(1)_{V,{\rm UV}}$ acts as a tensor product of operators on each local, unit cell Hilbert space. That is, UV group elements take the form $e^{i\theta \hat{Q}_V}\in U(1)_{V,{\rm UV}},$ where $\hat{Q}_V=\sum_{\mathbf{r}}\hat{n}_\mathbf{r}$, with the property that under $\rho$ it maps to $ U(1)_{V,{\rm IR}}\subseteq G_{\mathrm{IR}}$:
\begin{align}
    \rho(e^{i\theta \hat{Q}_V}):
    \begin{pmatrix}
        \psi_1 \\
        \psi_2
    \end{pmatrix}\mapsto e^{i\theta\tau^z}
    \begin{pmatrix}
        \psi_1 \\
        \psi_2
    \end{pmatrix}
    \quad.
    \label{eq:rhomapU(1)}
\end{align}
Physically, this condition amounts to assuming the presence of the usual on-site electric charge symmetry $U(1)_{V,{\rm UV}}$ on the lattice, and that the charge in the UV is the same as in the IR, i.e., a unit charge in this case. The on-site nature also implies that the UV theory always has a well-defined procedure to gauge $U(1)_{V,{\rm UV}}$ which will be important for our proof below.
\item We assume the presence of a time-reversal symmetry in $G_{\rm UV}$,
$$\mathbb{Z}_4^T\subset G_{\rm UV}\quad.$$
We will denote the symmetry operator as $\Theta$, and assume that it acts in an on-site manner and obeys $\Theta_{\rm UV}^2=(-1)^{\hat{F}}$, where $\hat{F}$ is the fermion parity operator.
\item We will also assume that time-reversal $\Theta_{\rm UV}$ commutes with the $U(1)_{V,{\rm UV}}$ charge generator $\hat{Q}_V$ such that
\begin{align}
    \Theta_{\rm UV}\hat{Q}_V \Theta_{\rm UV}^{-1}=\hat{Q}_V \quad .
    \label{eq:TRcommU1}
\end{align}
In particular, in combination with Eq.~\ref{eq:rhomapU(1)}, this means that the IR symmetries must obey
$$\rho(\Theta_{\rm UV})\tau^z\rho(\Theta_{\rm UV})^{-1}=\tau^z\quad ,$$
where the emanant IR time-reversal symmetry will be denoted $\Theta\equiv\rho(\Theta_{\rm UV})$. An example of an emanant time-reversal that satisfies the above condition is the conventional one in Eq.~\ref{eq:exampleTR}.
\item All other symmetries in $G_{\rm UV}$ are denoted as $G_{\rm other}$, which can include on-site internal symmetries $G_{\rm internal}$ or crystalline symmetries $G_{\rm space}$. We assume compactness for internal symmetries.
\item Additionally, we assume time-reversal commutes with all crystalline symmetry elements $g_{\mathrm{crystal}}\in G_{\rm space}$, e.g., translation symmetry, rotation symmetry, reflection, non-symmorphic symmetries, and inversion symmetry, up to an internal symmetry:
\begin{align}
    \Theta_{\rm UV} g_{\mathrm{crystal}}\Theta_{\rm UV}^{-1}=g_{\rm int}\,g_{\mathrm{crystal}}\quad,
    \label{eq:crystalTheta}
\end{align}
where $g_{\rm int} \in G_{\rm internal}$ and $\Theta_{\rm UV} g_{\rm int} \Theta_{\rm UV}^{-1}=g_{\rm int}^{-1}$ which follows since $\Theta_{\rm UV}^2=(-1)^{\hat{F}}$. Note that in the usual free-fermionic language $\Theta_{\rm UV}$ is chosen such that $g_{\rm int}=1$. However, more generally, one may choose $\Theta_{\rm UV}$ such that crystalline symmetries commute with it only up to an internal symmetry, e.g., $\Theta_{{\rm UV},j}=-i(\sigma^z_j)^{j}s^y_j\mathcal{K}$ ($j$ is a site index) commutes with translation symmetry only up to an internal group element $\prod_j\sigma_j^z$.
Time-reversal may or may not commute with elements of $G_{\rm internal}$, i.e.,
\begin{align}
    \Theta_{\rm UV} g_{\mathrm{int}}\Theta_{\rm UV}^{-1}=\tilde{g}_{\rm int}\quad,
    \label{eq:timereversalinternal}
\end{align}
for some $\tilde{g}_{\rm int}\in G_{\rm internal}$.
\end{enumerate}
We will now consider two separate cases in Sections~\ref{sec:case1symm} and \ref{sec:case2} of additional assumptions on the structure of $G_{\mathrm{UV}}$ which will require separate analysis and arguments to determine the obstruction to lattice regularizations of symmetry-protected, time-reversal invariant single Dirac nodes.

\subsection{Case 1: The $U(1)_{V,{\rm UV}}$ charge generator lies in the center of $G_{\rm UV}$}
\label{sec:case1symm}

In Case 1 we will assume that all elements of $G_{\rm UV}$ commute with the $U(1)_{V,{\rm UV}}$ charge generator $\hat{Q}_V$, such that
\begin{align}
    g \hat{Q}_V g^{-1}=\hat{Q}_V\quad,
    \label{eq:center}
\end{align}
$\forall g\in G_{\rm UV}$. Examples of $G_{\mathrm{UV}}$ that obey this condition include all crystalline space-groups and time-reversal symmetry, but not particle-hole symmetry which anticommutes with the $U(1)_{V,{\rm UV}}$ charge generator.

With condition~\ref{eq:center} in hand, we may place some restrictions on the UV to IR map $\rho$ in Eq.~\ref{eq:rho}. Since all elements of $G_{\rm UV}$ commute with $\hat{Q}_V,$ and $\rho$ is a group homomorphism (which preserves group laws), it follows that~\footnote{Note that, for simplicity, we employ an abuse of notation where $\rho$ represents both the group homomorphism as well as the adjoint-action/pushforward on the Lie algebra context dependent.}
$$\rho(\hat{Q}_V)=\rho(g\hat{Q}_V g^{-1})=\rho(g)\rho(\hat{Q}_V)\rho(g)^{-1}\quad,$$
$\forall g\in G_{\rm UV}$. From Eq.~\ref{eq:rhomapU(1)} $\rho(\hat{Q}_V)=\tau^z$, which means that all emanant IR symmetries must obey
$$\tau^z=\rho(g)
\tau^z\rho(g)^{-1}\quad.$$ This means that we should consider the centralizer of $U(1)_V$ in the unitary part of $\bar{G}_{\rm IR}$ (i.e., the internal symmetry group $U(2)$):
$$C_{U(2)}(U(1)_V)=\frac{U(1)_A\times U(1)_V}{\mathbb{Z}_2}\quad,$$
with $C_{G}(S)=\{g\in G|\forall s\in S, gsg^{-1}=s\}$ being the centralizer, i.e., all elements of $G$ that commute with the subset $S$.

Putting these observations together, we conclude that the emanant IR symmetry group $\bar\rho(G_{\mathrm{UV}})$ lies in a subgroup of $\bar{G}_{\rm IR}$ from Eq.~\ref{eq:GIRbar}, given by 
\begin{align}
\bar\rho(G_{\mathrm{UV}})\subseteq
\frac{\frac{U(1)_A\times U(1)_V}{\mathbb{Z}_2}\rtimes \mathbb{Z}_4^T}{\mathbb{Z}_2}\subset \bar{G}_{\rm IR}\quad,
\label{eq:prot}
\end{align}
where the the first part of the RHS is the centralizer of $U(1)_V$ in the unitary part of $\bar{G}_{\rm IR},$ the $Z_4^T$ represents time-reversal symmetry which we have already assumed to commute with the $U(1)_V$ charge generator in Eq.~\ref{eq:TRcommU1}, and the upper $\mathbb{Z}_2$ quotient is quotienting by $(e^{i\pi},e^{i\pi\tau^z})\in U(1)_A\times U(1)_V$~\footnote{Note that we have two $\mathbb{Z}_2$ quotients since there are multiple fermion parity elements that need to be identified: first $(e^{i\pi},e^{i\pi\tau^z})\in U(1)_A\times U(1)_V$, and then $(-1,-1)\in \frac{U(1)_A\times U(1)_V}{\mathbb{Z}_2}\rtimes \mathbb{Z}_4^T$.}.

So far, our analysis has shown that the condition in Eq. \ref{eq:center} greatly reduces the possible UV to IR symmetry maps, and, for example, rules out mass-forbidding mechanisms granted from the non-Abelian nature of $SU(2)$ (or other non-trivial non-Abelian subgroups such as the dihedral subgroup). Intuitively, we can see that this is the case since the IR generator for $U(1)_{V,{\rm IR}}$ in our notation is $\tau^z$, which does not commute with the other $SU(2)$ generators $\tau^x,\tau^y.$ These are the generators that would produce the non-Abelian symmetry operations, but they do not reside in Eq.~\ref{eq:prot} due to the non-commutation with $\tau^z$.
From Table~\ref{tab:symtomass} we see that the reduced IR symmetry group in Eq.~\ref{eq:prot} is still sufficient to symmetry-protect the Dirac fermion, i.e., there is still a possibility of symmetry protection so further work needs to be done to prove a no-go theorem. Next we will place further restrictions on the possible emanant symmetry group from the presence of time-reversal symmetry.

\subsubsection{Time-reversal and symmetry-protection}
\label{sec:UVtoIRsymm}

Given condition~\ref{eq:center}, under $\bar\rho$ all elements $g_{\mathrm{other}}\in G_{\mathrm{other}}\subset G_{\mathrm{UV}}$ map into 
$$\bar\rho(G_{\mathrm{other}})\subseteq\frac{U(1)_A\times U(1)_V}{\mathbb{Z}_2}\quad,$$
such that its image of its elements under $\bar\rho$ can be expressed as
\begin{align}
    \bar\rho(g_{\mathrm{other}})=e^{i\left(\alpha\mathds{1}+\beta\tau^z\right)}\quad,
    \label{eq:barcrystal}
\end{align}
for some $\alpha,\beta\in \mathbb{R}/2\pi\mathbb{Z}$.
As long as there exists an element $g_{\rm other}\in G_{\mathrm{other}}$ such that $\alpha\neq 0\mod \pi$, the Dirac fermion is symmetry-protected, since all fermionic masses are then prohibited due to the presence of a non-trivial subgroup of $U(1)_A$.

In the presence of time-reversal symmetry, there may be additional restrictions on Eq.~\ref{eq:barcrystal}. Consider two separate cases
\begin{enumerate}
    \item $g_{\rm other}=g_{\rm int}$ for some $g_{\rm int}\in G_{\rm internal}$.
    \item $g_{\rm other}=g_{\rm crystal}$ for some $g_{\rm crystal}\in G_{\rm crystal}$.
\end{enumerate}

In the first case, where $g_{\rm other}$ is an internal symmetry element $g_{\rm int}$, we have
\begin{align*}
    \bar\rho(\Theta_{\rm UV} g_{\mathrm{int}}\Theta_{\rm UV}^{-1})&=\Theta e^{i(\alpha\mathds{1}+\beta\tau^z)}\Theta^{-1}\\
    &=e^{-i(\alpha\mathds{1}+\beta\tau^z)}=\rho(g_{\rm int}^{-1})\quad,
\end{align*}
which is consistent with Eq.~\ref{eq:timereversalinternal} without having to restrict $\alpha$ or $\beta.$ Thus, time-reversal does not further restrict the $\rho$ map for cases when $g_{\rm other}=g_{\rm int}$. In order to have symmetry protection we will assume that for the case of an internal symmetry $\bar\rho(g_{\mathrm{other}})=e^{-iq\mathds{1}},$ where we have left off the $\tau^z$ generator since that can be removed via a $U(1)_{V,{\rm IR}}$ transformation. We will apply this result below after considering the crystal symmetry case.

In the second case, $g_{\rm other}=g_{\rm crystal}$, we have
$$\bar\rho(\Theta_{\rm UV} g_{\mathrm{crystal}}\Theta_{\rm UV}^{-1})=\Theta e^{i(\alpha\mathds{1}+\beta\tau^z)}\Theta^{-1}=e^{-i(\alpha\mathds{1}+\beta\tau^z)}\,.$$
Recall that from Eq.~\ref{eq:crystalTheta} we have the relationship
$$\bar\rho(\Theta_{\rm UV} g_{\mathrm{crystal}}\Theta_{\rm UV}^{-1})=\bar\rho(g_{\mathrm{int}}\,g_{\mathrm{crystal}})=\bar\rho(g_{\mathrm{int}})\bar\rho(g_{\mathrm{crystal}})\,.$$
Without loss of generality, we may assume that $\bar\rho(g_{\mathrm{int}})=1$ since a non-trivial map for $g_{\mathrm{int}}$ reduces to the first case.
Since the group elements are complex exponentials, their arguments are defined modulo $2\pi$. This means that the relationship
$$\bar\rho(\Theta_{\rm UV} g_{\mathrm{crystal}}\Theta_{\rm UV}^{-1})=\bar\rho(g_{\mathrm{crystal}})\quad,$$
implies
$$\alpha\pm\beta=0\mod \pi\quad.$$ This equation is determined by noting that the argument of the exponential is a diagonal matrix (necessarily from the centralizer condition), and both signs have to be simultaneously satisfied. 

The only solution that protects the Dirac node is when
\begin{align}
    \alpha=\pm\beta=\pm\frac{\pi}{2}\quad,
\end{align}
(here the $\pm$ are not correlated) which implies that 
\begin{align}
    \bar\rho(g_{\mathrm{crystal}})=
    \begin{pmatrix}
        1 & 0 \\
        0 & e^{iq_c\pi}
    \end{pmatrix}\quad,
    \label{eq:nom}
\end{align}
where $q_c\in \{0,1\}$ with $q_c=0$ corresponding to a trivial solution that does not have symmetry-protection, and $q_c=1$ corresponds to the symmetry-protected solution where, without loss of generality, we have chosen $\alpha=\frac{\pi}{2},\beta=-\frac{\pi}{2}$. This IR symmetry assigns the different flavors of Weyl fermions with different eigenvalues $\pm 1$ of $\tau^z$. Thus, the IR symmetry generates the group $\mathbb{Z}_2^R$ representing the unitary operation that takes $\psi_1\mapsto \psi_1$, $\psi_2\mapsto -\psi_2$. The symmetry-protection can be observed by the fact that the two Weyl fermions comprising the Dirac fermion have different charges cannot be coupled to open a gap. The above analysis implies that, in the case of crystalline symmetries, the only emanant symmetry group that symmetry-protects the Dirac fermion and is consistent with time-reversal symmetry is given by
\begin{align}
    \rho(G_{\mathrm{UV}})=\frac{\left(\mathbb{Z}_2^R\times U(1)_{V}\right)\rtimes \mathbb{Z}_4^T}{\mathbb{Z}_2}\quad,
    \label{eq:centerIRgroup}
\end{align}
where $\mathbb{Z}_2^R$ is sufficient to forbid the mass terms mentioned in Table~\ref{tab:symtomass}, and the $\mathbb{Z}_2$ quotient is of the element $(e^{i\pi\tau^z},-1)\in U(1)_V\rtimes \mathbb{Z}_4^T$.

For both the internal and crystal symmetry cases, the symmetry protection mechanism for the gapless Dirac fermion involves $\bar{\rho}$ mapping into non-trivial elements of the chiral $U(1)_A$ group. However, we must be careful because this group possesses an associated chiral anomaly, which means that a lattice model possessing a single gapless symmetry-protected Dirac fermion obeying the condition~\ref{eq:center}, will also always enjoy an emanant IR chiral anomaly. Such an anomaly will have to be matched in the UV. We will explore this anomaly in the next section, and our analysis will allow us to derive the relevant no-go theorems.

\subsubsection{IR anomaly associated to emanant symmetries}
\label{sec:IRanomalies}

Assuming condition~\ref{eq:center}, let us now discuss the IR anomaly. The IR anomaly of the emanant symmetry group $\frac{U(1)_V\times U(1)_A}{\mathbb{Z}_2}$ (or a non-trivial subgroup, such as in Eq.~\ref{eq:centerIRgroup}) for a 3+1d Dirac fermion consists of a
$\frac{U(1)_V\times U(1)_A}{\mathbb{Z}_2}$ chiral anomaly and a pure $U(1)_A$ gravitational anomaly. This anomaly is generally written in terms of the gauge fields of the left $U(1)_L$ and right $U(1)_R$ symmetries in $U(2)$, which act as
\begin{align}
    U(1)_R\times U(1)_L:\begin{pmatrix}
        \psi_1 \\
        \psi_2
    \end{pmatrix}\mapsto \begin{pmatrix}
        e^{i\theta_L} & 0 \\
        0 & e^{-i\theta_R}
    \end{pmatrix}
    \begin{pmatrix}
        \psi_1 \\
        \psi_2
    \end{pmatrix}\,,
\end{align}
since, strictly speaking, $U(1)_V\times U(1)_A$ requires a $\mathbb{Z}_2$ quotient that causes the associated gauge fields to depend on each other.

As in our discussion of the Nielsen-Ninomiya theorem above, it is convenient to encode these anomalies on the boundary of a $U(1)_R\times U(1)_L$ symmetry-protected topological (SPT) insulator in 4+1d via the bulk-boundary correspondence~\cite{CALLAN1985427}. In this language, the anomaly is characterized by the topological response action~\cite{RevModPhys.88.035001}
\begin{align}
    \Omega_{\mathrm{IR}}=&\frac{1}{2} \int \frac{1}{3}\left[A_R\wedge\left(\frac{dA_R}{2\pi}\right)^2-A_L\wedge\left(\frac{dA_L}{2\pi}\right)^2\right]\nonumber\\
    &+\frac{1}{2}\int\frac{1}{96\pi^2}(A_R-A_L)\wedge\mathrm{Tr}\left[R\wedge R\right]\quad,
    \label{eq:Omega}
\end{align}
where $A_{R,L}$ are the one-form gauge fields associated with the $U(1)_{R,L}$ symmetries, $R$ is the Riemann curvature, and the trace is over the Lorentz indices of the Riemann curvature~\cite{10.21468/SciPostPhysLectNotes.62}. The first two terms in Eq. \ref{eq:Omega} are associated with the $U(1)_R\times U(1)_L$ symmetries, while the third term is the mixed axial-gravitational anomaly term for which $\frac{1}{2}\int_\mathcal{M}\frac{1}{96\pi^2}\mathrm{Tr}[R\wedge R]\in 2\mathbb{Z}$ on a closed four-dimensional spin manifold $\mathcal{M}$~\cite{RevModPhys.88.035001}.

To implement such a symmetry-protected Dirac theory in the UV, we must use what we learned in Section~\ref{sec:UVtoIRsymm} on how the UV symmetries embed into the IR symmetries to match the IR and UV anomalies. It is in this procedure that we will find irreconcilable contradictions that result in the desired no-go theorems.

\subsubsection{Gauge field and anomaly matching}
\label{sec:gaugefieldmatch}

To match the anomalies, we must first match the UV gauge fields with their respective counterparts in the IR. In the UV, we will be dealing with the symmetry gauge fields associated with $G_{\rm UV}$. This includes, by assumption, the $U(1)_{V,{\rm UV}}$ gauge field $A$, as well as the gauge field $\tilde{z}$ associated to $g_{\rm other}$, which may be an internal symmetry or a crystalline symmetry.

In the more familiar case of an internal symmetry, i.e., $g_{\rm other}=g_{\rm int}$, the gauge field $\tilde{z}$ is a $G_{\rm other}$-valued gauge field. Wilson loops of such gauge fields $e^{i\oint \tilde{z}}=e^{i\alpha}\in G_{\rm other}$ measure the symmetry flux trapped in the loop. From Section~\ref{sec:UVtoIRsymm} we know that in a single Dirac node system that is protected by an internal symmetry there must exist an element $g_{\rm int}$ such that $\bar\rho(g_{\rm int})=e^{-iq\mathds{1}}$ for some $q\neq 0\mod \pi$. The flux  $\alpha\in \mathbb{R}/2\pi\mathbb{Z}$ if $g_{\rm int}$ is an element of a continuous symmetry group such that $q\in \mathbb{Z}$, and $\alpha\in \mathbb{Z}_N$ (for some $N\in \mathbb{N}\backslash\{0,1,2\}$ if we want symmetry protection) if $g_{\rm int}$ is an element of a discrete symmetry group such that $q\in\frac{2\pi \mathbb{Z}_N}{N}.$ 

To connect the UV and IR gauge fields we consider the mapping
\begin{align}
    \bar\rho( e^{i\theta \hat{Q}_V} g_{\rm int}^\alpha):
    \begin{pmatrix}
        \psi_1 \\
        \psi_2
    \end{pmatrix}\mapsto e^{i\left(\theta\tau^z-q\alpha\mathds{1}\right)}
     \begin{pmatrix}
        \psi_1 \\
        \psi_2
    \end{pmatrix}
    \quad.
    \label{eq:rhomapU(1)internal}
\end{align}
 Via Eqs.~\ref{eq:gaugefieldsmatch} and \ref{eq:gaugefieldrelation}, we may read off the relationship between the UV gauge fields $A$, $\tilde{z}$, and the IR gauge fields $A_{R,L}$ to be 
\begin{align}
    A_R= A+q\tilde{z}\quad, && A_L= A-q\tilde{z} \quad,
    \label{eq:matchinginternalA}
\end{align}
for the internal symmetry gauge field $\tilde{z}$.
Using Eqs.~\ref{eq:emergeability}, \ref{eq:matchinginternalA}, and \ref{eq:Omega} one can derive the form of the UV anomaly term $\Omega^{int}_{\rm UV}$ to be
\begin{align}
    \Omega^{int}_{\mathrm{UV}}
    =&q\int \tilde{z}\wedge\bigg[\left(\frac{dA}{2\pi}\right)^2+\frac{q^2}{3}\left(\frac{d\tilde{z}}{2\pi}\right)^2+\frac{1}{96\pi^2}\mathrm{Tr}[R\wedge R]\bigg],
    \label{eq:OmegaUVinternal}
\end{align}
where $q\neq0\mod\pi$ for a symmetry-protected system, and the quantization of $q$ depends on the symmetry group $G_{\rm internal}$, as discussed above. Notice that this term always represents a non-trivial higher-dimensional $U(1)_{V,{\rm UV}}\times G_{\rm internal}$ SPT bulk, and thus describes a quantized 3+1d non-trivial anomaly term on the boundary. We will see the consequences of this anomaly term below, after we consider the anomaly for the crystal symmetry case.

In the second case, where $g_{\rm other}=g_{\rm crystal}$ instead of an internal symmetry, one can similarly construct a crystalline gauge field $\tilde{z}$, corresponding to UV crystalline symmetry $g_{\rm crystal}$, by following the gauging procedure of space group symmetries, outlined in Ref.~\cite{PhysRevX.8.011040}. The crystalline gauge field $\tilde{z}$ is a $\mathbb{Z}$-valued gauge-field for translation or non-symmorphic crystalline symmetries, i.e., $\tilde{z}\in H^1(\mathcal{M},\mathbb{Z})$ for a manifold $\mathcal{M}.$ Alternatively, $\tilde{z}$ is a $\mathbb{Z}_n$ gauge field for point group symmetries, i.e., $\tilde{z}\in H^1(\mathcal{M},\mathbb{Z}_n)$. Physically, such a gauge field measures the associated crystalline defects (such as dislocations for translation symmetry or disclinations for rotation symmetry). These defects, also referred to as crystalline \textit{fluxes} $e^{i \oint \tilde{z}}\in G_{\rm crystal}$, analogous to how internal gauge fields such as $U(1)_{V,{\rm UV}}$ measure internal defects such as $U(1)_{V,{\rm UV}}$ magnetic fluxes. 

To connect the UV and IR gauge fields observe that, using Eq.~\ref{eq:nom}, we have
\begin{align}
    \bar\rho(e^{i\theta \hat{Q}_V}g_{\rm crystal}^\alpha):
    \begin{pmatrix}
        \psi_1 \\
        \psi_2
    \end{pmatrix}\mapsto 
    \begin{pmatrix}
        e^{i\theta} & 0 \\
        0 & e^{-i\left(\theta+q_c \pi \alpha\right)}
    \end{pmatrix}
    \begin{pmatrix}
        \psi_1 \\
        \psi_2
    \end{pmatrix}
    ,
    \label{eq:rhomapU(1)crystal}
\end{align}
for some $\alpha\in \mathbb{Z}$. Again, using Eqs.~\ref{eq:gaugefieldsmatch} and \ref{eq:gaugefieldrelation}, one can conclude
\begin{align}
    A_R= A+q_c \pi \tilde{z}\quad, && A_L= A \quad,
    \label{eq:matchingcrystalA}
\end{align}
for a crystalline symmetry gauge field $\tilde{z}$, which counts the enclosed number of $g_{\rm crystal}$ defects such that $\oint \tilde{z}=\alpha$. Recall that $q_c=1 (0)$ for a crystalline symmetry-protected (fine-tuned) system.

In combination with Eqs.~\ref{eq:emergeability} and \ref{eq:Omega}, one can derive the UV anomaly term, $\Omega^{crys}_{\mathrm{UV}}$, to be
\begin{align}
    \Omega^{crys}_{\mathrm{UV}}
    =& \frac{q_c\pi}{2}\int \tilde{z}\wedge\bigg[\left(\frac{dA}{2\pi}\right)^2+\frac{1}{12}\left(d\tilde{z}\right)^2+\frac{1}{2}d\tilde{z}\left(\frac{dA}{2\pi}\right)\bigg],
    \label{eq:OmegaUVsimp}
\end{align}
where we have removed the mixed $U(1)_A$-gravitational anomaly since it always takes value in $2\pi$ and is thus trivial. Just like in the internal symmetry case, this UV anomaly term is quantized and always non-trivial, representing a non-trivial $G_{\rm space}\times U(1)_{V,{\rm UV}}\rtimes\Theta$ SPT state in the higher-dimensional bulk.

Let us comment on how time-reversal affects the above relationship between IR and UV ($U(1)_{V,{\rm UV}}$ and crystalline) gauge fields. Under time-reversal the internal $U(1)_{f}$ gauge fields $A_{f}$, with $f\in\{R,L\}$, transform as~\cite{Weinberg_1995}
\begin{align}
    \Theta:\quad
    \begin{cases}
        A^0_{f}(t)\mapsto A^0_{f}(-t)\\
        A^j_{f}(t)\mapsto -A^j_{f}(-t)
    \end{cases}\quad,
    \label{eq:TRtransformation}
\end{align}
where $j\in\{1,2,3\}$, and one can observe that $A^0_{f}$ is the scalar potential and $A^j_{f}$ is the vector potential associated with $U(1)_{f}$. This notation may be related to the more-compact 1-form notation via $A_{f}:= A^0_{f}dt+A^1_{f} dx+A^2_{f} dy+A^3_{f} dz$. The time-reversal transformation leaves the $U(1)_R\times U(1)_L$ chiral anomaly in Eq.~\ref{eq:Omega} invariant~\footnote{Note that in existing literature, time-reversal is often said to act as $A_f\mapsto-A_f$, which is true upon the substitution of variable $t\mapsto-t$. Such a short-hand method is a convenient way to check for the presence or absence of a symmetry.}. In the UV, the $U(1)_{V,{\rm UV}}$ gauge field $A$ transforms in the same fashion as in Eq.~\ref{eq:TRtransformation}. However, the crystalline symmetry gauge field $\tilde{z}$ transforms under time-reversal as
\begin{align}
    \Theta:\quad
    \begin{cases} 
    \tilde{z}^0(t)\mapsto -\tilde{z}^0(-t)\\
    \tilde{z}^j(t)\mapsto \tilde{z}^j(-t)
    \end{cases}
    \quad,
\end{align}
where $\tilde{z}:=\tilde{z}^0 dt+\tilde{z}^1 dx+\tilde{z}^2 dy+ \tilde{z}^3 dz$. Thus, the matching between IR and UV gauge fields is time-reversal symmetric only when $\tilde{z}$ couples to a prefactor $0\mod 2\pi$ or $\pi\mod 2\pi$, as is the case in Eq.~\ref{eq:matchingcrystalA}.

\subsubsection{Obstructions resulting in the no-go}
\label{sec:nogo}

To interpret the physical implications and resulting obstructions of the actions in Eqs.~\ref{eq:OmegaUVinternal} and \ref{eq:OmegaUVsimp}, we can separately consider the possibilities of the gauge field $\tilde{z}$:

(a) \textit{On-site internal symmetry gauge field $\tilde{z}$}: Here, the system would be subject to the non-trivial UV anomaly given in Eq.~\ref{eq:OmegaUVinternal}. This anomaly would imply an obstruction to gauging both $U(1)_{V,{\rm UV}}$ and the $g_{\rm int}$ symmetry simultaneously in 3+1d. However, completely analogous to the Nielsen-Ninomiya case, all on-site symmetries on the lattice have a well-defined gauging procedure, and are thus automatically gaugeable. We arrive at a contradiction unless $q=0\mod 2\pi.$ Hence, to avoid the contradiction we must choose $q=0\mod 2\pi,$ thus the Dirac fermion is fine-tuned, i.e., not symmetry-protected.

(b) \textit{$\mathbb{Z}_{2N}$ crystalline gauge field $\tilde{z}$ with $N\in\{1,2,3\}$~\footnote{On lattices, the only possible point groups are $\mathbb{Z}_{2,3,4,6}$. A consequence of our analysis in Section~\ref{sec:UVtoIRsymm} is that in time-reversal symmetric systems, a single Dirac node cannot possess non-trivial $3$-fold rotation symmetry charge if it lies on a 3-fold invariant axis~\cite{Zaheerthesis}. This is because the $\bar\rho(g_{\rm crystal})\in \mathbb{Z}_2^R$ such that the only possible map of $\mathbb{Z}_3$ that is a group homomorphism is the trivial map (all elements map to the identity). So it suffices to just consider $\mathbb{Z}_{2N}$ point group gauge fields.}}: If $\tilde{z}$ is a $\mathbb{Z}_{2N}$ point group gauge field, like $2N$-fold rotation, then we must instead consider the fixed points, e.g., rotation center for rotation symmetries, of the symmetries where an analogous argument holds. Fixed points correspond to choosing $\int d\tilde{z}=2N$~\footnote{Recall that for a lattice without disclination defects $\frac{d\tilde z}{2N}\in H^{2}(\mathcal{M},\mathbb{Z})$, where $\mathcal{M}$ is the manifold.} such that the first term of the anomaly in Eq.~\ref{eq:OmegaUVsimp} would predict an integer quantum Hall term
\begin{align}
    \frac{q_c N}{2}\int A\wedge\frac{dA}{2\pi}=\frac{q_c N}{4\pi}\int d^3x\, \epsilon^{\mu\nu\rho}A_\mu\partial_\nu A_\rho\quad,
    \label{eq:disclinationterm}
\end{align}
which is a non-trivial bulk $U(1)_{V,{\rm UV}}$ SPT. Again, this would imply an obstruction to gauging the $U(1)_{V,{\rm UV}}$ symmetry on the boundary (corresponding to the rotation center), which leads to a contradiction unless $q_c=0$, since on-site symmetries are always gaugeable. Thus, we see the system is necessarily fine-tuned. Another physical way to observe the contradiction is to notice that when $q_c=1$ this term requires $N$ chiral modes at the boundary, i.e., along the  rotation center, which is not possible with on-site $U(1)_{V,{\rm UV}}$~\cite{Nielsen83}. Eq.~\ref{eq:disclinationterm} also describes a gauge anomaly on the 1+1d edge which is an obstruction to gauging the $U(1)_{V,{\rm UV}}$ symmetry. However, once again, in a lattice system $U(1)_{V,{\rm UV}}$ is fully gaugeable and results in a contradiction, unless the Dirac fermion is fine-tuned.

(c) \textit{$\mathbb{Z}$ crystalline gauge field $\tilde{z}$}: If $\tilde{z}$ is a $\mathbb{Z}$-valued crystalline gauge field, like in the case of translation or a non-symmorphic symmetry, then we see that the first 4+1d SPT term in Eq.~\ref{eq:OmegaUVsimp}~\footnote{The second and third terms describe $U(1)_{V,{\rm UV}}$ and momentum filling along dislocation defects~\cite{PhysRevResearch.3.043067,PhysRevX.12.031007,PhysRevB.98.085140}. However, these terms will not be important for the following analysis of the no-go theorem.} describes $\hat{z}$-layers~\footnote{This holds for both the usual translation as well as non-symmorphic symmetries which can be thought of as a half (or fractional) translation in this context.} of the 3+1d `t Hooft anomaly
\begin{align}
    \frac{q_c\pi}{2}\int\frac{dA}{2\pi}\wedge \frac{dA}{2\pi}=\frac{q_c\pi}{8\pi^2}\int d^4x\,\epsilon^{\mu\nu\rho\kappa}\partial_\mu A_\nu \partial_\rho A_\kappa\quad,
    \label{eq:layeranomaly}
\end{align}
which is the topological response of the 3+1d $U(1)_{V,{\rm UV}}\rtimes \Theta_{\rm UV}$ fermionic SPT. Hence, the 4+1d action represents a translation-invariant stack of 3+1d topological insulators. The boundary of such a system will be a stack of surfaces of these 3+1d topological insulators. Again, the question reduces to whether this boundary can be realized in an on-site $U(1)_{V,{\rm UV}}\rtimes \Theta_{\rm UV}$ symmetric system. However, it is well-known that the surface of a single 3+1d topological insulator corresponds to a parity anomaly (also sometimes referred to as a T-anomaly). 

Connecting to our previous discussion of anomalies, the logic that anomalies represent obstructions to gauging still holds --- however, since we are dealing with a global anomaly in this case (as opposed to a perturbative anomaly), it is more subtle to deal with and easier to use arguments such as an obstruction to a symmetric system.
Indeed, this anomaly implies that there is an obstruction to having both an on-site time-reversal and on-site $U(1)_{V,{\rm UV}}$ symmetric theory in 2+1d~\cite{RevModPhys.88.035001}. A stack of such systems will still harbor this anomaly. Hence, there is a contradiction unless $q_c=0$, which means that the single Dirac fermion system is necessarily fine-tuned and, in fact, uncharged under translation or any other $\mathbb{Z}$ crystalline symmetry.

Putting the three cases together, we derive Theorem~\ref{thm:1}, which states that under the assumed assumptions all single Dirac fermionic systems are fine-tuned.\hfill$\blacksquare$

\subsubsection{Corollaries regarding semimetallic systems}

We may derive the following two corollaries regarding Dirac and Weyl semimetals that are relevant for the condensed matter context. Here we consider systems where the low-energy theory consists only of a single 3+1d Dirac fermion, i.e., two 3+1d Weyl fermions.

(i) Consider when the left- and right-handed Weyl fermions (i.e., Weyl nodes) possess the same crystalline momentum, such that they form a single Dirac node. We know from our previous analysis that this system can be symmetry-protected only if there exists another symmetry (such as an internal, rotation, or screw symmetry) that protects the Dirac node. However, as we have previously deduced, this is not possible under the assumptions of Eq.~\ref{eq:center}.
\begin{corollary}
    In a 3+1d time-reversal and on-site $U(1)_{V,{\rm UV}}$-symmetric fermionic system, any single Dirac node system is fine-tuned, i.e., cannot be symmetry-protected even in the presence of extra microscopic symmetries, obeying condition~\ref{eq:center}. In particular, it follows that all free-fermionic systems obeying Eq.~\ref{eq:center} have no such lone symmetry-protected Dirac nodes.
    \label{thm:nodirac}
\end{corollary}
(ii) Now, consider when the left and right-handed Weyl nodes lie on different crystalline momenta. As deduced in Sec.~\ref{sec:UVtoIRsymm}, in order to be consistent with time-reversal, one node must be at $\pi$, and the other at $0$ along a momentum direction associated to a preserved translation symmetry. This translation symmetry, along with $U(1)_{V,{\rm UV}}$ automatically symmetry-protects the Weyl fermions as their momenta represent the translation symmetry charges. However, anomaly matching gives rise to a contradiction, as outlined in Section~\ref{sec:nogo}, which implies that the following theorem regarding Weyl semimetallic systems.
\begin{corollary}
    In a 3+1d time-reversal and on-site $U(1)_{V,{\rm UV}}$-symmetric fermionic system, a single pair of right and left-handed Weyl fermion, situated at different crystalline momenta, does not exist; not even as a fine-tuned system.
    \label{thm:noweyl}
\end{corollary}
This is well-known in the free fermion limit of Weyl semimetals since time-reversal invariance requires Weyl fermions at opposite crystalline momenta to be of the same handed-ness while the Weyl anomaly requires the presence of an equal amount of chiral left and right Weyl nodes, thus resulting in at least four Weyl nodes in time-reversal symmetric systems. However, the above theorem also holds even for strongly-interacting systems, demonstrating the non-perturbative power of anomaly arguments.

\subsection{Case 2: $U(1)_{V,{\rm UV}}\triangleleft G_{\mathrm{UV}}$}
\label{sec:case2}

A generalization of Case 1 is to instead assume that $U(1)_{V,{\rm UV}}$ is a normal subgroup of $G_{\mathrm{UV}}$, written 
\begin{align}
    U(1)_{V,{\rm UV}}\triangleleft G_{\mathrm{UV}}\quad,
    \label{eq:normal}
\end{align}
which implies that all other elements $g\in G_{\mathrm{UV}}$ act on $a\in U(1)_{V,{\rm UV}}$ as $g:a\mapsto g a g^{-1}\in U(1)_{V,{\rm UV}}$. This condition is strictly broader than our previous condition in Section~\ref{sec:case1symm}, where we demanded that all UV symmetry elements commute with the charge generator of $U(1)_{V,{\rm UV}}.$ The normal subgroup condition implies that elements of $G_{\mathrm{UV}}$ can change elements of $U(1)_{V,{\rm UV}}$ but have to leave them in $U(1)_{V,{\rm UV}}.$ Examples of UV symmetry groups that obey this normal subgroup condition include charge conjugation (which was not allowed in Case 1), time-reversal, and all crystalline space-groups. Charge conjugation symmetry takes $U(1)_{V,{\rm UV}}$ elements to their inverses, while time-reversal and crystalline symmetries we have assumed to leave $U(1)_{V,{\rm UV}}$ elements unmodified. Some UV symmetry groups excluded by this condition include the symmetries generated by the Onsager algebra~\cite{Vernier_2019,chatterjee2025quantized,gioia2025exactchiralsymmetries31d} which was recently used to realize an emergent $SU(2)$ anomaly on the lattice~\cite{gioia2025exactchiralsymmetries31d}, or those that have a `non-compact' $U(1)_{V,{\rm UV}}$ symmetry, i.e., the charge generator is not quantized.

Using the condition in Eq.~\ref{eq:normal}, we may once again place restrictions on how the emanant symmetry group $\bar{\rho}(G_{\rm UV})$ falls in $\bar{G}_{\rm IR}$, given by Eq.~\ref{eq:GIRbar}. Since $U(1)_{V,{\rm UV}}$ is a normal subgroup of $G_{\mathrm{UV}}$ it follows that $\bar\rho(U(1)_{V,{\rm UV}})$ is also a normal subgroup of $\bar\rho(G_{\mathrm{UV}}).$ This is possible only if the image of $G_{\mathrm{UV}}$, i.e., $\bar\rho(G_{\mathrm{UV}})$, lies in the normalizer of $U(1)_V$ in $\bar{G}_{\mathrm{IR}}$, as given by
\begin{align}
\bar\rho(G_{\mathrm{UV}})\subseteq N_{\bar{G}_{\mathrm{IR}}}(U(1)_V)=\frac{U(1)_A\times \mathrm{Pin}^-(2)}{\mathbb{Z}_2},
\end{align}
where $N_G(S)=\{g\in G|gSg^{-1}=S\}$ is the normalizer of a set $S$ in a group $G$, the RHS was explicitly computed, $\mathrm{Pin}^-(2)$ is generated by $U(1)_V=e^{i\theta \tau^z}$ and a swap of the two flavors of Weyls via $i\tau^x$~\footnote{$\mathrm{Pin}^-(2)$ can be thought of as the double cover of $U(1)_V\rtimes \mathbb{Z}_2^C\cong O(2)$.}, and the $\mathbb{Z}_2$ quotient refers to quotienting by $(e^{i\pi},-1)\in U(1)_A\times \mathrm{Pin}^-(2)$. 

Other than the Weyl flavor swap symmetry, this group is almost identical to the IR symmetry group studied in Case 1. Hence, to find a scenario not already covered by Case 1 we need to consider symmetry protection mechanisms that use the flavor-swap symmetry in an essential way. The flavor swap symmetry \emph{anti-commutes} with the $U(1)_{V,{\rm IR}}$ generator $\tau^z$ and hence acts as a charge conjugation symmetry. Thus, we want to consider symmetry-protection using \emph{on-site} charge-conjugation type symmetries, whose group generator $g_{cc}\in G_{\rm UV}$ obeys the property
\begin{align}
    g_{cc}\hat{Q}_V g_{cc}^{-1} =-\hat{Q}_V\,, && \bar\rho(g_{cc})=e^{i\left(\alpha\frac{\tau^x}{2}+\beta\frac{\tau^y}{2}\right)}h\,,
    \label{eq:O2}
\end{align}
where either $(\alpha,\beta)=(\pi,0)$ or $(0,\pi),$ $\bar\rho$ was defined under Eq.~\ref{eq:barcrystal}, and $h\in \frac{U(1)_A\times U(1)_V}{\mathbb{Z}_2}$. As expected, we can see that charge-conjugation maps non-trivially into $\mathrm{Pin}^-(2)$, instead of just a subgroup of $\frac{U(1)_A\times U(1)_V}{\mathbb{Z}_2}$ (which was already covered in Section~\ref{sec:case1symm}).
We note that we have excluded UV symmetry groups that do not obey the assumption in Eq.~\ref{eq:normal} and the on-site condition. Such excluded groups include situations where charge-conjugation and a crystalline symmetry are broken, but their combination is preserved in $G_\mathrm{UV}$, since these are not on-site.

The potential use of the emanant $\mathrm{Pin}^-(2)$ group is a different mechanism than the symmetry-protection via an emanant $U(1)_A$ subgroup and its associated chiral anomaly, as was used in Section~\ref{sec:case1symm}. Although we expect that a Nielsen-Ninomiya-like anomaly matching argument, as in Section~\ref{sec:NNthm} and \ref{sec:case1symm}, should work, it is still unclear how to express the relevant anomaly as a topological term. Therefore, we will resort to a different argument to show an obstruction to a single time-reversal invariant, symmetry-protected Dirac node.

\subsubsection{Chern insulator pump}

In this section, we will employ a different, physically intuitive, albeit less rigorous, argument to show there is a contradiction if we assume a symmetry-protected single Dirac fermion, $U(1)_{V,{\rm UV}}\triangleleft G_{\mathrm{UV}}$, and that the $G_{\mathrm{UV}}$ elements responsible for the symmetry-protecting mechanism are on-site. To observe the contradiction, consider the parameter-dependent mass term
\begin{align}
    M(\alpha,\beta)=\Psi^\dag \alpha\left(\cos\beta \,\gamma^0+\sin \beta \,i\gamma^0\gamma^5\right)\Psi\quad,
    \label{eq:massparameter}
\end{align}
where $\alpha$ and $\beta$ are parameters. This term preserves $U(1)_{V,{\rm UV}}$, but for generic parametrization breaks the chiral, charge conjugation, and time-reversal symmetries. A special case of this parametrization comes in the form of a vortex line in the $\hat{z}$-direction when we choose $\alpha=\Delta \tanh(r)$ and $\beta=\theta$, with $r=\sqrt{x^2+y^2}$, and $\theta$ is the angular component in the $xy$ plane, such that
\begin{align}
    M(\theta)=\Psi^\dag \Delta\tanh(r)\left(\cos\theta \,\gamma^0+\sin \theta \,i\gamma^0\gamma^5\right)\Psi\quad,
    \label{eq:massvortex}
\end{align}
where $\Delta>0$. It is well-known that such a mass vortex for a massive 3+1d Dirac fermion traps a single 1+1d complex chiral fermion in the center~\cite{CALLAN1985427,Bagherian2024}.

Let us first demonstrate that this is the case using a Chern insulator pumping argument. Recall from Section~\ref{sec:IRanomalies} that a massless Dirac fermion enjoys a $U(1)_R\times U(1)_L$ mixed chiral anomaly term
\begin{align}
    \Omega_{\mathrm{IR}}=&\frac{1}{2} \int \frac{1}{3}\left[A_R\wedge\left(\frac{dA_R}{2\pi}\right)^2-A_L\wedge\left(\frac{dA_L}{2\pi}\right)^2\right]\quad,\nonumber\\
    =& \int_{\mathcal{M}} A_A \wedge\left(\frac{dA_V}{2\pi}\right)^2+(...)\mod 2\pi\quad,\label{eq:diracpumpanomaly}
\end{align}
where we have rewritten the anomaly in terms of the $U(1)_V$ and $U(1)_A$ gauge fields $A_V$ and $A_A$, via $A_{R}=A_V+ A_A$, $A_{L}=A_V- A_A$, with $\mathcal{M}$ being a 4+1d manifold. The $(...)$ includes all the other terms upon such a substitution that can be safely ignored.

Adding a mass vortex of the form in Eq.~\ref{eq:massvortex} breaks the $U(1)_A$ symmetry. This means that the $U(1)_A$ gauge field becomes a Higgsed gauge field, with the Landau-Ginsburg action
$$S_{\rm LG}=\int d^4x |\alpha|^2(\partial_\mu \beta - 2A_{A,\mu})^2\quad,$$
such that the mass vortex in Eq.~\ref{eq:massvortex} is accompanied by a flux in $U(1)_A$~\cite{Zee_book}, i.e.,
\begin{align}
    2A_A=d\beta\quad.\label{eq:Ahiggs}
\end{align}
The factor of 2 comes from the fact that the mass has charge 2 under $U(1)_A$. Using this relationship alongside Eq. \ref{eq:diracpumpanomaly}, the anomaly term becomes via Stokes theorem
\begin{align}
    \Omega_{\mathrm{IR}}=&\frac{1}{2} \int_{\partial \mathcal{M}} \frac{d\beta}{2\pi}\wedge A_V\wedge\frac{dA_V}{2\pi}\quad,
    \label{eq:phiIR}
\end{align}
where $\partial \mathcal{M}$ is the 3+1d boundary of $\mathcal{M}$. Choosing the vortex configuration, mentioned above, with $\beta=\theta$, this term physically means that a Chern insulator with Chern number 1 is pumped throughout the system as we sweep from $\theta=0$ to $2\pi$. Recall that in the polar coordinate system there exists a singularity at $x=y=0$ such that
$$d(d\theta)=2\pi \delta^{(2)}(x,y)dx\wedge dy\quad,$$ 
where $\delta^{(2)}$ is the 2d Dirac delta function. Using this relationship, upon a gauge transformation $A_V\rightarrow A_V+d\phi$, the variation of Eq.~\ref{eq:phiIR} is given by
\begin{align}
    \delta\Omega_{\mathrm{IR}}=& \int \frac{d\theta}{2\pi} \wedge d\phi\wedge \frac{dA_V}{2\pi}=\int d^4x\,\delta^{(2)}(x,y)\phi\frac{E_z}{2\pi},
    \label{eq:vortexgaugenoninv}
\end{align}
where $E_z=\partial_t A_{Vz}-\partial_z A_{Vt}$ is the electric field in $\hat{z}$. We see that the system is not gauge-invariant at the vortex center $x=y=0$, corresponding exactly to the gauge non-invariance of a single 1+1d chiral fermion. This argument shows that there is a 1+1d chiral fermion at the center of the mass vortex which is in agreement with the results of References~\cite{CALLAN1985427,Bagherian2024}.

How will this lead to a contradiction with our assumptions? First, suppose that $U(1)_{V,{\rm UV}}\triangleleft G_{\mathrm{UV}}$. We will now show that each unitary $g \in G_{\mathrm{UV}}$ acts as a shift $\beta \mapsto \beta + \xi(g)$ in Eq.~\ref{eq:massparameter}. Indeed, for $h\in U(1)_{V,{\rm UV}}$ and $g\in G_{\mathrm{UV}}$ we have
$$g^{-1} h g= \tilde{h}\quad, $$
for some $\tilde{h}\in U(1)_{V,{\rm UV}}$. Thus, if we act on a $U(1)_{V,{\rm UV}}$ invariant mass with $g$, we get another $U(1)_{V,{\rm UV}}$ invariant mass
\begin{align*}
    h g M(\alpha,\beta) g^{-1} h^{-1}&= g \tilde{h} M(\alpha,\beta) \tilde{h}^{-1}g^{-1}\quad, 
    \\
    &=g M(\alpha,\beta) g^{-1}\quad,
\end{align*}
for all $h,\tilde{h}\in U(1)_{V,{\rm UV}}$.
The two components of $M(\alpha,\beta)$ cover the entire family of $U(1)_{V,{\rm UV}}$ invariant masses, and $g$ acts unitarily on them, yielding at most a rotation of $\beta$, i.e., $gM(\alpha,\beta)g^{-1}=M(\alpha,\beta+\xi(g))$.

In order for our Dirac fermion to be symmetry-protected, we know that there must exist some unitary element\footnote{If there is a second anti-unitary time-reversal symmetry, we can combine it with our usual time-reversal symmetry $\Theta_{\rm UV}$ to get this protecting unitary symmetry.} $g_{\mathrm{sym}}\in G_{\mathrm{UV}}$ which acts non-trivially on $M(\alpha,\beta)$, such that $\xi(g_{\rm sym}) \neq 0\mod 2\pi$.
Now, let us assume such a symmetry-protected single Dirac fermionic system, obeying the assumptions in Section~\ref{sec:case2}. This implies that there exists a $g_{\mathrm{sym}}\in G_{\mathrm{UV}}$ with the property
\begin{align}
    g_{\mathrm{sym}}:\quad M(\alpha,\beta)\mapsto M(\alpha,\beta+\xi)\quad,
\end{align}
where $\xi=\frac{2\pi}{N}$ for some $N\in\{2,3,4,...\}$~\footnote{Finding such an element $g_{\rm sym}\in G_{\rm UV}$ with a rational $\xi$ is always possible since $G_{\rm UV}$ is assumed to be compact, i.e., there always exists some $\mathbb{Z}_N\subseteq G_{\rm UV}$ with $N\in\{2,3,4,...\}$.}.
Observe that $g_{\mathrm{sym}}$ is an on-site symmetry that relates states at $\beta$ with $\beta+\xi$. This means that the gapped phases at the two points $\beta$ and $\beta+\xi$ are equivalent upon gauging $g_{\mathrm{sym}}$~\footnote{The gauging procedure may be performed by assuming that the symmetry-breaking masses actually arise from spontaneous symmetry-breaking of $g_{\mathrm{sym}}$ rather than explicit symmetry breaking.}, i.e., they are related by pumping of an invertible state with Chern number $k$~\footnote{A Chern number is well-defined even when gauging charge conjugation (which anti-commutes with the $U(1)_{V,{\rm UV}}$ charge) since the term is quadratic in $A_V$, i.e. invariant under charge conjugation. Invertibility follows since pumping from $\beta+\xi\rightarrow \beta$ inverts the pumping from $\beta\rightarrow \beta+\xi$.}, with $k\in\mathbb{Z}$. It follows that as we wind from $0$ to $2\pi$, this process happens $N$ times, pumping $Nk$ Chern insulators. This corresponds to
\begin{align}
    \Omega_{\mathrm{IR}}=&\frac{Nk}{2} \int_{\partial \mathcal{M}} \frac{d\beta}{2\pi}\wedge A_V\wedge\frac{dA_V}{2\pi}\quad.
    \label{eq:gaugequant}
\end{align}
However, since $N > 1$, then $|Nk| > 1$,\footnote{We assumed $k\neq 0.$ If $k=0$ we still find a contradiction since it would imply zero chiral fermions in the mass vortex core.} and hence this term contradicts Eq.~\ref{eq:phiIR}, since, following the same calculations below Eq.~\ref{eq:phiIR}, the core of a mass vortex (Eq.~\ref{eq:massvortex}) would possess $|Nk|$ chiral fermions as opposed to $1$, thus creating a contradiction with the known results of References~\cite{CALLAN1985427,Bagherian2024}.

This argument shows that such a $g_{\mathrm{sym}}$ does not exist, which means that all lattice systems having a single Dirac fermion are fine-tuned under condition~\ref{eq:normal} with the extra assumption that $g_{\mathrm{sym}}$ is on-site.
Note that in this argument, it is crucial for $g_{\mathrm{sym}}$ to be on-site such that it can be gauged, leading to the derivation of Eq.~\ref{eq:gaugequant}. This means that combinations of particle-hole with crystalline symmetries are not covered by this argument, even though there are no concrete time-reversal invariant lattice examples of a single Dirac node protected by such a symmetry. However, this analysis allows us to rule out on-site particle-hole symmetry mechanisms to protect single Dirac nodes, in addition to the symmetries covered in Case 1.

We will discuss some of the possible realizations of a single Dirac node that are converses of these theorems by (1) breaking a relevant symmetry (e.g., time-reversal broken Weyl semimetal), (2) realizing a symmetry in a non-on-site and non-compact manner (e.g., almost-local Dirac node), or (3) a fine-tuned single Dirac node situation such that no symmetry protects the gaplessness.

\section{Converse examples}
\label{sec:circumvent}

In the following section, we will construct 3+1d free-fermion Hamiltonian examples of single (or sometimes multiple) Dirac nodes that are \textit{converse} to the no-go theorems in Sec.~\ref{sec:proof} by breaking certain assumptions. Such examples will help show that the theorems are tight and aid the reader in distinguishing known Hamiltonians and how they fall within the context of our framework.

To facilitate our discussion, we remind the reader that a free-fermionic second-quantized Hamiltonian with on-site $U(1)_{V,{\rm UV}}$ symmetry takes the form
\begin{align}
    \mathcal{H}=\sum_{ \mathbf{r},\mathbf{r}', s,s'}c_{\mathbf{r},s}^\dag H(\mathbf{r},\mathbf{r}')_{ss'}c_{\mathbf{r}',s'}\quad,
\end{align}
where $c_i^\dag$ and $c_i$ are fermionic creation and annihilation operators of the state $i$, obeying the relationships $\{c_i^\dag,c_j\}=\delta_{ij}$ and $\{c_i^\dag,c_j^\dag\}=\{c_i,c_j\}=0.$ The coordinates $\mathbf{r},\mathbf{r}'$ label the real-space position, $s,s'$ label additional unit cell degrees of freedom such as orbital and spin, and the Hamiltonian coefficients $H(\mathbf{r},\mathbf{r}')_{ss'}$ obey the Hermiticity condition $H(\mathbf{r},\mathbf{r}')_{ss'}=H(\mathbf{r}',\mathbf{r})_{s's}^*$. 

In the presence of translation symmetry, we have $H(\mathbf{r},\mathbf{r}')_{ss'}=H(\mathbf{r}-\mathbf{r}')_{ss'}$, such that we may perform a Fourier transform
\begin{align}
    c_{\mathbf{r},s}=\frac{1}{\sqrt{N_{\mathrm{tot}}}}
    \sum_{\mathbf{k}}e^{i\mathbf{k}\cdot\mathbf{r}} c_{\mathbf{k},s}\,,&& c^
    \dag_{\mathbf{r},s}=\frac{1}{\sqrt{N_{\mathrm{tot}}}}
    \sum_{\mathbf{k}}e^{-i\mathbf{k}\cdot\mathbf{r}} c^
    \dag_{\mathbf{k},s}\,,\nonumber
\end{align}
where $N_{\mathrm{tot}}$ is the total number of lattice sites. Using this basis transformation, we arrive at the Hamiltonian
\begin{align}
    \mathcal{H}=\sum_{ \mathbf{k}, s,s'}c_{\mathbf{k},s}^\dag H(\mathbf{k})_{ss'}c_{\mathbf{k},s'}\quad,
\end{align}
where $H(\mathbf{k})=\sum_{\mathbf{r}-\mathbf{r}'}H(\mathbf{r}-\mathbf{r}')e^{-i\mathbf{k}\cdot (\mathbf{r}-\mathbf{r}')}$ is the Hermitian matrix known as the Bloch Hamiltonian, with $\mathbf{k}$ taking values in the Brillouin zone (BZ). We will generally describe our Hamiltonians directly in this momentum-space, Bloch Hamiltonian language.

\subsection{Example: Fine-tuned single Dirac node}
\label{sec:finetuned}

As our first example, we will construct a 3+1d time-reversal symmetric free-fermionic system with a single, fine-tuned Dirac node. An example is the following Bloch Hamiltonian
\begin{align}
    H_{\mathrm{FT}}(\mathbf{k})=&\sin k_x \sigma^z s^x+\sin k_y \sigma^z s^y\nonumber\\
    &\quad+\sin k_z \sigma^z s^z+ m(\mathbf{k})\sigma^x\quad,
    \label{eq:finetuned}
\end{align}
where $m(\mathbf{k})=3-\cos k_x-\cos k_y-\cos k_z.$ This Hamiltonian is symmetric under the time-reversal operator $\Theta=-i s^y\mathcal{K}$ (here, $\mathcal{K}$ is complex conjugation), and has an energy dispersion
\begin{align}
    E_{\pm}^{\alpha}(\mathbf{k})=\pm\sqrt{\left(\sin k_x\right)^2+\left(\sin k_y\right)^2+\left(\sin k_z\right)^2+m(\mathbf{k})^2}.
\end{align}
If we expand the dispersion around the $\Gamma$-point we can see that there is a single Dirac point at $\mathbf{k}=0$ with the low-energy IR theory given by Eq.~\ref{eq:IRaction}. However, such a system is always fine-tuned, even in the presence of time-reversal symmetry, as one can always add a time-reversal invariant mass term such as $\epsilon\sigma^x$ which will perturbatively open a gap of $O(|\epsilon|)$ at the $\Gamma$-point.

This Hamiltonian has $U(1)_{V,{\rm UV}}$ implemented in an on-site manner, and obeys condition~\ref{eq:center}. Hence, Theorem~\ref{thm:nodirac} shows that there is no consistent way to implement crystalline symmetries to prevent such a mass while preserving time-reversal symmetry.

\subsection{Example: Almost-local $U(1)$ symmetry with a single symmetry-protected Dirac node}
\label{sec:nononsiteU(1)}

In this example, we construct a single Dirac node protected by time-reversal and rotation, but with a not-on-site $U(1)$ in analogy to an approach in Ref.~\onlinecite{gioia2025exactchiralsymmetries31d}.
We begin with a model on a cubic lattice with eight Dirac nodes given by the Bloch Hamiltonian
\begin{align}
    H_{\mathrm{8DSM}}(\mathbf{k})=\sin k_x\sigma^z s^x+\sin k_y \sigma^z s^z+\sin k_z  \sigma^z s^y.
    \label{eq:8DSM}
\end{align} The spectrum of this model harbors Dirac nodes at all eight time-reversal-invariant-momenta (TRIM)\footnote{TRIM are momentum values that transform to themselves (up to reciprocal lattice vectors $\mathbf{G}$) under time-reversal symmetry $\mathbf{k}_\mathrm{TRIM}\rightarrow \mathbf{k}_\mathrm{TRIM}\mod\mathbf{G}$.} of the BZ.
This model possesses time-reversal symmetry, given by $\Theta=-is^y\mathcal{K}$, and $\pi$-rotation symmetry around the $k_z$ axis, given by operator $C_2=e^{i\frac{\pi}{2}\sigma^z s^y}$.  

In the IR, this $\pi$-rotation symmetry maps into $\bar{G}_{\mathrm{IR}}$ as
$$\bar\rho:\quad C_2\rightarrow  \mathbb{Z}_2^{R}\subseteq \bar{G}_{\mathrm{IR}}\quad,$$ i.e.,  physically this means that $C_2$ assigns a different ($0$ or $\pi$) rotational charge to opposing flavors of Weyl nodes at a given momentum. Notice that such an assignment of rotational charges is consistent with time-reversal symmetry where the $C_2$ charge generator ($C_2$ rotation) anti-commutes (commutes) with time-reversal symmetry. We note that this is the same IR symmetry group as in Eq.~\ref{eq:centerIRgroup}, which we saw is sufficient to symmetry-protect the Dirac nodes in the presence of time-reversal symmetry and $U(1)_{V,{\rm UV}}$. Hence, all the Dirac nodes are symmetry-protected. 

Our theorem does not forbid eight symmetry-protected Dirac nodes, so we want to reduce the eight nodes to just a single symmetry-protected node. In order to gap out all the Dirac nodes except the one at, say,  $\mathbf{k}=0$, we must sacrifice one of these symmetries and we will consider several cases. To consider $U(1)_{V,{\rm UV}}$-breaking terms we rewrite the Hamiltonian in the Bogoliubov-de-Gennes (BdG) formalism with the basis $$d^\dag_\mathbf{k}\equiv(c^\dag_{\mathbf{k}A\uparrow},c^\dag_{\mathbf{k}A\downarrow},c^\dag_{\mathbf{k}B\uparrow},c^\dag_{\mathbf{k}B\downarrow},c_{-\mathbf{k}A\uparrow},c_{-\mathbf{k}A\downarrow},c_{-\mathbf{k}B\uparrow},c_{-\mathbf{k}B\downarrow}),$$ giving the second-quantized BdG Hamiltonian
\begin{align}
    \mathcal{H}=\sum_{\mathbf{k}}d^\dag_\mathbf{k} H^{\mathrm{BdG}}(\mathbf{k})d_\mathbf{k}\quad,
\end{align}
such that the Bloch Hamiltonian in Eq.~\ref{eq:8DSM} becomes
\begin{align}
    H^{\mathrm{BdG}}_{\mathrm{8DSM}}(\mathbf{k})=\frac{1}{2}[&\sin k_x\sigma^z s^x+\sin k_y \sigma^z s^z+\sin k_z \mu^z \sigma^z s^y],
\end{align}
which has a particle-hole redundancy such that all Hamiltonian terms must satisfy
\begin{align}
    \mu^x (H^{\mathrm{BdG}}(\mathbf{k}))^T \mu^x= -H^{\mathrm{BdG}}(-\mathbf{k})\quad.
\end{align}
In this formalism, the $\mu^a$ Pauli matrices act on the particle/hole degrees of freedom, and hence $U(1)_{V,\rm{UV}}=e^{i\theta\mu^z}$ and $C_2^{\mathrm{BdG}}=e^{i\frac{\pi}{2}\sigma^z s^y}$.

We will now add a $U(1)_{V,{\rm UV}}$ symmetry-breaking, but time-reversal preserving, term that will gap out all Dirac nodes except the one at $\mathbf{k=0}$. It is convenient to describe the momentum dependence of this term via a bump function given by
\begin{align}
    B(\mathbf{k},w)=
    \begin{cases}
    e^{\frac{|\mathbf{k}|^2}{|\mathbf{k}|^2-w^2}} & \mathrm{for}\,\,|\mathbf{k}|<w\\
    0 & \mathrm{for}\,\,|\mathbf{k}|\geq w
    \end{cases}\quad,
\end{align}
where $w>0$ determines the width of the bump. This function is smooth but non-analytic. We now add a $U(1)_V$ breaking term such that the total Hamiltonian is given by
\begin{align}
    H^{\mathrm{BdG}}_{\rm 1DSM}(\mathbf{k})=& H^\text{BdG}_{\mathrm{8DSM}}(\textbf{k})+  \sum_{\alpha}B\left(\mathbf{k}-\mathbf{k}_\alpha,\frac{\pi}{2}\right)\mu^y s^y,
    \label{eq:almostlocalD}
\end{align}
where the $\alpha$ sum runs over all the TRIM points except $\Gamma$, and we chose $w=\pi/2$ so that the bump functions have the most local Fourier transform in position space and do not overlap in momentum space. This term was chosen because it gaps all Dirac nodes by adding the time-reversal and rotation invariant superconducting mass term $\mu^y s^y$ around each TRIM point except the $\Gamma$-point ($\mathbf{k}=0$). In order to symmetry-protect the Dirac node at the $\Gamma$-point we need to impose a symmetry that resembles our original $U(1)_{V,{\rm UV}}$ symmetry in addition to maintaining the time-reversal and rotation symmetries. We will consider two such examples in Sections~\ref{example:1} and \ref{example:2}. Then, we will consider a case where $C_2$ symmetry is modified in a non-trivial way in Section~\ref{example:3}.

\subsubsection{Not on-site, not compact $U(1)$ example 1}
\label{example:1}

For Sections~\ref{example:1} and \ref{example:2}, the charge generator of our not on-site symmetry in the second-quantized language takes the form of
\begin{align}
    \hat{Q}_j=\sum_{\mathbf{k}}d_{\mathbf{k}}^\dag S_j(\mathbf{k}) d_{\mathbf{k}}\quad,
\end{align}
where $j\in\{1,2\}$ is the index of the two examples, and $S_j(\bf k)$ is the charge operator in the single particle Bloch form.

In example 1, consider a not on-site and non-compact $U(1)$ symmetry, described by
\begin{align}
\label{eqnbumpchiral}
    S_1(\mathbf{k})=B\left(\mathbf{k},\frac{\pi}{2}\right)\mu^z
    \quad.
\end{align}
If we were to Fourier transform this generator into position space, one would see that there are smaller than polynomial decaying tails in position-space, i.e., the generator is almost local, but crucially is not on-site. Observe that the charge $\hat{Q}_1$ is not quantized, i.e., its spectrum is not quantized, and therefore we call the associated symmetry group non-compact.

In momentum space, this symmetry effectively acts as an IR $U(1)_V$ symmetry on a patch of the BZ surrounding the $\Gamma$-point. As such, the set of symmetries $\{\Theta, C_2,e^{i\theta \hat{Q}_1}\}\subseteq G_{\rm UV}$ fully protects the single Dirac node at $\mathbf{k}=0$, since
$$\bar\rho:C_2\times e^{i\theta \hat{Q}_1}\rightarrow  \mathbb{Z}_2^R\times U(1)_V\subseteq \bar{G}_{\mathrm{IR}}\quad,$$
is a large enough subgroup of $\bar{G}_{\mathrm{IR}}$ to prevent all the mass terms, as shown in Table~\ref{tab:symtomass} and Eq.~\ref{eq:centerIRgroup}. However, at all the other TRIM points (which correspond to massive Dirac fermions), we have
$$\bar\rho:C_2\times e^{i\theta \hat{Q}_1}\rightarrow  \mathbb{Z}_2^R\subseteq \bar{G}_{\mathrm{IR}}\quad,$$
which no longer provides adequate symmetry protection against $U(1)_V$ breaking terms, as can be seen in Table~\ref{tab:symtomass}.

One may wonder how this example is consistent with the heuristic argument presented in the introduction, where turning on a rotation symmetry-breaking mass allows the rotation operation to seemingly toggle between a trivial and non-trivial topological insulator --- something that would be impossible. Let us see how this conundrum is resolved: we may add a rotational symmetry breaking term $M\mu^z\sigma^x$, which preserves all other symmetries. This mass term gaps the system into one of two possible insulating states, described by the topological insulator term
$$\frac{\Phi}{2}\int\frac{dA}{2\pi}\wedge \frac{dA}{2\pi}\quad,$$
where $\Phi\in\{0,\pi\}$\footnote{We use the symbol $\Phi$ instead of the conventional $\Theta$ to distinguish it from our notation for the time-reversal operator.} with $A$ being the gauge field associated with the $e^{i\theta\hat{Q}_1}$ symmetry. Rotations change the sign of the Dirac fermion mass at the $\Gamma$ point, which also changes $\Phi\rightarrow \Phi+\pi$. 

However, notice that our microscopic $U(1)$ symmetry is non-compact. To proceed, let us assume that our microscopic gauge field $A$ is actually an $\mathbb{R}$ gauge field instead of the usual $U(1)$ gauge field. This would imply that the topological insulator term is always trivial since $dA=0$ for an $\mathbb{R}$ gauge field on any closed manifold. Thus, this resolves the conundrum as rotations just take you from a trivial insulator to another trivial insulator state in the same phase.
A physical reason to see why the above term is always trivial is to observe that any single 2+1d Dirac node that would live on the boundary of the putative 3+1d topological insulator (whose symmetry-protected stability is a hallmark of a real topological insulator) can always be trivialized via the time-reversal and $\hat{Q}_1$ symmetric $\Delta\sigma^y s^z$ terms that move the resulting 2+1d Dirac node to a momentum point in $\hat{z}$ that allows us to gap it symmetrically.

\subsubsection{Not on-site $U(1)$ example 2}
\label{example:2}

In this example, we construct another interesting $U(1)$ symmetry which is not on-site, but is in fact compact and has quantized charge.
To do so, let us first define a smooth transition function $g(\mathbf{k},\alpha,\omega)$ that at $|\mathbf{k}|=\alpha$ starts to transition between $0$ and $1$ over a width of $\omega$, given by
\begin{align}
    g(\mathbf{k},\alpha,\omega)=
    \begin{cases}
    1 & \mathrm{for}\,\,|\mathbf{k}|\geq \alpha+\omega\\
    \frac{e^{-\frac{1}{|\mathbf{k}|-\alpha}}}{e^{-\frac{1}{|\mathbf{k}|-\alpha}}+e^{-\frac{1}{\omega-|\mathbf{k}|+\alpha}}} & \mathrm{for}\,\,\alpha+\omega>|\mathbf{k}|>\alpha\\
    0 &  \mathrm{for}\,\,|\mathbf{k}|\leq\alpha
    \end{cases}\quad,
\end{align}
where $\alpha,\omega>0$. With this function in hand, we can define an almost-local charge generator $S_2$ that is quantized, where
\begin{align}
\label{eqnbumpchiral2}
    S_2(\mathbf{k})=&\sqrt{1-g\left(\mathbf{k},\frac{\pi}{2}-\epsilon,\epsilon\right)^2}\mu^z
    \nonumber\\
    &\quad+ g\left(\mathbf{k},\frac{\pi}{2}-\epsilon,\epsilon\right)\mu^y\sigma^x s^y\quad.
\end{align}
This new set of symmetries $\{\Theta, C_2, e^{i\theta\hat{Q}_2}\}\subseteq G_{\rm UV}$ also protects the Dirac node at the $\Gamma$-point, with $\hat{Q}_2$ being fully quantized (as seen by the eigenvalues of $S_2(\mathbf{k})$ taking values of $\pm 1$ since it squares to the identity).

Notice that if $S_2(\mathbf{k})$ is in the UV symmetry algebra, then so is the term
\begin{align}
    S_2(\mathbf{k})+C_2 S_2(\mathbf{k}) C_2^\dag=2\sqrt{1-g\left(\mathbf{k},\frac{\pi}{2}-\epsilon,\epsilon\right)^2}\mu^z\,,
\end{align}
which is non-compact. Therefore, our UV microscopic group contains non-compact elements and is in fact very similar to the previous example. As in example 1, one may wonder how this construction is consistent with the argument in the introduction. Again, the reason boils down to the non-compact nature of the microscopic symmetry group, as demonstrated in Section~\ref{example:1}.

\subsubsection{Modified Rotation Operator}
\label{example:3}

Instead of modifying the $U(1)_{V,{\rm UV}}$ generator, we may instead choose to redefine our $C_2$ rotation operator to be
\begin{align}
    \tilde{C}_2=e^{i\frac{\pi}{2}\left(\sqrt{1-g\left(\mathbf{k},\frac{\pi}{2}-\epsilon,\epsilon\right)^2} \sigma^z s^y+g\left(\mathbf{k},\frac{\pi}{2}-\epsilon,\epsilon\right) \mu^x\sigma^y\right)}\quad,
\end{align}
for some small $\epsilon>0$. A $\tilde{C}_2$ symmetric Hamiltonian with a single Dirac node is 
\begin{align}
    H^{\mathrm{BdG}}_{\rm 1DSM}(\mathbf{k})=& H^\text{BdG}_{\mathrm{8DSM}}(\textbf{k})+  \sum_{\alpha}B\left(\mathbf{k}-\mathbf{k}_\alpha,\frac{\pi}{2}\right)\mu^z \sigma^x,
    \label{eq:almostlocalDother}
\end{align}
where the $\alpha$ sum again runs over all TRIM points except $\Gamma$. Combining $\tilde{C}_2$ with the usual $U(1)_{V,{\rm UV}}$ with generator
\begin{align}
    Q_{U(1)_{V,{\rm UV}}}=\mu^z\quad,
\end{align}
and time-reversal symmetry $\Theta=-is^y\mathcal{K}$, will protect the single Dirac node at the $\Gamma$-point, with $\{\Theta,\tilde{C}_2,U(1)_{V,{\rm UV}}\}\subseteq G_{\rm UV}$.

Notice here that $U(1)_{V,{\rm UV}}$ is not a normal subgroup of the $G_{\mathrm{UV}}$ since a conjugation action by $\tilde{C}_2$ takes $U(1)_{V,{\rm UV}}$ out of $U(1)_{V,{\rm UV}}$. The modified rotation operator is not quite a rotation symmetry. Indeed, along the axis/center of rotation this operator acts almost-locally, i.e., possesses sub-polynomial tails such that along rotation centers there is an internal symmetry action that involves these tails. This, along with breaking condition~\ref{eq:normal},  allows the system to bypass Theorem~\ref{thm:1} and therefore symmetry-protect a single time-reversal invariant Dirac fermion.

\subsection{Example: Broken time-reversal resulting in single symmetry-protected Dirac fermion}
\label{sec:nonsymdirac}
Let us now consider two cases where we violate the assumption of time-reversal symmetry.

\subsubsection{Time-reversal broken Weyl semimetal}
\label{sec:TbrokenWeyl}

We start with the minimal example of a time-reversal broken Weyl semimetal with two low-energy Weyl nodes of opposite chirality. A simple Bloch Hamiltonian model can be written as
\begin{align}
    H_{\mathrm{WSM}}(\mathbf{k})=&\sin k_x s^x+\sin k_y s^y\nonumber\\
    &+\left(\sin Q-\sin k_z+m_W(\mathbf{k})\right)s^z,
    \label{eq:TbrokenWeyl}
\end{align}
where $m_W(\mathbf{k})=2-\cos k_x-\cos k_y$. This system possesses a Weyl node of negative chirality at $\mathbf{k}=(0,0,Q)$ and one of positive chirality at $\mathbf{k}=(0,0,\pi-Q)$~\footnote{In the presence of a $U(1)_{V,{\rm UV}}$ symmetry, this model can be related to the usual magnetic Weyl semimetal with nodes at $k_z=\pm \frac{Q}{2}$ via a large gauge-transformation in $U(1)_{V,{\rm UV}}$.}. Such a system has an IR theory with a single  Dirac fermion, albeit having broken Lorentz symmetry as in Eq.~\ref{eq:IRactionbroken}. Such a system breaks time-reversal symmetry, since the time-reversal operator $\Theta=-is^y\mathcal{K}$ anticommutes with the Pauli matrices, and thus does not allow terms such as $m_W(\mathbf{k})s^z$. This is not surprising since it is well-known that time-reversal would require more than two Weyl nodes in this system~\footnote{This is generally argued via Berry curvature $\mathbf{F}$ in free-fermionic systems, where time-reversal symmetry requires $\mathbf{F}(\mathbf{k})=-\mathbf{F}(-\mathbf{k})$. However, as we have shown, such a statement also holds for interacting systems.}, i.e., a minimum of four Weyl nodes which we will explore in Section~\ref{sec:TinvariantWeyl}. In the presence of translation symmetry in the $\hat{z}$-direction and $U(1)_{V,{\rm UV}}$, the two Weyl nodes in this model are perturbatively stable, i.e., symmetry-protected, as long as the nodes sit at different $k_z$ momenta. For the following section, take $\{U(1)_{V,{\rm UV}}, T_z\}\subseteq G_{\rm UV}$, where $T_z$ represents the $\hat{z}$-translation symmetry group.

One may wonder why such a single symmetry-protected Dirac fermion is even possible in the framework of anomaly-matching in Section~\ref{sec:proof} --- in the absence of time-reversal symmetry, how do the arguments change? To understand the differences, we may use the anomaly analysis in Section~\ref{sec:proof}, but without the extra time-reversal condition. Since translation still maps via $\bar\rho$ into a subgroup of $\frac{U(1)_A\times U(1)_V}{\mathbb{Z}_2}$, we once again have to consider an emanant chiral anomaly (Eq.~\ref{eq:Omega}) in the IR, which must be matched in the UV. In this case, the matching of the IR and UV gauge fields is
$$A_R=A+q_R \tilde{z}\,,\quad A_L=A+q_L \tilde{z}\,,$$
with $\tilde{z}$ being the translation gauge field and $q_{R,L}$ the momentum positions of the respective Weyl node with $q_R=\pi-Q$ and $q_L=Q$. Notice that in the absence of time-reversal symmetry, the values of $q_{R,L}$ are not restricted to specific momentum points. The anomaly matching procedure requires the UV to possess the following 4+1d term
\begin{align}
     \Omega_{\mathrm{UV}}
     =&\frac{1}{2}\int \tilde{z}\wedge\bigg[(q_R-q_L)\left(\frac{dA}{2\pi}\right)^2+\frac{q_R^3-q_L^3}{3}\left(\frac{d\tilde{z}}{2\pi}\right)^2\nonumber\\
     &+(q_R^2-q_L^2)\,\frac{d\tilde{z}}{2\pi}\,\frac{dA}{2\pi}+\frac{q_R-q_L}{96\pi^2}\mathrm{Tr}[R\wedge R]\bigg],
     \label{eq:OmegaUV}
\end{align}
where it is crucial to notice that none of these topological terms represent a non-trivial 4+1d SPT since the coefficient can be symmetrically tuned to triviality (e.g., when $q_R=q_L$) by changing $q_R$ and $q_L$. Recall that this is unlike the time-reversal invariant case where the coefficient was quantized for crystalline symmetries due to the analysis in Section~\ref{sec:UVtoIRsymm}. This non-quantization means that the boundary of a theory, described by Eq.~\ref{eq:OmegaUV}, does not possess an anomaly (an obstruction to gauging) in the UV~\footnote{Note that such a term has been previously referred to as an `unquantized' anomaly~\cite{PhysRevResearch.3.043067}. However, it does not obey the standard anomaly properties, such as an obstruction to gauging.}. This, in turn, means that there is no contradiction thus far, and that such a time-reversal broken but symmetry-protected system can (and does) exist.

Additionally, we can use the anomaly matching procedure and Eq.~\ref{eq:OmegaUV} to deduce some well-known properties of the magnetic Weyl semimetal, which lives on the boundary of Eq.~\ref{eq:OmegaUV}. The first boundary theory term is
\begin{align}
    S_{\mathrm{UV}}=\frac{1}{2}\frac{\pi-2Q}{4\pi^2}\int \tilde{z}\wedge A \wedge dA\quad,
\end{align}
which describes the anomalous Hall conductivity and the chiral magnetic effect of a magnetic Weyl semimetal~\cite{PhysRevLett.124.096603}. Varying with respect to $A^\mu$, allows us to deduce the current expectation value $\langle j^\mu\rangle$ to be
$$\langle j^\mu\rangle=\sigma_H\epsilon^{\mu\nu\alpha\beta}\tilde{z}_{\nu}\partial_\alpha A_\beta \quad,$$
where $\sigma_H$ is the Hall conductivity per $\hat{z}$-layer (recall that $\int \tilde{z}_z dz =L_z$ is the length of the system in the $\hat{z}$ direction), given by the prefactor
$$\sigma_H=\frac{\pi-2Q}{4\pi^2}\quad.$$
Of course, such a system with a Hall conductance breaks time-reversal symmetry, thereby circumventing Theorem~\ref{thm:1}.
Although we will not elaborate further in this paper, the other boundary terms also describe known phenomena in magnetic Weyl semimetal, such as the thermal Hall conductivity from the boundary of the last term in Eq.~\ref{eq:OmegaUV}.

\subsubsection{Time-reversal broken Dirac semimetal}
\label{sec:nonsymDirac}

We can also construct a symmetry-protected, but time-reversal broken, single Dirac semimetal model where the two Weyl nodes lie at the same momentum. This can be achieved by taking the previous Weyl semimetal model at $Q=0$ (i.e., Weyl nodes at $k_z\in\{0,\pi\}$) and breaking $\hat{z}$ translation from $\mathbb{Z}\rightarrow 2\mathbb{Z}$ and $C_2$ rotation symmetry, but keeping the combination of translation and rotation which turns into a non-symmorphic $\hat{z}$ screw symmetry $C_2^z$ in the new Brillouin zone. The detailed procedure is outlined in Appendix~\ref{app:nonsymDirac}, where we derive the single Dirac node Hamiltonian of the form
\begin{align}
H_{\mathrm{DSM}}(\mathbf{k})&=\sin k_x\,
s^x
+\sin k_y\,s^y+m_W(\mathbf{k})s^z \label{eq:nonsymdirac}\\
+\frac{1}{2}(\cos & \,k_z -1)\sigma^y s^z
+\frac{1}{2}\sin k_z  \,\sigma^x
s^z
+\sin k_x 
\sigma^z s^z\,,\nonumber
\end{align}
where the $\sigma$ Pauli matrices denote the sublattice degree of freedom that resulted from breaking the original (now half) translation symmetry. Inherited from Eq.~\ref{eq:TbrokenWeyl}, time-reversal symmetry is also broken due to the $m_W(\mathbf{k}) s^z$ term.

The low-energy physics of this system consists of a single Dirac node at $\mathbf{k}=0$. The UV symmetries of this model are $U(1)_{V,{\rm UV}}$, and the non-symmorphic symmetry that acts as
\begin{align}
C^z_{2}(\mathbf{k})
=i\begin{pmatrix}
0 & 1\\
e^{ik_z} & 0
\end{pmatrix}_\sigma s^z\quad,
\end{align}
which rotates $k_x\rightarrow -k_x$ and $k_y\rightarrow - k_y$. Taking $\{U(1)_{V,{\rm UV}}, C^z_{2}\}\subseteq G_{\rm UV}$, the low energy single Dirac fermion is symmetry-protected.

Analogous to the magnetic Weyl semimetal case, here the non-symmorphic symmetry maps into the IR chiral symmetry such that we have the chiral anomaly described in Eq.~\ref{eq:OmegaUV} with gauge field $\tilde{z}$ denoting the non-symmorphic symmetry gauge field. The associated charges are $k_R=\pi$, $k_L=0$ for the Hamiltonian in Eq.~\ref{eq:nonsymdirac}. However, there exist IR Lorentz symmetry-breaking but $G_{\rm UV}$ symmetric terms such as $\delta s^z$ that are able to split the Dirac node into two Weyl nodes without affecting the IR chiral anomaly, which shows that the Dirac node is symmetry-protected but not pinned in momentum space. More details can be found in Appendix~\ref{app:nonsymDirac}.

\subsection{Example: Two or more low-energy Dirac nodes}
\label{sec:twoormore}

After considering systems that had broken or non-on-site symmetries, let us now consider symmetry-protected Dirac semimetals that have on-site $U(1)_{V,{\rm UV}}$ and preserve time-reversal. Conventional Type I and II Dirac semimetals are both examples of systems that obey the assumptions of Theorem~\ref{thm:nodirac} in the presence of time-reversal symmetry: and hence they must have two or more low-energy Dirac fermions. Type I Dirac semimetals possess Dirac nodes as a result of band-inversion along a symmetry-protected rotational axis. Perhaps unsurprisingly, this procedure automatically results in multiple ($> 1$) low-energy Dirac fermions. Examples of such semimetals include Na$_3$Bi~\cite{Fang12} and Ca$_3$As$_2$~\cite{Fang13}, which have been studied in the context of anomalies and their associated topological response in Refs.~\onlinecite{PhysRevResearch.3.043067,PhysRevB.110.155110}.

In contrast, Type II Dirac semimetals can have Dirac nodes pinned at certain TRIM where four-dimensional irreducible representations of the symmetry group are supported. Well-known examples of Type II Dirac semimetals include the distorted Dirac spinel model~\cite{PhysRevLett.112.036403} (space group 74), which has two Dirac nodes (one at the $T$-point and and one at the $T'$-point), and the diamond lattice model~\cite{FKM,Kane12} (space group 227), which has three Dirac nodes (one at each of the 3 $X$ points); both examples will be explored in detail in the following sections. Naturally, one may wonder how three symmetry-protected Dirac nodes are permitted, when a single lone Dirac node is not. The resolution is similar to the general Nielsen-Ninomiya theorem where an odd number of Weyl fermions are permitted if they are appropriately charged under $U(1)_{\rm UV}$ (see the end of Section~\ref{sec:NNthm}). Analogously, our Dirac no-go theorems are not strict ``doubling" theorems (unless extra assumptions are made), and an odd number of Dirac nodes can be consistent in the UV depending on how the individual nodes are charged under which protecting symmetries; one example is the diamond lattice model discussed below.

\subsubsection{Distorted Dirac spinel model}
\label{sec:spinel}

\begin{figure}[htb]
    \centering
    \vspace{0.3cm}
        \includegraphics[width=0.45\columnwidth]{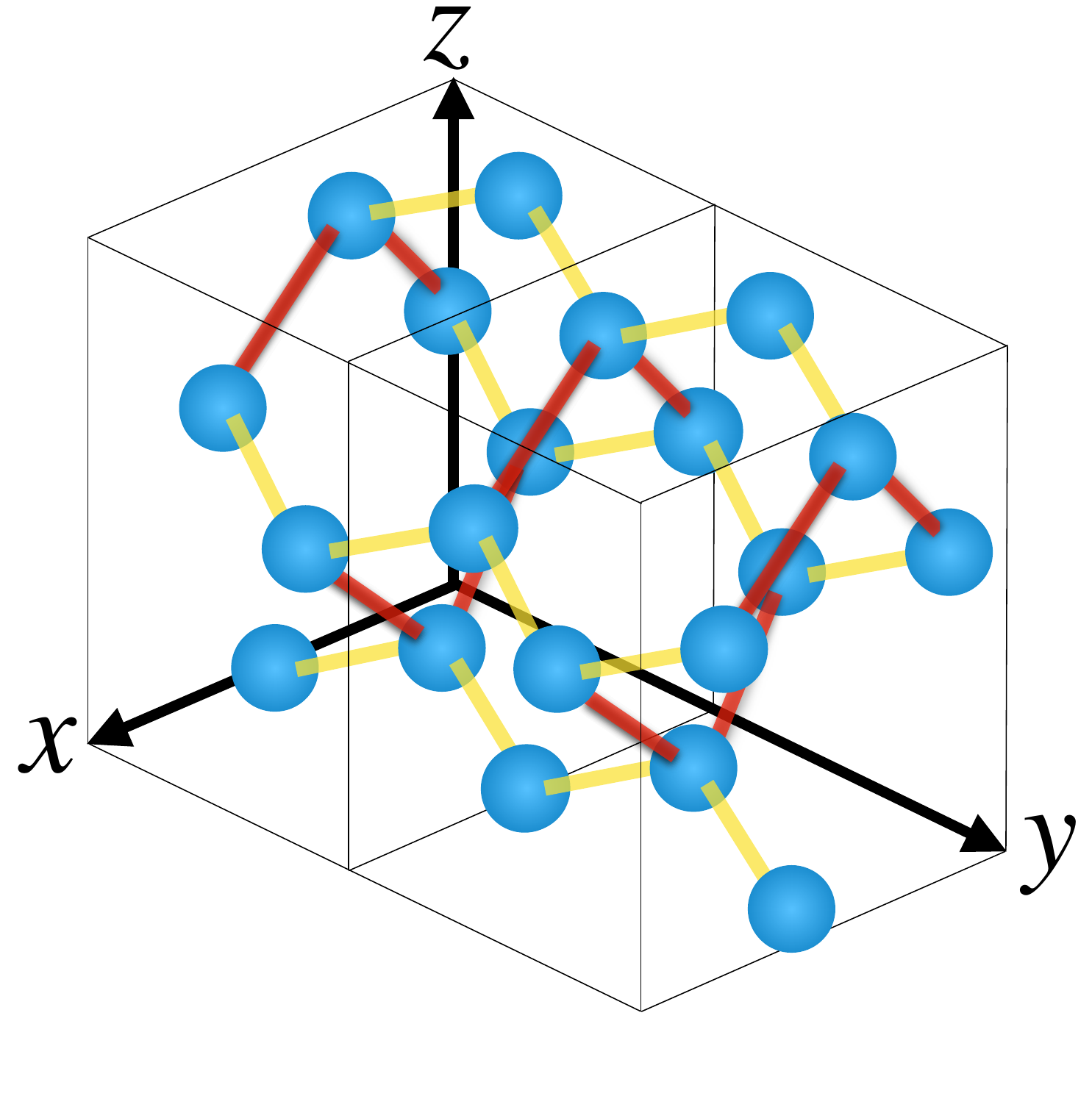}
        \vspace{0.2cm}
        \includegraphics[width=0.45\columnwidth]{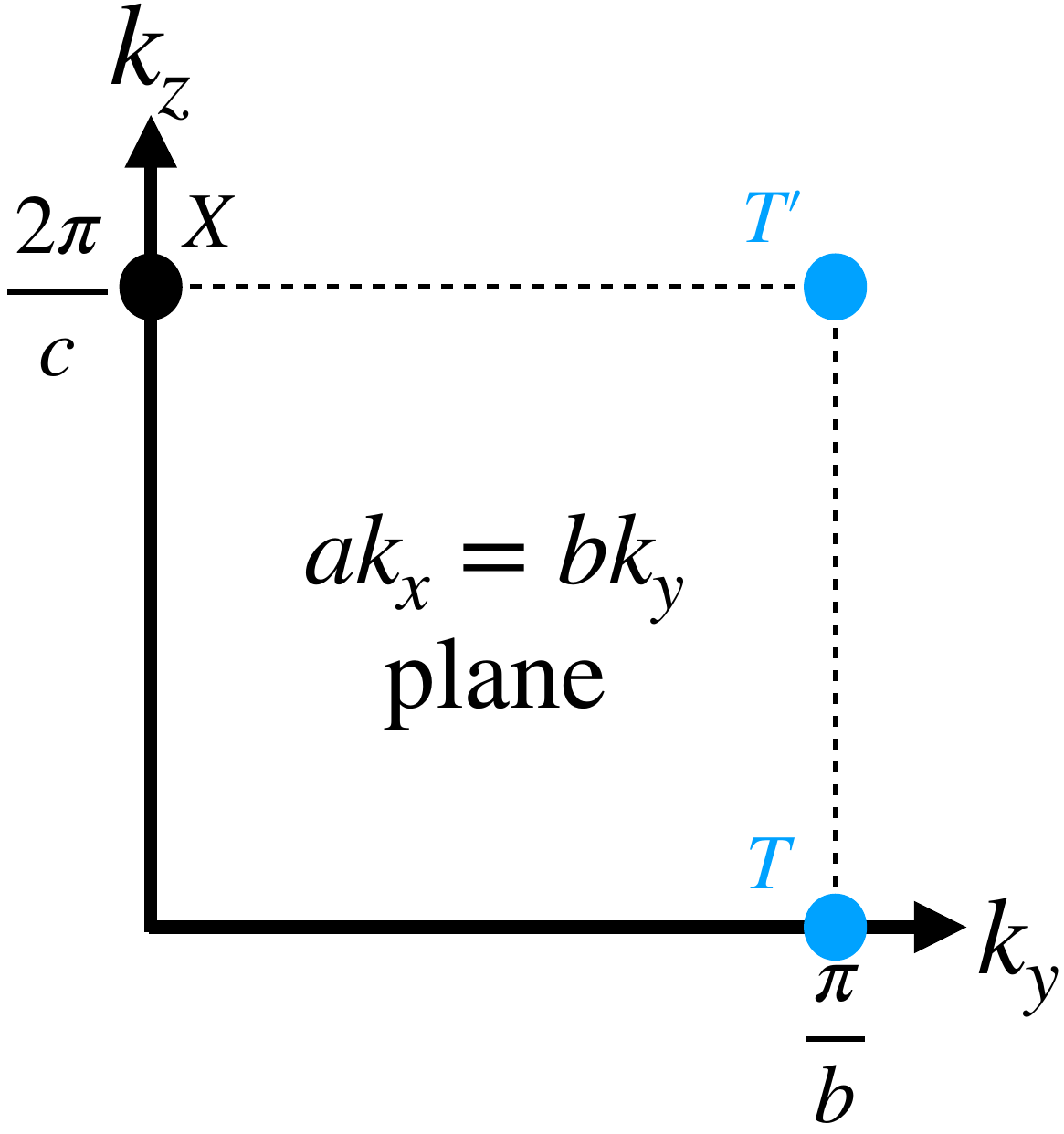}
    \caption{\label{fig:distortedlattice}Here we have depicted the distorted Dirac spinel lattice (a) and the momentum space depiction of the $T$ and $T'$ high-symmetry points on the $ak_x=bk_y$ plane (b). Note that along the $k_x=\frac{\pi}{a},\,k_y=\frac{\pi}{b}$ axis we have a $\frac{4\pi}{c}$ periodicity of the Brillouin zone rather than the usual $\frac{2\pi}{c}$.}
\end{figure}

To study the Dirac spinel example, we build a tight-binding model of space group 74 based on an s-orbitals with a 2-atom basis placed on a body-centered orthorhombic lattice with Hamiltonian~\cite{PhysRevLett.112.036403}
\begin{align}
\mathcal{H}_{\mathrm{dDS}}=\sum_{\langle i,j\rangle,s}t_{ij}c_{i,s}^\dag c_{j,s}+\sum_{\langle\langle i,j\rangle\rangle}i\lambda_{ij}c_i^\dag\boldsymbol{s}\cdot\left(\mathbf{d}_{ij}^1\times\mathbf{d}_{ij}^2\right)c_j,
\label{eq:Dspinel}
\end{align}
where $\boldsymbol{s}=\{s^x,s^y,s^z\}$ are Pauli matrices for spin. This model uses the same lattice as Steinberg \textit{et al.}~\cite{PhysRevLett.112.036403}, but we choose a simpler 4-band model, from sublattice ($A$, $B$) and spin ($\uparrow$, $\downarrow$) degrees of freedom, to demonstrate the main physics without loss of generality.
The lattice coordinates corresponding to the nearest neighbors of the $A$ sublattice at Bravais lattice position $\mathbf{r}=\mathbf{0}$ are given by
$\mathbf{t}_0=(0,0,0)$, $\mathbf{t}_1=(a,0,0)$, $\mathbf{t}_2=(0,b,0)$, $\mathbf{t}_3=(\frac{a}{2},\frac{b}{2},\frac{c}{2})$ (see Figure~\ref{fig:distortedlattice}(a) for visualization of the lattice). For the next-nearest-neighbor (spin-orbit coupling) terms we have specific bond directions from $A$ to $B$ sublattice sites at $\mathbf{0}$, which take the form $\mathbf{d}^{1\pm}=(\pm \frac{a}{2},0,[4\gamma-1]\frac{c}{2})$ and $\mathbf{d}^{2\pm}=(0,\pm \frac{b}{2},2\gamma c)$, where $\gamma$ quantifies the relative length distortion of the two types of bonds away from a regular diamond lattice. When $a=b=\frac{c}{\sqrt{2}}$ and $\gamma=\frac{1}{8}$, this represents the undistorted diamond lattice.

It is convenient to solve this Hamiltonian in momentum space with reciprocal lattice vectors $\mathbf{g}_1=2\pi\left(\frac{1}{a},\frac{1}{b},0\right)$, $\mathbf{g}_2=2\pi\left(0,\frac{1}{b},\frac{1}{c}\right)$, $\mathbf{g}_3=2\pi\left(\frac{1}{a},0,\frac{1}{c}\right)$ that span the BZ. Now we can expand the Hamiltonian around the following high-symmetry $\mathbf{k}$ points
\begin{align*}
X=(0,0,\frac{2\pi}{c}), && T=(\frac{\pi}{a},\frac{\pi}{b},0), && T'=(\frac{\pi}{a},-\frac{\pi}{b},0),
\end{align*}
where we note that $T$ and $T'$ are distinct high symmetry (TRIM) points in the BZ, i.e., unrelated by a reciprocal lattice vector. Their respective low-energy Bloch Hamiltonians around these points take the forms
\small
\begin{align}
H_X=&\left(\bar t\,q_y+t_l q_z\right)\sigma^y+ \left((1-4 \gamma ) \lambda _{\text{sl}}+4 \gamma  \lambda _{\text{ss}}\right)\,q_y\, \sigma^z s^x\nonumber\\
&+\left((1-4 \gamma ) \lambda _{\text{ll}}+4 \gamma  \lambda _{\text{sl}}\right)\,q_x\,\sigma^z s^y+2\bar t\,\sigma^x\quad,\label{eq:HX}\\ \nonumber\\
H_T=&-\left(t_sq_y+t_lq_x\right)\sigma^y+ \left(4\gamma \lambda_{sl}\,q_y-(1-4\gamma)\lambda_{ll}\,q_x\right)\sigma^z s^y\nonumber\\
&-\left(4\gamma \lambda_{ss}\,q_y-(1-4\gamma)\lambda_{sl}\,q_x\right)\sigma^z s^x-\lambda_{sl}\,q_z\,\sigma^z s^z, \label{eq:HT}\\\nonumber\\
H_{T'}=&-\left(t_sq_y-t_lq_x\right)\sigma^y- \left(4\gamma \lambda_{sl}\,q_y+(1-4\gamma)\lambda_{ll}\,q_x\right)\sigma^z s^y\nonumber\\
&- \left(4\gamma \lambda_{ss}\,q_y+(1-4\gamma)\lambda_{sl}\,q_x\right)\sigma^z s^x+\lambda_{sl}\,q_z\,\sigma^z s^z,\label{eq:HTp}
\end{align}
\normalsize
where $\boldsymbol{\sigma}=(\sigma^x,\sigma^y,\sigma^z)$ are Pauli matrices for the sublattice degree of freedom, the subscripts $s$ and $l$ indicate the amount of consecutive short and/or long bond jumps, $\bar t=t_s-t_l$, $\mathbf{q}$ is the difference between $\mathbf{k}$ and the respective high-symmetry point, and we have set the lattice constants to unity from here on unless explicitly present. From the spectra of these Hamiltonians we find that Dirac nodes are present at $T$ and $T'$, while the $X$ point is gapped by the elongation of bonds. We also note that the similarity between $H_{T}$ and $H_{T'}$ comes from space group 74's $C_2$ rotation symmetry around the $x$-axis ($C_{2x}(\mathbf{k})=i\sigma^x s^x$) that relates these two points.

A detailed symmetry analysis can be found in Appendix~\ref{app:distorteddirac}, and here we recapitulate the relevant information. The gaplessness of the Dirac nodes at $T$ and $T'$ is protected by the UV symmetry group
\begin{align}
    \langle\{U(1)_{V,\rm{UV}},C^y_{2z},T_{\mathbf{t}_3}\}\rangle\subseteq G_{\mathrm{UV}}\quad,
\end{align}
where $\langle S\rangle$ is the group generated by the set $S$, $C^y_{2z}$ is a non-symmorphic symmetry combining a $\pi$-rotation around $\hat{z}$ and a half-translation in $\hat{y}$, and $T_{\mathbf{t}_3}$ is the translation symmetry in $\mathbf{t}_3$.

How does this system evade the anomaly-matching issues that haunted the single time-reversal invariant Dirac fermion? Once again we can appropriate the anomaly analysis in Section~\ref{sec:proof}, but this time with the two copies of the chiral anomaly, since there are two Dirac fermions in the IR. During the matching of the UV and IR gauge fields, the non-symmorphic gauge field $\tilde{z}$ of the $C^y_{2z}$ symmetry enters the UV anomaly in Eq.~\ref{eq:OmegaUV}. However, since we have two low-energy Dirac fermions, associated with two different TRIM points, each fermion contributes a term of the form in Eq.~\ref{eq:OmegaUV} with different symmetry charges: $\{q_R^T,q_L^T\}$ for the Dirac fermion at $T$, and $\{q_R^{T'},q_L^{T'}\}$ for the Dirac fermion at $T'$. In the distorted spinel model, we have $q_R^T=q_L^{T'}$ and $q_L^T=q_R^{T'}$, which allow for cancellations of the anomaly terms in the UV.

A consequence of this cancellation can be observed on a $C^y_{2z}$ disclination. Recall that a single Dirac node would generate time-reversal breaking chiral modes on disclination-like defects (see Eq.~\ref{eq:disclinationterm}) which would result in a contradiction. However, with the two Dirac nodes in this model, we avoid this issue by having opposite $C^y_{2z}$ symmetry eigenvalues for the two Dirac nodes such that along a $C^y_{2z}$ disclination the anomalies cancel, i.e.,
$$\frac{1}{2}\int A\wedge\frac{dA}{2\pi}-\frac{1}{2}\int A\wedge\frac{dA}{2\pi}=0\quad,$$
thus resulting in no contradiction. Analogous resolutions hold for other protecting symmetries as we show in Appendix~\ref{app:distorteddirac}.

\subsubsection{Diamond Lattice Fu-Kane-Mele model}
\label{sec:FKM}
Next we consider the Fu-Kane-Mele model, which is a type II Dirac semimetal on a diamond lattice that possesses three symmetry-protected Dirac nodes, an example of an odd number of nodes. A tight-binding model of this diamond lattice (space group 227) can be constructed using a four band model, with spin and sublattice degrees of freedom, and a Hamiltonian given by~\cite{FKM}
\begin{align}
\mathcal{H}_{\mathrm{FKM}}=t\sum_{\langle i,j\rangle,s}c_{i,s}^\dag c_{j,s}+\frac{8i\lambda_{\mathrm{SO}}}{a^2}\sum_{\langle\langle i,j\rangle\rangle}c_i^\dag\boldsymbol{s}\cdot\left(\mathbf{d}_{ij}^1\times\mathbf{d}_{ij}^2\right)c_j\,,
\end{align}
where $\boldsymbol{s}=\{s^x,s^y,s^z\}$ are the spin Pauli matrices.
The Bravais lattice positions corresponding to the nearest neighbors of $(x,y,z)=\frac{a}{4}(1,1,1)$ are given by
$\mathbf{t}_0=(0,0,0)$, 
$\mathbf{t}_1=\frac{a}{2}(0,1,1)$,
$\mathbf{t}_2=\frac{a}{2}(1,0,1)$, and
$\mathbf{t}_3=\frac{a}{2}(1,1,0)$.
The corresponding reciprocal lattice vectors are
$\mathbf{g}_1=\frac{2\pi}{a}(-1,1,1)$, 
$\mathbf{g}_2=\frac{2\pi}{a}(1,-1,1)$, and
$\mathbf{g}_3=\frac{2\pi}{a}(1,1,-1)$.
The specific bond directions from $A$ to $B$ sublattice, where $A$ is situated at $\frac{a}{4}(1,1, 1)$, take the form
$\mathbf{d}^{1\pm}=\frac{a}{4}(\pm1,\pm1, -1)$, 
$\mathbf{d}^{2\pm}=\frac{a}{4}(\mp1,\pm1, 1)$.

Now, we expand the Bloch Hamiltonian around the three high-symmetry $X$ points, given by momentum space coordinates $X^{\hat{r}}=\frac{2\pi \hat{r}}{a}$,
\begin{align}
H_{X^x}&=-t a \,q_x\sigma^y-4\lambda_{\mathrm{SO}}aq_y\sigma^z s^y+4\lambda_{\mathrm{SO}}aq_z\sigma^z s^z,\\
H_{X^y}&=-t a \,q_y\sigma^y+4\lambda_{\mathrm{SO}}aq_x\sigma^z s^x-4\lambda_{\mathrm{SO}}aq_z\sigma^z s^z,\\
H_{X^z}&=-t a \, q_z\sigma^y-4\lambda_{\mathrm{SO}}aq_x\sigma^z s^x+4\lambda_{\mathrm{SO}}aq_y\sigma^z s^y,
\end{align}
where the $\boldsymbol{\sigma}=\{\sigma^x,\sigma^y,\sigma^z\}$ are Pauli matrices for the sublattice degrees of freedom $A$ and $B$.
Note that our Hamiltonians are multiplied by a negative sign compared to the original Fu-Kane-Mele because we use a different basis (i.e., we have swapped our labeling the $A$ and $B$ sublattice). As expected, the Hamiltonians give rise to three Dirac nodes, one at each of the  $X^{\hat{r}}$ points.

An example of the set of symmetries that protect the gaplessness of each $X$ point is given by
\begin{enumerate}
\item
Gaplessness of $H_{X^x}$ is protected if we have either $\{U(1)_{V,{\rm UV}},C_{2y},T_{\mathbf{t}_1}\}$ or $\{U(1)_{V,{\rm UV}},C_{2z},T_{\mathbf{t}_1}\}$,
\item
Gaplessness of $H_{X^y}$ is protected if we have either $\{U(1)_{V,{\rm UV}},C_{2x},T_{\mathbf{t}_2}\}$ or $\{U(1)_{V,{\rm UV}},C_{2z},T_{\mathbf{t}_2}\}$,
\item
Gaplessness of $H_{X^z}$ is protected if we have either $\{U(1)_{V,{\rm UV}},C_{2x},T_{\mathbf{t}_3}\}$ or $\{U(1)_{V,{\rm UV}},C_{2y},T_{\mathbf{t}_3}\}$,
\end{enumerate}
where $C_{2\alpha}$ ($\alpha\in\{x,y,z\}$) is a $\pi$ rotation around the axis $\alpha$.  In momentum space, the rotation operators are given by
\begin{align*}
C_{2x}(\mathbf{k})=i  \begin{pmatrix}
e^{i\mathbf{k}\cdot\mathbf{t}_1} & 0\\
0 & 1
\end{pmatrix}_\sigma s^x \quad,\\
C_{2y}(\mathbf{k})=i  \begin{pmatrix}
e^{i\mathbf{k}\cdot\mathbf{t}_2} & 0\\
0 & 1
\end{pmatrix}_\sigma s^y\quad,\\
C_{2z}(\mathbf{k})=i  \begin{pmatrix}
e^{i\mathbf{k}\cdot\mathbf{t}_3} & 0\\
0 & 1
\end{pmatrix}_{\sigma}s^z\quad.
\end{align*}
We can define a UV gauge field for each of the rotation symmetries: $\tilde{z}_\alpha,$ and consider their contribution to the disclination anomaly term in Eqs. \ref{eq:OmegaUV} and \ref{eq:disclinationterm}. The key point is to note that each $C_{2\alpha}$ symmetry protects exactly two nodes at the same time. Thus, like the distorted Dirac spinel case, we have cancellation of all anomalous disclination terms such as those in Eq.~\ref{eq:disclinationterm}, i.e., each anomaly term cancels with a Dirac partner of the opposite rotation charge. This mechanism shows that we do not have to have an even number of Dirac nodes, just that their anomalies must cancel pairwise. A more exhaustive symmetry analysis can be found in Appendix~\ref{app:FKM}.

In connection with the previous section, we can modify this Fu-Kane-Mele Hamiltonian to generate the Dirac spinel model. This is done by stretching the bonds $t\rightarrow t+\delta t_p$, where $p\in\{1,2,3,4\}$, to obtain a mass perturbation of $$m^{\hat{r}}\sigma^x=\sum_p\delta t_p\mathrm{sgn}[\mathbf{d}^p\cdot\hat{r}]\sigma^x\quad,$$
where $\mathbf{d}^p$ is the $p$th nearest neighbor bond vector. When the $\mathbf{d}^{2\pm}$ bonds are equally elongated ($\delta t_{2\pm}<0$), and the basis rotated to match the orthorhombic lattice, we obtain the spinel model expanded at the $X$, $T$, and $T'$ points, as given in Eqs.~\ref{eq:HX}, \ref{eq:HT}, and \ref{eq:HTp}.

\subsubsection{Time-reversal invariant Weyl semimetal}
\label{sec:TinvariantWeyl}

Finally, we also briefly consider a third model, known as the time-reversal invariant Weyl semimetal, that preserves time-reversal symmetry, on-site $U(1)_{V,{\rm UV}}$, and two low-energy emergent Dirac fermions (albeit Lorentz symmetry-breaking) that are symmetry-protected. To obtain this model, take any model with two Dirac nodes (e.g., distorted Dirac spinel or Na$_3$Bi), and add a time-reversal symmetric term that breaks the emergent Lorentz symmetry of the Dirac fermions such that the two Dirac fermions separate into four Weyl fermions in momentum space. Such a minimal model on a cubic lattice is given by the Bloch Hamiltonian
\begin{align}
    H_\text{TRSWSM}(\mathbf{k})=\sin k_x \sigma^z & s^x+\sin k_y \sigma^z s^y\nonumber\\
    &+m(\mathbf{k})\sigma^x+\delta\sin k_z s^z,
    \label{eq:TinvWeyl}
\end{align}
where $m(\mathbf{k})=2-\cos k_x-\cos k_y +\cos k_z-\cos Q$, $Q<\pi$, $0<\delta\ll1$, $\{s^x,s^y,s^z\}$ are spin Pauli matrices, $\{\sigma^x,\sigma^y,\sigma^z\}$ are Pauli matrices for a sublattice degree of freedom, and time-reversal is given by $\Theta=-is^y\mathcal{K}$. For $\delta\ll 1$ there are four Weyl nodes at $\pm\mathbf{k}_{\pm}$ with $\mathbf{k}_\pm\equiv(0,0,Q\pm \delta),$ and Weyl nodes at opposite momenta have the same chirality. This system has symmetry protection provided by $U(1)_{V,{\rm UV}}$ and translation symmetry in the $\hat{z}$-direction, but evades the anomaly associated with the layering argument in Eq.~\ref{eq:layeranomaly} since it has two copies of the anomaly term in Eq.~\ref{eq:OmegaUV} with the property that the associated charges (in this case, momentum) obey $q_R^1=-q_R^2$ and $q_L^1=-q_L^2$, thus resulting in a vanishing anomaly term.

In summary, we see that both the time-reversal invariant Dirac and Weyl semimetal models evade Theorems~\ref{thm:nodirac} and \ref{thm:noweyl} because they contain a higher number of symmetry-protected Dirac nodes in the IR.

\section{Discussion}
\label{sec:discussion}

In this paper, we have presented a detailed method to map between the UV and IR symmetries of a single 3+1d Dirac fermion. We used this mapping to show that all time-reversal invariant lattice models having a single 3+1d single Dirac fermion in the IR are necessarily fine-tuned, given some natural assumptions. Such a method can be straightforwardly generalized to other IR theories of various numbers of Weyl and Dirac fermions. Using this method, we can justify the presence of emanant IR quantum anomalies: specifically, the emanant chiral anomaly for a lattice 3+1d Dirac node when there exists an on-site microscopic $U(1)_{V,{\rm UV}}$ that is in the center (or a normal subgroup under certain additional assumptions) of the UV symmetry group. In the presence of a time-reversal symmetry where $\Theta=(-1)^{\hat{F}},$ this chiral anomaly becomes quantized in the UV such that we find no-go theorems which forbid the lattice realization of a single symmetry-protected time-reversal invariant Dirac node. Such a statement applies to most models used in the condensed-matter literature such as free-fermion Dirac and Weyl semimetal models, as well as strongly-interacting generalizations.

We also explored a handful of converse examples, where certain assumptions are broken, such as the magnetic Weyl semimetal and an almost-local single Dirac fermion model, which demonstrate the tightness of our theorems and allow us to explicitly show which assumptions are violated in these examples. We note that the Hamiltonian models of the converses that we present are far from complete. For example, we have not discussed theories with other low-energy gapless degrees of freedom (besides Dirac fermions) such as gapless scalar fields or bosonic fields, as well as other strongly interacting models.

Many questions and potential generalizations remain for future study. Most pressingly is the question of a possible generalization of Theorem~\ref{thm:1} that involves less restrictions on the microscopic symmetry group, i.e., what happens if $U(1)_{V,\rm{UV}}$ is just a subgroup of $G_{\rm UV}$ without any extra conditions? It would seem that a more general theorem should hold via some anomaly argument involving the matching of an emanant IR $U(2)$ (or subgroup) anomaly. However, the details remain to be determined. A smaller goal would be to remove the assumption in Section~\ref{sec:case2} that the protecting charge-conjugation symmetry is on-site. In this case, could one also prove the theorem for not on-site charge conjugation that acts in the $\mathrm{Pin}^-(2)$ symmetry group of the IR Dirac fermion? We also assumed that the time-reversal symmetry commutes with the $U(1)_{V,{\rm UV}}$ charge generator, as is the case in most condensed-matter contexts. However, if we relax this assumption, is it possible to create a lattice model of a single time-reversal invariant symmetry-protected Dirac fermion? These questions are important to answer for both the condensed matter and high-energy physics communities.

One thing we touch upon briefly in the main text is that Lorentz-invariance breaking terms act as symmetric perturbations to the system and thus do not alter the anomaly of the Dirac node. This means that by solely looking at the anomaly there is no way to tell whether a Dirac fermion is purely symmetry-protected, or also pinned. The pinning is instead a direct result of the mapping between the UV and IR symmetries, which is the subject of a companion paper~\cite{leitaylorryan}.

Some other questions also remain: is there a model, similar to the one in Sec.~\ref{sec:nononsiteU(1)}, that uses a non-on-site $U(1)_{V,{\rm UV}}$ that is finite range (compact support) rather than almost local? It does not seem that there is a specific reason for why our symmetry requires sub-polynomial tails, as is the case in Section~\ref{sec:nononsiteU(1)}. Are there ways to implement non-Abelian symmetries at low-energies by defying the assumption that $U(1)_{V,{\rm UV}}$ obeys the condition~\ref{eq:center} while maintaining time-reversal symmetry? Such methods have been used in Ref.~\cite{gioia2025exactchiralsymmetries31d} to implement the $SU(2)$ symmetry in the IR for a time-reversal broken Weyl semimetal; however, a similar construction for a time-reversal invariant single Dirac node seems to be obstructed. Perhaps there is some non-trivial interplay between this anomaly and the Lieb-Schultz-Mattis type theorems and approaches~\cite{10.21468/SciPostPhys.15.2.051}? We discuss some implications on LSM theorems from the perspective of flux threading in the Appendix~\ref{sec:pottingsoil}. A generalization of this sort of argument for other dimensions may also yield interesting possibilities or obstructions. Finally, we can ask how we can generalize these results to develop theorems that will apply to other symmetry classes of fermionic SPTs in all spacetime dimensions.

\textit{Acknowledgements:}  The authors thank Ryan Thorngren, Chong Wang, Ashvin Vishwanath, Charlie Kane, and Julian May-Mann for insightful discussions. The authors are grateful to the Kavli Institute for Theoretical Physics for hosting us during the KITP program: A Quantum Universe in a Crystal supported in part by grant NSF PHY-2309135. LG acknowledges support from the Walter Burke Institute for Theoretical Physics at Caltech and the Caltech Institute for Quantum Information and Matter. 
AAB was supported by the Natural Sciences and Engineering Research Council (NSERC) of Canada 
and by the Center for Advancement of Topological Semimetals, an Energy Frontier Research Center funded by the U.S. Department of Energy Office of Science, Office of Basic Energy Sciences, through the Ames Laboratory under contract DE-AC02-07CH11358. 
Research at Perimeter Institute is supported in part by the Government of Canada through the Department of Innovation, Science and Economic Development and by the Province of Ontario through the Ministry of Economic Development, Job Creation and Trade. TLH thanks ARO MURI W911NF2020166 for support.

\bibliography{arxiv}

\begin{thebibliography}{47}%
\makeatletter
\providecommand \@ifxundefined [1]{%
 \@ifx{#1\undefined}
}%
\providecommand \@ifnum [1]{%
 \ifnum #1\expandafter \@firstoftwo
 \else \expandafter \@secondoftwo
 \fi
}%
\providecommand \@ifx [1]{%
 \ifx #1\expandafter \@firstoftwo
 \else \expandafter \@secondoftwo
 \fi
}%
\providecommand \natexlab [1]{#1}%
\providecommand \enquote  [1]{``#1''}%
\providecommand \bibnamefont  [1]{#1}%
\providecommand \bibfnamefont [1]{#1}%
\providecommand \citenamefont [1]{#1}%
\providecommand \href@noop [0]{\@secondoftwo}%
\providecommand \href [0]{\begingroup \@sanitize@url \@href}%
\providecommand \@href[1]{\@@startlink{#1}\@@href}%
\providecommand \@@href[1]{\endgroup#1\@@endlink}%
\providecommand \@sanitize@url [0]{\catcode `\\12\catcode `\$12\catcode `\&12\catcode `\#12\catcode `\^12\catcode `\_12\catcode `\%12\relax}%
\providecommand \@@startlink[1]{}%
\providecommand \@@endlink[0]{}%
\providecommand \url  [0]{\begingroup\@sanitize@url \@url }%
\providecommand \@url [1]{\endgroup\@href {#1}{\urlprefix }}%
\providecommand \urlprefix  [0]{URL }%
\providecommand \Eprint [0]{\href }%
\providecommand \doibase [0]{https://doi.org/}%
\providecommand \selectlanguage [0]{\@gobble}%
\providecommand \bibinfo  [0]{\@secondoftwo}%
\providecommand \bibfield  [0]{\@secondoftwo}%
\providecommand \translation [1]{[#1]}%
\providecommand \BibitemOpen [0]{}%
\providecommand \bibitemStop [0]{}%
\providecommand \bibitemNoStop [0]{.\EOS\space}%
\providecommand \EOS [0]{\spacefactor3000\relax}%
\providecommand \BibitemShut  [1]{\csname bibitem#1\endcsname}%
\let\auto@bib@innerbib\@empty
\bibitem [{\citenamefont {Nielsen}\ and\ \citenamefont {Ninomiya}(1981{\natexlab{a}})}]{NIELSEN198120}%
  \BibitemOpen
  \bibfield  {author} {\bibinfo {author} {\bibfnamefont {H.}~\bibnamefont {Nielsen}}\ and\ \bibinfo {author} {\bibfnamefont {M.}~\bibnamefont {Ninomiya}},\ }\bibfield  {title} {\bibinfo {title} {Absence of neutrinos on a lattice: (i). proof by homotopy theory},\ }\href {https://doi.org/https://doi.org/10.1016/0550-3213(81)90361-8} {\bibfield  {journal} {\bibinfo  {journal} {Nuclear Physics B}\ }\textbf {\bibinfo {volume} {185}},\ \bibinfo {pages} {20} (\bibinfo {year} {1981}{\natexlab{a}})}\BibitemShut {NoStop}%
\bibitem [{\citenamefont {Nielsen}\ and\ \citenamefont {Ninomiya}(1981{\natexlab{b}})}]{NIELSEN1981173}%
  \BibitemOpen
  \bibfield  {author} {\bibinfo {author} {\bibfnamefont {H.}~\bibnamefont {Nielsen}}\ and\ \bibinfo {author} {\bibfnamefont {M.}~\bibnamefont {Ninomiya}},\ }\bibfield  {title} {\bibinfo {title} {Absence of neutrinos on a lattice: (ii). intuitive topological proof},\ }\href {https://doi.org/https://doi.org/10.1016/0550-3213(81)90524-1} {\bibfield  {journal} {\bibinfo  {journal} {Nuclear Physics B}\ }\textbf {\bibinfo {volume} {193}},\ \bibinfo {pages} {173} (\bibinfo {year} {1981}{\natexlab{b}})}\BibitemShut {NoStop}%
\bibitem [{\citenamefont {Friedan}(1982)}]{Friedan82}%
  \BibitemOpen
  \bibfield  {author} {\bibinfo {author} {\bibfnamefont {D.}~\bibnamefont {Friedan}},\ }\bibfield  {title} {\bibinfo {title} {A proof of the nielsen-ninomiya theorem},\ }\bibfield  {journal} {\bibinfo  {journal} {Communications in Mathematical Physics}\ }\textbf {\bibinfo {volume} {85}},\ \href {https://projecteuclid.org/journals/communications-in-mathematical-physics/volume-85/issue-4/A-proof-of-the-Nielsen-Ninomiya-theorem/cmp/1103921543.full?tab=ArticleLink} {} (\bibinfo {year} {1982})\BibitemShut {NoStop}%
\bibitem [{\citenamefont {Fidkowski}\ and\ \citenamefont {Xu}(2023)}]{Fidkowski_2023}%
  \BibitemOpen
  \bibfield  {author} {\bibinfo {author} {\bibfnamefont {L.}~\bibnamefont {Fidkowski}}\ and\ \bibinfo {author} {\bibfnamefont {C.}~\bibnamefont {Xu}},\ }\bibfield  {title} {\bibinfo {title} {A no-go result for implementing chiral symmetries by locality-preserving unitaries in a three-dimensional hamiltonian lattice model of fermions},\ }\bibfield  {journal} {\bibinfo  {journal} {Physical Review Letters}\ }\textbf {\bibinfo {volume} {131}},\ \href {https://doi.org/10.1103/physrevlett.131.196601} {10.1103/physrevlett.131.196601} (\bibinfo {year} {2023})\BibitemShut {NoStop}%
\bibitem [{\citenamefont {O'Brien}\ \emph {et~al.}(2017)\citenamefont {O'Brien}, \citenamefont {Beenakker},\ and\ \citenamefont {Adagideli}}]{PhysRevLett.118.207701}%
  \BibitemOpen
  \bibfield  {author} {\bibinfo {author} {\bibfnamefont {T.~E.}\ \bibnamefont {O'Brien}}, \bibinfo {author} {\bibfnamefont {C.~W.~J.}\ \bibnamefont {Beenakker}},\ and\ \bibinfo {author} {\bibfnamefont {i.~d.~I.}\ \bibnamefont {Adagideli}},\ }\bibfield  {title} {\bibinfo {title} {Superconductivity provides access to the chiral magnetic effect of an unpaired weyl cone},\ }\href {https://doi.org/10.1103/PhysRevLett.118.207701} {\bibfield  {journal} {\bibinfo  {journal} {Phys. Rev. Lett.}\ }\textbf {\bibinfo {volume} {118}},\ \bibinfo {pages} {207701} (\bibinfo {year} {2017})}\BibitemShut {NoStop}%
\bibitem [{\citenamefont {Gioia}\ and\ \citenamefont {Thorngren}(2025)}]{gioia2025exactchiralsymmetries31d}%
  \BibitemOpen
  \bibfield  {author} {\bibinfo {author} {\bibfnamefont {L.}~\bibnamefont {Gioia}}\ and\ \bibinfo {author} {\bibfnamefont {R.}~\bibnamefont {Thorngren}},\ }\href {https://arxiv.org/abs/2503.07708} {\bibinfo {title} {Exact chiral symmetries of 3+1d hamiltonian lattice fermions}} (\bibinfo {year} {2025}),\ \Eprint {https://arxiv.org/abs/2503.07708} {arXiv:2503.07708 [cond-mat.str-el]} \BibitemShut {NoStop}%
\bibitem [{\citenamefont {Meng}\ and\ \citenamefont {Balents}(2012)}]{Meng12}%
  \BibitemOpen
  \bibfield  {author} {\bibinfo {author} {\bibfnamefont {T.}~\bibnamefont {Meng}}\ and\ \bibinfo {author} {\bibfnamefont {L.}~\bibnamefont {Balents}},\ }\bibfield  {title} {\bibinfo {title} {Weyl superconductors},\ }\href {https://doi.org/10.1103/PhysRevB.86.054504} {\bibfield  {journal} {\bibinfo  {journal} {Phys. Rev. B}\ }\textbf {\bibinfo {volume} {86}},\ \bibinfo {pages} {054504} (\bibinfo {year} {2012})}\BibitemShut {NoStop}%
\bibitem [{\citenamefont {Meng}\ and\ \citenamefont {Budich}(2019)}]{Meng19}%
  \BibitemOpen
  \bibfield  {author} {\bibinfo {author} {\bibfnamefont {T.}~\bibnamefont {Meng}}\ and\ \bibinfo {author} {\bibfnamefont {J.~C.}\ \bibnamefont {Budich}},\ }\bibfield  {title} {\bibinfo {title} {Unpaired weyl nodes from long-ranged interactions: Fate of quantum anomalies},\ }\href {https://doi.org/10.1103/PhysRevLett.122.046402} {\bibfield  {journal} {\bibinfo  {journal} {Phys. Rev. Lett.}\ }\textbf {\bibinfo {volume} {122}},\ \bibinfo {pages} {046402} (\bibinfo {year} {2019})}\BibitemShut {NoStop}%
\bibitem [{\citenamefont {Qi}\ \emph {et~al.}(2008)\citenamefont {Qi}, \citenamefont {Hughes},\ and\ \citenamefont {Zhang}}]{Qi08}%
  \BibitemOpen
  \bibfield  {author} {\bibinfo {author} {\bibfnamefont {X.-L.}\ \bibnamefont {Qi}}, \bibinfo {author} {\bibfnamefont {T.~L.}\ \bibnamefont {Hughes}},\ and\ \bibinfo {author} {\bibfnamefont {S.-C.}\ \bibnamefont {Zhang}},\ }\bibfield  {title} {\bibinfo {title} {Topological field theory of time-reversal invariant insulators},\ }\href {https://doi.org/10.1103/PhysRevB.78.195424} {\bibfield  {journal} {\bibinfo  {journal} {Phys. Rev. B}\ }\textbf {\bibinfo {volume} {78}},\ \bibinfo {pages} {195424} (\bibinfo {year} {2008})}\BibitemShut {NoStop}%
\bibitem [{\citenamefont {Yang}\ and\ \citenamefont {Nagaosa}(2014)}]{Nagaosa14}%
  \BibitemOpen
  \bibfield  {author} {\bibinfo {author} {\bibfnamefont {B.-J.}\ \bibnamefont {Yang}}\ and\ \bibinfo {author} {\bibfnamefont {N.}~\bibnamefont {Nagaosa}},\ }\bibfield  {title} {\bibinfo {title} {Classification of stable three-dimensional dirac semimetals with nontrivial topology},\ }\href {http://dx.doi.org/10.1038/ncomms5898} {\bibfield  {journal} {\bibinfo  {journal} {Nat Commun}\ }\textbf {\bibinfo {volume} {5}} (\bibinfo {year} {2014})}\BibitemShut {NoStop}%
\bibitem [{\citenamefont {Armitage}\ \emph {et~al.}(2018)\citenamefont {Armitage}, \citenamefont {Mele},\ and\ \citenamefont {Vishwanath}}]{RevModPhys.90.015001}%
  \BibitemOpen
  \bibfield  {author} {\bibinfo {author} {\bibfnamefont {N.~P.}\ \bibnamefont {Armitage}}, \bibinfo {author} {\bibfnamefont {E.~J.}\ \bibnamefont {Mele}},\ and\ \bibinfo {author} {\bibfnamefont {A.}~\bibnamefont {Vishwanath}},\ }\bibfield  {title} {\bibinfo {title} {Weyl and dirac semimetals in three-dimensional solids},\ }\href {https://doi.org/10.1103/RevModPhys.90.015001} {\bibfield  {journal} {\bibinfo  {journal} {Rev. Mod. Phys.}\ }\textbf {\bibinfo {volume} {90}},\ \bibinfo {pages} {015001} (\bibinfo {year} {2018})}\BibitemShut {NoStop}%
\bibitem [{\citenamefont {Fu}\ \emph {et~al.}(2007)\citenamefont {Fu}, \citenamefont {Kane},\ and\ \citenamefont {Mele}}]{FKM}%
  \BibitemOpen
  \bibfield  {author} {\bibinfo {author} {\bibfnamefont {L.}~\bibnamefont {Fu}}, \bibinfo {author} {\bibfnamefont {C.~L.}\ \bibnamefont {Kane}},\ and\ \bibinfo {author} {\bibfnamefont {E.~J.}\ \bibnamefont {Mele}},\ }\bibfield  {title} {\bibinfo {title} {Topological insulators in three dimensions},\ }\href {https://doi.org/10.1103/PhysRevLett.98.106803} {\bibfield  {journal} {\bibinfo  {journal} {Phys. Rev. Lett.}\ }\textbf {\bibinfo {volume} {98}},\ \bibinfo {pages} {106803} (\bibinfo {year} {2007})}\BibitemShut {NoStop}%
\bibitem [{\citenamefont {Steinberg}\ \emph {et~al.}(2014)\citenamefont {Steinberg}, \citenamefont {Young}, \citenamefont {Zaheer}, \citenamefont {Kane}, \citenamefont {Mele},\ and\ \citenamefont {Rappe}}]{PhysRevLett.112.036403}%
  \BibitemOpen
  \bibfield  {author} {\bibinfo {author} {\bibfnamefont {J.~A.}\ \bibnamefont {Steinberg}}, \bibinfo {author} {\bibfnamefont {S.~M.}\ \bibnamefont {Young}}, \bibinfo {author} {\bibfnamefont {S.}~\bibnamefont {Zaheer}}, \bibinfo {author} {\bibfnamefont {C.~L.}\ \bibnamefont {Kane}}, \bibinfo {author} {\bibfnamefont {E.~J.}\ \bibnamefont {Mele}},\ and\ \bibinfo {author} {\bibfnamefont {A.~M.}\ \bibnamefont {Rappe}},\ }\bibfield  {title} {\bibinfo {title} {Bulk dirac points in distorted spinels},\ }\href {https://doi.org/10.1103/PhysRevLett.112.036403} {\bibfield  {journal} {\bibinfo  {journal} {Phys. Rev. Lett.}\ }\textbf {\bibinfo {volume} {112}},\ \bibinfo {pages} {036403} (\bibinfo {year} {2014})}\BibitemShut {NoStop}%
\bibitem [{\citenamefont {Cho}\ \emph {et~al.}(2017)\citenamefont {Cho}, \citenamefont {Hsieh},\ and\ \citenamefont {Ryu}}]{PhysRevB.96.195105}%
  \BibitemOpen
  \bibfield  {author} {\bibinfo {author} {\bibfnamefont {G.~Y.}\ \bibnamefont {Cho}}, \bibinfo {author} {\bibfnamefont {C.-T.}\ \bibnamefont {Hsieh}},\ and\ \bibinfo {author} {\bibfnamefont {S.}~\bibnamefont {Ryu}},\ }\bibfield  {title} {\bibinfo {title} {Anomaly manifestation of lieb-schultz-mattis theorem and topological phases},\ }\href {https://doi.org/10.1103/PhysRevB.96.195105} {\bibfield  {journal} {\bibinfo  {journal} {Phys. Rev. B}\ }\textbf {\bibinfo {volume} {96}},\ \bibinfo {pages} {195105} (\bibinfo {year} {2017})}\BibitemShut {NoStop}%
\bibitem [{\citenamefont {Xu}\ \emph {et~al.}(2015)\citenamefont {Xu}, \citenamefont {Belopolski}, \citenamefont {Alidoust}, \citenamefont {Neupane}, \citenamefont {Bian}, \citenamefont {Zhang}, \citenamefont {Sankar}, \citenamefont {Chang}, \citenamefont {Yuan}, \citenamefont {Lee}, \citenamefont {Huang}, \citenamefont {Zheng}, \citenamefont {Ma}, \citenamefont {Sanchez}, \citenamefont {Wang}, \citenamefont {Bansil}, \citenamefont {Chou}, \citenamefont {Shibayev}, \citenamefont {Lin}, \citenamefont {Jia},\ and\ \citenamefont {Hasan}}]{Xu613}%
  \BibitemOpen
  \bibfield  {author} {\bibinfo {author} {\bibfnamefont {S.-Y.}\ \bibnamefont {Xu}}, \bibinfo {author} {\bibfnamefont {I.}~\bibnamefont {Belopolski}}, \bibinfo {author} {\bibfnamefont {N.}~\bibnamefont {Alidoust}}, \bibinfo {author} {\bibfnamefont {M.}~\bibnamefont {Neupane}}, \bibinfo {author} {\bibfnamefont {G.}~\bibnamefont {Bian}}, \bibinfo {author} {\bibfnamefont {C.}~\bibnamefont {Zhang}}, \bibinfo {author} {\bibfnamefont {R.}~\bibnamefont {Sankar}}, \bibinfo {author} {\bibfnamefont {G.}~\bibnamefont {Chang}}, \bibinfo {author} {\bibfnamefont {Z.}~\bibnamefont {Yuan}}, \bibinfo {author} {\bibfnamefont {C.-C.}\ \bibnamefont {Lee}}, \bibinfo {author} {\bibfnamefont {S.-M.}\ \bibnamefont {Huang}}, \bibinfo {author} {\bibfnamefont {H.}~\bibnamefont {Zheng}}, \bibinfo {author} {\bibfnamefont {J.}~\bibnamefont {Ma}}, \bibinfo {author} {\bibfnamefont {D.~S.}\ \bibnamefont {Sanchez}}, \bibinfo {author} {\bibfnamefont {B.}~\bibnamefont {Wang}}, \bibinfo {author} {\bibfnamefont {A.}~\bibnamefont {Bansil}},
  \bibinfo {author} {\bibfnamefont {F.}~\bibnamefont {Chou}}, \bibinfo {author} {\bibfnamefont {P.~P.}\ \bibnamefont {Shibayev}}, \bibinfo {author} {\bibfnamefont {H.}~\bibnamefont {Lin}}, \bibinfo {author} {\bibfnamefont {S.}~\bibnamefont {Jia}},\ and\ \bibinfo {author} {\bibfnamefont {M.~Z.}\ \bibnamefont {Hasan}},\ }\bibfield  {title} {\bibinfo {title} {Discovery of a weyl fermion semimetal and topological fermi arcs},\ }\href {https://doi.org/10.1126/science.aaa9297} {\bibfield  {journal} {\bibinfo  {journal} {Science}\ }\textbf {\bibinfo {volume} {349}},\ \bibinfo {pages} {613} (\bibinfo {year} {2015})},\ \Eprint {https://arxiv.org/abs/https://science.sciencemag.org/content/349/6248/613.full.pdf} {https://science.sciencemag.org/content/349/6248/613.full.pdf} \BibitemShut {NoStop}%
\bibitem [{\citenamefont {Burkov}\ and\ \citenamefont {Balents}(2011)}]{Burkov11-1}%
  \BibitemOpen
  \bibfield  {author} {\bibinfo {author} {\bibfnamefont {A.~A.}\ \bibnamefont {Burkov}}\ and\ \bibinfo {author} {\bibfnamefont {L.}~\bibnamefont {Balents}},\ }\bibfield  {title} {\bibinfo {title} {Weyl semimetal in a topological insulator multilayer},\ }\href {https://doi.org/10.1103/PhysRevLett.107.127205} {\bibfield  {journal} {\bibinfo  {journal} {Phys. Rev. Lett.}\ }\textbf {\bibinfo {volume} {107}},\ \bibinfo {pages} {127205} (\bibinfo {year} {2011})}\BibitemShut {NoStop}%
\bibitem [{\citenamefont {{Burkov}}(2018)}]{Burkov_ARCMP}%
  \BibitemOpen
  \bibfield  {author} {\bibinfo {author} {\bibfnamefont {A.~A.}\ \bibnamefont {{Burkov}}},\ }\bibfield  {title} {\bibinfo {title} {{Weyl Metals}},\ }\href@noop {} {\bibfield  {journal} {\bibinfo  {journal} {Annual Review of Condensed Matter Physics}\ }\textbf {\bibinfo {volume} {9}},\ \bibinfo {pages} {359} (\bibinfo {year} {2018})}\BibitemShut {NoStop}%
\bibitem [{\citenamefont {Yan}\ and\ \citenamefont {Felser}(2017)}]{Felser_ARCMP}%
  \BibitemOpen
  \bibfield  {author} {\bibinfo {author} {\bibfnamefont {B.}~\bibnamefont {Yan}}\ and\ \bibinfo {author} {\bibfnamefont {C.}~\bibnamefont {Felser}},\ }\bibfield  {title} {\bibinfo {title} {Topological materials: Weyl semimetals},\ }\href@noop {} {\bibfield  {journal} {\bibinfo  {journal} {Annual Review of Condensed Matter Physics}\ }\textbf {\bibinfo {volume} {8}} (\bibinfo {year} {2017})}\BibitemShut {NoStop}%
\bibitem [{\citenamefont {Wang}\ and\ \citenamefont {Sau}(2024)}]{wang2024interactionrobustnesschiralanomaly}%
  \BibitemOpen
  \bibfield  {author} {\bibinfo {author} {\bibfnamefont {S.}~\bibnamefont {Wang}}\ and\ \bibinfo {author} {\bibfnamefont {J.~D.}\ \bibnamefont {Sau}},\ }\href {https://arxiv.org/abs/2401.09409} {\bibinfo {title} {Interaction robustness of the chiral anomaly in weyl semimetals and luttinger liquids from a mixed anomaly approach}} (\bibinfo {year} {2024}),\ \Eprint {https://arxiv.org/abs/2401.09409} {arXiv:2401.09409 [cond-mat.str-el]} \BibitemShut {NoStop}%
\bibitem [{\citenamefont {Yi}\ \emph {et~al.}(2023)\citenamefont {Yi}, \citenamefont {Ying}, \citenamefont {Gioia},\ and\ \citenamefont {Burkov}}]{PhysRevB.107.115147}%
  \BibitemOpen
  \bibfield  {author} {\bibinfo {author} {\bibfnamefont {J.}~\bibnamefont {Yi}}, \bibinfo {author} {\bibfnamefont {X.}~\bibnamefont {Ying}}, \bibinfo {author} {\bibfnamefont {L.}~\bibnamefont {Gioia}},\ and\ \bibinfo {author} {\bibfnamefont {A.~A.}\ \bibnamefont {Burkov}},\ }\bibfield  {title} {\bibinfo {title} {Topological order in interacting semimetals},\ }\href {https://doi.org/10.1103/PhysRevB.107.115147} {\bibfield  {journal} {\bibinfo  {journal} {Phys. Rev. B}\ }\textbf {\bibinfo {volume} {107}},\ \bibinfo {pages} {115147} (\bibinfo {year} {2023})}\BibitemShut {NoStop}%
\bibitem [{\citenamefont {Cheng}\ and\ \citenamefont {Seiberg}(2023)}]{10.21468/SciPostPhys.15.2.051}%
  \BibitemOpen
  \bibfield  {author} {\bibinfo {author} {\bibfnamefont {M.}~\bibnamefont {Cheng}}\ and\ \bibinfo {author} {\bibfnamefont {N.}~\bibnamefont {Seiberg}},\ }\bibfield  {title} {\bibinfo {title} {{Lieb-Schultz-Mattis, Luttinger, and 't Hooft - anomaly matching in lattice systems}},\ }\href {https://doi.org/10.21468/SciPostPhys.15.2.051} {\bibfield  {journal} {\bibinfo  {journal} {SciPost Phys.}\ }\textbf {\bibinfo {volume} {15}},\ \bibinfo {pages} {051} (\bibinfo {year} {2023})}\BibitemShut {NoStop}%
\bibitem [{\citenamefont {Thorngren}\ \emph {et~al.}(2021)\citenamefont {Thorngren}, \citenamefont {Vishwanath},\ and\ \citenamefont {Verresen}}]{PhysRevB.104.075132}%
  \BibitemOpen
  \bibfield  {author} {\bibinfo {author} {\bibfnamefont {R.}~\bibnamefont {Thorngren}}, \bibinfo {author} {\bibfnamefont {A.}~\bibnamefont {Vishwanath}},\ and\ \bibinfo {author} {\bibfnamefont {R.}~\bibnamefont {Verresen}},\ }\bibfield  {title} {\bibinfo {title} {Intrinsically gapless topological phases},\ }\href {https://doi.org/10.1103/PhysRevB.104.075132} {\bibfield  {journal} {\bibinfo  {journal} {Phys. Rev. B}\ }\textbf {\bibinfo {volume} {104}},\ \bibinfo {pages} {075132} (\bibinfo {year} {2021})}\BibitemShut {NoStop}%
\bibitem [{\citenamefont {Else}\ \emph {et~al.}(2021)\citenamefont {Else}, \citenamefont {Thorngren},\ and\ \citenamefont {Senthil}}]{PhysRevX.11.021005}%
  \BibitemOpen
  \bibfield  {author} {\bibinfo {author} {\bibfnamefont {D.~V.}\ \bibnamefont {Else}}, \bibinfo {author} {\bibfnamefont {R.}~\bibnamefont {Thorngren}},\ and\ \bibinfo {author} {\bibfnamefont {T.}~\bibnamefont {Senthil}},\ }\bibfield  {title} {\bibinfo {title} {Non-fermi liquids as ersatz fermi liquids: General constraints on compressible metals},\ }\href {https://doi.org/10.1103/PhysRevX.11.021005} {\bibfield  {journal} {\bibinfo  {journal} {Phys. Rev. X}\ }\textbf {\bibinfo {volume} {11}},\ \bibinfo {pages} {021005} (\bibinfo {year} {2021})}\BibitemShut {NoStop}%
\bibitem [{\citenamefont {Ye}\ \emph {et~al.}(2022)\citenamefont {Ye}, \citenamefont {Guo}, \citenamefont {He}, \citenamefont {Wang},\ and\ \citenamefont {Zou}}]{10.21468/SciPostPhys.13.3.066}%
  \BibitemOpen
  \bibfield  {author} {\bibinfo {author} {\bibfnamefont {W.}~\bibnamefont {Ye}}, \bibinfo {author} {\bibfnamefont {M.}~\bibnamefont {Guo}}, \bibinfo {author} {\bibfnamefont {Y.-C.}\ \bibnamefont {He}}, \bibinfo {author} {\bibfnamefont {C.}~\bibnamefont {Wang}},\ and\ \bibinfo {author} {\bibfnamefont {L.}~\bibnamefont {Zou}},\ }\bibfield  {title} {\bibinfo {title} {{Topological characterization of Lieb-Schultz-Mattis constraints and applications to symmetry-enriched quantum criticality}},\ }\href {https://doi.org/10.21468/SciPostPhys.13.3.066} {\bibfield  {journal} {\bibinfo  {journal} {SciPost Phys.}\ }\textbf {\bibinfo {volume} {13}},\ \bibinfo {pages} {066} (\bibinfo {year} {2022})}\BibitemShut {NoStop}%
\bibitem [{\citenamefont {Nakahara}(2003)}]{nakahara}%
  \BibitemOpen
  \bibfield  {author} {\bibinfo {author} {\bibfnamefont {M.}~\bibnamefont {Nakahara}},\ }\href@noop {} {\emph {\bibinfo {title} {Geometry, Topology and Physics}}},\ \bibinfo {edition} {2nd}\ ed.\ (\bibinfo  {publisher} {CRC Press},\ \bibinfo {year} {2003})\BibitemShut {NoStop}%
\bibitem [{\citenamefont {Witten}(2016)}]{RevModPhys.88.035001}%
  \BibitemOpen
  \bibfield  {author} {\bibinfo {author} {\bibfnamefont {E.}~\bibnamefont {Witten}},\ }\bibfield  {title} {\bibinfo {title} {Fermion path integrals and topological phases},\ }\href {https://doi.org/10.1103/RevModPhys.88.035001} {\bibfield  {journal} {\bibinfo  {journal} {Rev. Mod. Phys.}\ }\textbf {\bibinfo {volume} {88}},\ \bibinfo {pages} {035001} (\bibinfo {year} {2016})}\BibitemShut {NoStop}%
\bibitem [{\citenamefont {Arouca}\ \emph {et~al.}(2022)\citenamefont {Arouca}, \citenamefont {Cappelli},\ and\ \citenamefont {Hansson}}]{10.21468/SciPostPhysLectNotes.62}%
  \BibitemOpen
  \bibfield  {author} {\bibinfo {author} {\bibfnamefont {R.}~\bibnamefont {Arouca}}, \bibinfo {author} {\bibfnamefont {A.}~\bibnamefont {Cappelli}},\ and\ \bibinfo {author} {\bibfnamefont {T.~H.}\ \bibnamefont {Hansson}},\ }\bibfield  {title} {\bibinfo {title} {{Quantum Field Theory Anomalies in Condensed Matter Physics}},\ }\href {https://doi.org/10.21468/SciPostPhysLectNotes.62} {\bibfield  {journal} {\bibinfo  {journal} {SciPost Phys. Lect. Notes}\ ,\ \bibinfo {pages} {62}} (\bibinfo {year} {2022})}\BibitemShut {NoStop}%
\bibitem [{\citenamefont {Tong}(2018)}]{tonggaugetheory}%
  \BibitemOpen
  \bibfield  {author} {\bibinfo {author} {\bibfnamefont {D.}~\bibnamefont {Tong}},\ }\href {https://www.damtp.cam.ac.uk/user/tong/gaugetheory/gt.pdf} {\bibinfo {title} {Gauge theory}} (\bibinfo {year} {2018})\BibitemShut {NoStop}%
\bibitem [{\citenamefont {Bilal}(2008)}]{bilal2008lecturesanomalies}%
  \BibitemOpen
  \bibfield  {author} {\bibinfo {author} {\bibfnamefont {A.}~\bibnamefont {Bilal}},\ }\href {https://arxiv.org/abs/0802.0634} {\bibinfo {title} {Lectures on anomalies}} (\bibinfo {year} {2008}),\ \Eprint {https://arxiv.org/abs/0802.0634} {arXiv:0802.0634 [hep-th]} \BibitemShut {NoStop}%
\bibitem [{\citenamefont {Callan}\ and\ \citenamefont {Harvey}(1985)}]{CALLAN1985427}%
  \BibitemOpen
  \bibfield  {author} {\bibinfo {author} {\bibfnamefont {C.}~\bibnamefont {Callan}}\ and\ \bibinfo {author} {\bibfnamefont {J.}~\bibnamefont {Harvey}},\ }\bibfield  {title} {\bibinfo {title} {Anomalies and fermion zero modes on strings and domain walls},\ }\href {https://doi.org/https://doi.org/10.1016/0550-3213(85)90489-4} {\bibfield  {journal} {\bibinfo  {journal} {Nuclear Physics B}\ }\textbf {\bibinfo {volume} {250}},\ \bibinfo {pages} {427} (\bibinfo {year} {1985})}\BibitemShut {NoStop}%
\bibitem [{\citenamefont {Thorngren}\ and\ \citenamefont {Else}(2018)}]{PhysRevX.8.011040}%
  \BibitemOpen
  \bibfield  {author} {\bibinfo {author} {\bibfnamefont {R.}~\bibnamefont {Thorngren}}\ and\ \bibinfo {author} {\bibfnamefont {D.~V.}\ \bibnamefont {Else}},\ }\bibfield  {title} {\bibinfo {title} {Gauging spatial symmetries and the classification of topological crystalline phases},\ }\href {https://doi.org/10.1103/PhysRevX.8.011040} {\bibfield  {journal} {\bibinfo  {journal} {Phys. Rev. X}\ }\textbf {\bibinfo {volume} {8}},\ \bibinfo {pages} {011040} (\bibinfo {year} {2018})}\BibitemShut {NoStop}%
\bibitem [{\citenamefont {Weinberg}(1995)}]{Weinberg_1995}%
  \BibitemOpen
  \bibfield  {author} {\bibinfo {author} {\bibfnamefont {S.}~\bibnamefont {Weinberg}},\ }\href@noop {} {\emph {\bibinfo {title} {The Quantum Theory of Fields}}}\ (\bibinfo  {publisher} {Cambridge University Press},\ \bibinfo {year} {1995})\BibitemShut {NoStop}%
\bibitem [{\citenamefont {Zaheer}(2014)}]{Zaheerthesis}%
  \BibitemOpen
  \bibfield  {author} {\bibinfo {author} {\bibfnamefont {S.}~\bibnamefont {Zaheer}},\ }\emph {\bibinfo {title} {Three dimensional Dirac semimetals}},\ \href@noop {} {Ph.D. thesis},\ \bibinfo  {school} {University of Pennsylvania} (\bibinfo {year} {2014})\BibitemShut {NoStop}%
\bibitem [{\citenamefont {Nielsen}\ and\ \citenamefont {Ninomiya}(1983)}]{Nielsen83}%
  \BibitemOpen
  \bibfield  {author} {\bibinfo {author} {\bibfnamefont {H.}~\bibnamefont {Nielsen}}\ and\ \bibinfo {author} {\bibfnamefont {M.}~\bibnamefont {Ninomiya}},\ }\bibfield  {title} {\bibinfo {title} {The adler-bell-jackiw anomaly and weyl fermions in a crystal},\ }\href {https://doi.org/http://dx.doi.org/10.1016/0370-2693(83)91529-0} {\bibfield  {journal} {\bibinfo  {journal} {Physics Letters B}\ }\textbf {\bibinfo {volume} {130}},\ \bibinfo {pages} {389 } (\bibinfo {year} {1983})}\BibitemShut {NoStop}%
\bibitem [{\citenamefont {Gioia}\ \emph {et~al.}(2021)\citenamefont {Gioia}, \citenamefont {Wang},\ and\ \citenamefont {Burkov}}]{PhysRevResearch.3.043067}%
  \BibitemOpen
  \bibfield  {author} {\bibinfo {author} {\bibfnamefont {L.}~\bibnamefont {Gioia}}, \bibinfo {author} {\bibfnamefont {C.}~\bibnamefont {Wang}},\ and\ \bibinfo {author} {\bibfnamefont {A.~A.}\ \bibnamefont {Burkov}},\ }\bibfield  {title} {\bibinfo {title} {Unquantized anomalies in topological semimetals},\ }\href {https://doi.org/10.1103/PhysRevResearch.3.043067} {\bibfield  {journal} {\bibinfo  {journal} {Phys. Rev. Res.}\ }\textbf {\bibinfo {volume} {3}},\ \bibinfo {pages} {043067} (\bibinfo {year} {2021})}\BibitemShut {NoStop}%
\bibitem [{\citenamefont {Gioia}\ and\ \citenamefont {Wang}(2022)}]{PhysRevX.12.031007}%
  \BibitemOpen
  \bibfield  {author} {\bibinfo {author} {\bibfnamefont {L.}~\bibnamefont {Gioia}}\ and\ \bibinfo {author} {\bibfnamefont {C.}~\bibnamefont {Wang}},\ }\bibfield  {title} {\bibinfo {title} {Nonzero momentum requires long-range entanglement},\ }\href {https://doi.org/10.1103/PhysRevX.12.031007} {\bibfield  {journal} {\bibinfo  {journal} {Phys. Rev. X}\ }\textbf {\bibinfo {volume} {12}},\ \bibinfo {pages} {031007} (\bibinfo {year} {2022})}\BibitemShut {NoStop}%
\bibitem [{\citenamefont {Metlitski}\ and\ \citenamefont {Thorngren}(2018)}]{PhysRevB.98.085140}%
  \BibitemOpen
  \bibfield  {author} {\bibinfo {author} {\bibfnamefont {M.~A.}\ \bibnamefont {Metlitski}}\ and\ \bibinfo {author} {\bibfnamefont {R.}~\bibnamefont {Thorngren}},\ }\bibfield  {title} {\bibinfo {title} {Intrinsic and emergent anomalies at deconfined critical points},\ }\href {https://doi.org/10.1103/PhysRevB.98.085140} {\bibfield  {journal} {\bibinfo  {journal} {Phys. Rev. B}\ }\textbf {\bibinfo {volume} {98}},\ \bibinfo {pages} {085140} (\bibinfo {year} {2018})}\BibitemShut {NoStop}%
\bibitem [{\citenamefont {Vernier}\ \emph {et~al.}(2019)\citenamefont {Vernier}, \citenamefont {O’Brien},\ and\ \citenamefont {Fendley}}]{Vernier_2019}%
  \BibitemOpen
  \bibfield  {author} {\bibinfo {author} {\bibfnamefont {E.}~\bibnamefont {Vernier}}, \bibinfo {author} {\bibfnamefont {E.}~\bibnamefont {O’Brien}},\ and\ \bibinfo {author} {\bibfnamefont {P.}~\bibnamefont {Fendley}},\ }\bibfield  {title} {\bibinfo {title} {Onsager symmetries in $u(1)$ -invariant clock models},\ }\href {https://doi.org/10.1088/1742-5468/ab11c0} {\bibfield  {journal} {\bibinfo  {journal} {Journal of Statistical Mechanics: Theory and Experiment}\ }\textbf {\bibinfo {volume} {2019}},\ \bibinfo {pages} {043107} (\bibinfo {year} {2019})}\BibitemShut {NoStop}%
\bibitem [{\citenamefont {Chatterjee}\ \emph {et~al.}(2025)\citenamefont {Chatterjee}, \citenamefont {Pace},\ and\ \citenamefont {Shao}}]{chatterjee2025quantized}%
  \BibitemOpen
  \bibfield  {author} {\bibinfo {author} {\bibfnamefont {A.}~\bibnamefont {Chatterjee}}, \bibinfo {author} {\bibfnamefont {S.~D.}\ \bibnamefont {Pace}},\ and\ \bibinfo {author} {\bibfnamefont {S.-H.}\ \bibnamefont {Shao}},\ }\bibfield  {title} {\bibinfo {title} {Quantized axial charge of staggered fermions and the chiral anomaly},\ }\href@noop {} {\bibfield  {journal} {\bibinfo  {journal} {Physical Review Letters}\ }\textbf {\bibinfo {volume} {134}},\ \bibinfo {pages} {021601} (\bibinfo {year} {2025})}\BibitemShut {NoStop}%
\bibitem [{\citenamefont {Bagherian}\ \emph {et~al.}(2024)\citenamefont {Bagherian}, \citenamefont {Fraser}, \citenamefont {Homiller},\ and\ \citenamefont {Stout}}]{Bagherian2024}%
  \BibitemOpen
  \bibfield  {author} {\bibinfo {author} {\bibfnamefont {H.}~\bibnamefont {Bagherian}}, \bibinfo {author} {\bibfnamefont {K.}~\bibnamefont {Fraser}}, \bibinfo {author} {\bibfnamefont {S.}~\bibnamefont {Homiller}},\ and\ \bibinfo {author} {\bibfnamefont {J.}~\bibnamefont {Stout}},\ }\bibfield  {title} {\bibinfo {title} {Zero modes of massive fermions delocalize from axion strings},\ }\href {https://doi.org/10.1007/JHEP05(2024)079} {\bibfield  {journal} {\bibinfo  {journal} {Journal of High Energy Physics}\ }\textbf {\bibinfo {volume} {2024}},\ \bibinfo {pages} {79} (\bibinfo {year} {2024})}\BibitemShut {NoStop}%
\bibitem [{\citenamefont {Zee}(2003)}]{Zee_book}%
  \BibitemOpen
  \bibfield  {author} {\bibinfo {author} {\bibfnamefont {A.}~\bibnamefont {Zee}},\ }\href@noop {} {\emph {\bibinfo {title} {Quantum field theory in a nutshell}}}\ (\bibinfo  {publisher} {Princeton University Press},\ \bibinfo {year} {2003})\BibitemShut {NoStop}%
\bibitem [{\citenamefont {Wang}\ \emph {et~al.}(2020)\citenamefont {Wang}, \citenamefont {Gioia},\ and\ \citenamefont {Burkov}}]{PhysRevLett.124.096603}%
  \BibitemOpen
  \bibfield  {author} {\bibinfo {author} {\bibfnamefont {C.}~\bibnamefont {Wang}}, \bibinfo {author} {\bibfnamefont {L.}~\bibnamefont {Gioia}},\ and\ \bibinfo {author} {\bibfnamefont {A.~A.}\ \bibnamefont {Burkov}},\ }\bibfield  {title} {\bibinfo {title} {Fractional quantum hall effect in weyl semimetals},\ }\href {https://doi.org/10.1103/PhysRevLett.124.096603} {\bibfield  {journal} {\bibinfo  {journal} {Phys. Rev. Lett.}\ }\textbf {\bibinfo {volume} {124}},\ \bibinfo {pages} {096603} (\bibinfo {year} {2020})}\BibitemShut {NoStop}%
\bibitem [{\citenamefont {Wang}\ \emph {et~al.}(2012)\citenamefont {Wang}, \citenamefont {Sun}, \citenamefont {Chen}, \citenamefont {Franchini}, \citenamefont {Xu}, \citenamefont {Weng}, \citenamefont {Dai},\ and\ \citenamefont {Fang}}]{Fang12}%
  \BibitemOpen
  \bibfield  {author} {\bibinfo {author} {\bibfnamefont {Z.}~\bibnamefont {Wang}}, \bibinfo {author} {\bibfnamefont {Y.}~\bibnamefont {Sun}}, \bibinfo {author} {\bibfnamefont {X.-Q.}\ \bibnamefont {Chen}}, \bibinfo {author} {\bibfnamefont {C.}~\bibnamefont {Franchini}}, \bibinfo {author} {\bibfnamefont {G.}~\bibnamefont {Xu}}, \bibinfo {author} {\bibfnamefont {H.}~\bibnamefont {Weng}}, \bibinfo {author} {\bibfnamefont {X.}~\bibnamefont {Dai}},\ and\ \bibinfo {author} {\bibfnamefont {Z.}~\bibnamefont {Fang}},\ }\bibfield  {title} {\bibinfo {title} {Dirac semimetal and topological phase transitions in ${A}_{3}$bi ($a=\text{Na}$, k, rb)},\ }\href {https://doi.org/10.1103/PhysRevB.85.195320} {\bibfield  {journal} {\bibinfo  {journal} {Phys. Rev. B}\ }\textbf {\bibinfo {volume} {85}},\ \bibinfo {pages} {195320} (\bibinfo {year} {2012})}\BibitemShut {NoStop}%
\bibitem [{\citenamefont {Wang}\ \emph {et~al.}(2013)\citenamefont {Wang}, \citenamefont {Weng}, \citenamefont {Wu}, \citenamefont {Dai},\ and\ \citenamefont {Fang}}]{Fang13}%
  \BibitemOpen
  \bibfield  {author} {\bibinfo {author} {\bibfnamefont {Z.}~\bibnamefont {Wang}}, \bibinfo {author} {\bibfnamefont {H.}~\bibnamefont {Weng}}, \bibinfo {author} {\bibfnamefont {Q.}~\bibnamefont {Wu}}, \bibinfo {author} {\bibfnamefont {X.}~\bibnamefont {Dai}},\ and\ \bibinfo {author} {\bibfnamefont {Z.}~\bibnamefont {Fang}},\ }\bibfield  {title} {\bibinfo {title} {Three-dimensional dirac semimetal and quantum transport in cd${}_{3}$as${}_{2}$},\ }\href {https://doi.org/10.1103/PhysRevB.88.125427} {\bibfield  {journal} {\bibinfo  {journal} {Phys. Rev. B}\ }\textbf {\bibinfo {volume} {88}},\ \bibinfo {pages} {125427} (\bibinfo {year} {2013})}\BibitemShut {NoStop}%
\bibitem [{\citenamefont {May-Mann}\ \emph {et~al.}(2024)\citenamefont {May-Mann}, \citenamefont {Hirsbrunner}, \citenamefont {Gioia},\ and\ \citenamefont {Hughes}}]{PhysRevB.110.155110}%
  \BibitemOpen
  \bibfield  {author} {\bibinfo {author} {\bibfnamefont {J.}~\bibnamefont {May-Mann}}, \bibinfo {author} {\bibfnamefont {M.~R.}\ \bibnamefont {Hirsbrunner}}, \bibinfo {author} {\bibfnamefont {L.}~\bibnamefont {Gioia}},\ and\ \bibinfo {author} {\bibfnamefont {T.~L.}\ \bibnamefont {Hughes}},\ }\bibfield  {title} {\bibinfo {title} {Crystalline axion electrodynamics in charge-ordered dirac semimetals},\ }\href {https://doi.org/10.1103/PhysRevB.110.155110} {\bibfield  {journal} {\bibinfo  {journal} {Phys. Rev. B}\ }\textbf {\bibinfo {volume} {110}},\ \bibinfo {pages} {155110} (\bibinfo {year} {2024})}\BibitemShut {NoStop}%
\bibitem [{\citenamefont {Young}\ \emph {et~al.}(2012)\citenamefont {Young}, \citenamefont {Zaheer}, \citenamefont {Teo}, \citenamefont {Kane}, \citenamefont {Mele},\ and\ \citenamefont {Rappe}}]{Kane12}%
  \BibitemOpen
  \bibfield  {author} {\bibinfo {author} {\bibfnamefont {S.~M.}\ \bibnamefont {Young}}, \bibinfo {author} {\bibfnamefont {S.}~\bibnamefont {Zaheer}}, \bibinfo {author} {\bibfnamefont {J.~C.~Y.}\ \bibnamefont {Teo}}, \bibinfo {author} {\bibfnamefont {C.~L.}\ \bibnamefont {Kane}}, \bibinfo {author} {\bibfnamefont {E.~J.}\ \bibnamefont {Mele}},\ and\ \bibinfo {author} {\bibfnamefont {A.~M.}\ \bibnamefont {Rappe}},\ }\bibfield  {title} {\bibinfo {title} {Dirac semimetal in three dimensions},\ }\href {https://doi.org/10.1103/PhysRevLett.108.140405} {\bibfield  {journal} {\bibinfo  {journal} {Phys. Rev. Lett.}\ }\textbf {\bibinfo {volume} {108}},\ \bibinfo {pages} {140405} (\bibinfo {year} {2012})}\BibitemShut {NoStop}%
\bibitem [{\citenamefont {Gioia}\ \emph {et~al.}()\citenamefont {Gioia}, \citenamefont {Hughes},\ and\ \citenamefont {Thorngren}}]{leitaylorryan}%
  \BibitemOpen
  \bibfield  {author} {\bibinfo {author} {\bibfnamefont {L.}~\bibnamefont {Gioia}}, \bibinfo {author} {\bibfnamefont {T.~L.}\ \bibnamefont {Hughes}},\ and\ \bibinfo {author} {\bibfnamefont {R.}~\bibnamefont {Thorngren}},\ }\href@noop {} {\bibinfo {title} {upcoming work}}\BibitemShut {NoStop}%
\end{thebibliography}%

\appendix

\section{Dirac masses and symmetries in the Hamiltonian language}
\label{app:masses}

The Dirac fermion with the action in Eq.~\ref{eq:IRaction} can be written in the second-quantized Hamiltonian form as
\begin{align}
    \mathcal{H}&=\int d^3r \begin{pmatrix}
        \Psi^\dag(\mathbf{r}) & \Psi^T(\mathbf{r}) 
    \end{pmatrix}
    H(\mathbf{r})
    \begin{pmatrix}
        \Psi(\mathbf{r}) \\
        \Psi^*(\mathbf{r}) 
    \end{pmatrix}\quad,
\end{align}
where the kinetic term of the massless Dirac fermion is given by
\begin{align}
    H(\mathbf{r})&=\frac{1}{2}\begin{pmatrix}
        \gamma^0\gamma^j  & 0 \\
        0 & \left(\gamma^0\gamma^j\right)^T
    \end{pmatrix}_{\tau}i\partial_j\quad,\nonumber\\\
    &=\frac{1}{2}\left[\sigma^z s^x i\partial_x+\tau^z\sigma^z s^y i\partial_y+\sigma^z s^z i\partial_z\right]\quad,
    \label{eq:apphamdirac}
\end{align}
with the $\tau$ labeling the Nambu space (distinguishing between $\Psi$ and $\Psi^*$) degree of freedom. The second line of Eq.~\ref{eq:apphamdirac} is derived using the Weyl basis in Eq.~\ref{eq:weylbasis}. Recall that in this Nambu basis, all Hamiltonian terms must satisfy the redundancy relationship
\begin{align}
    \tau^x H(\mathbf{r})^* \tau^x= -H(\mathbf{r})\quad.
    \label{eq:apphamcond}
\end{align}
In this framework, the mass terms in Eq.~\ref{eq:3dmasses} are given by the Nambu Hamiltonian
\begin{align}
    H_M(\mathbf{r})=&M_1 \tau^z\sigma^x+M_2\sigma^y+M_3\tau^y\sigma^z s^y\nonumber\\
    &+M_4\tau^y s^y+M_5 \tau^x\sigma^z s^y+M_6\tau^x s^y\quad,
\end{align}
where we can see that each term mutually anticommutes with all the terms in Eq.~\ref{eq:apphamdirac}, thereby giving the Dirac fermion a mass gap. One can check that these are the only possible masses available to gap the Dirac fermion.

\subsection{Unitary symmetries}
\label{app:symmDirac}

We may also analyze internal unitary symmetries, mentioned in the main text, in this language. The $U(2)$ symmetry group that acts non-trivially on the masses is generated by
\begin{align}
    \tau^z\,, && \tau^z\sigma^z\,, && \tau^x \sigma^x s^y\,, && \tau^y \sigma^x s^y\,,
\end{align}
which are the only matrices that commute with Eq.~\ref{eq:apphamdirac}, while also satisfying condition~\ref{eq:apphamcond}. The first term is the usual $U(1)_V$ generator and the second term is $\gamma^5$, which is the chiral $U(1)_A$ generator, while the third and fourth terms can be thought of as generators of different charge conjugations. One can see that this forms the decomposition (in Eq.~\ref{eq:U(2)decomp})
\begin{align}
    U(2)= \frac{U(1)_A\times SU(2)}{\mathbb{Z}_2}\quad,
\end{align}
where $U(1)_A$ is generated by $\tau^z\sigma^z$ while $SU(2)$ is generated by $\{\tau^z,\tau^x \sigma^x s^y,\tau^y \sigma^x s^y\}$.

\subsection{Anti-unitary symmetries}

Finally, our discussion would not be complete without touching upon anti-unitary symmetries, specifically time-reversal symmetry.

In the Hamiltonian formulation, time-reversal operator $\Theta$, with $\Theta^2=(-1)^{\hat{F}}$, must commute with Eq.~\ref{eq:apphamdirac}, while acting anti-linearly on components such that $i\mapsto -i$. An operator that satisfies this relationship is given by
\begin{align}
    \Theta=-is^y \mathcal{K}\quad,
\end{align}
where $\mathcal{K}$ is the complex conjugation operator. This $\Theta$ also satisfies the additional condition that it commutes with the $U(1)_V$ charge generator $\tau^z$, as required in the main text Eq.~\ref{eq:TRcommU1}. Notice that our choice of $\Theta$, satisfying the above conditions, is not unique, since we may redefine the operator via rotations of
$$g\in \frac{U(1)_V\times U(1)_A}{\mathbb{Z}_2}\quad,$$
such that we may define a new time-reversal operator
$$\tilde{\Theta}=g \Theta\quad.$$
This redefinition will be inconsequential to the arguments and physics in the main text.

\section{Non-symmorphic Dirac semimetal model}
\label{app:nonsymDirac}

We will aim to build a symmetry-protected Dirac semimetal with a single Dirac node. We do this by taking a simple Weyl semimetal and then break translational symmetry and $\pi$ rotation symmetry but keep the conjoint symmetry, which turns into a non-symmorphic screw rotation symmetry. In this section we also demonstrate that a symmetry-protected DSM with a single Dirac node is only possible when the protecting symmetry involves a non-symmorphic symmetry, which gives rise to a non-vanishing chiral anomaly, and thus a non-integer Hall conductivity.

\subsubsection{Motivation}

Take the Bloch Hamiltonian of a simple magnetic Weyl semimetal as given in Eq.~\ref{eq:TbrokenWeyl} with $Q=0$
\begin{align}
\mathcal{H}_{\mathrm{WSM}}=&\sum_{\mathbf{k}}\begin{pmatrix}
c^\dag_{A\mathbf{k}} & c^\dag_{B\mathbf{k}}
\end{pmatrix}\bigg[
\sin k_x\, s^x+\sin k_y\, s^y \nonumber\\
+&\left(m_W(\mathbf{k})-\sin k_z a\right)s^z\bigg]\begin{pmatrix}
c_{A\mathbf{k}} \\ c_{B\mathbf{k}}
\end{pmatrix},
\label{eq:HWSM}
\end{align}
which has two Weyl nodes at $(k_x,k_y,k_z)=\mathbf{0}$ and $(0,0,\frac{\pi}{a})$, with $s$ representing a two band degree of freedom on the lattice, and $\Psi=(c_{A\mathbf{k}},c_{B\mathbf{k}})^T$. Here (and for the rest of this appendix) we have re-included the lattice constant $a$ in $\hat{z}$ in order to facilitate the discussion regarding the eventual doubling of the Brillouin zone.
The gaplessness of this model is purely protected by translational symmetry in the $\hat{z}$ direction and the $U(1)$ charge conservation symmetry. This local stability is a result of the well-known chiral anomaly with translation symmetry acting as the low energy realization of the chiral symmetry, described the term in Eq.~\ref{eq:IRactionbroken}.

A simple single-node Dirac semimetal can be derived as follows: add a translational symmetry breaking term
\begin{align}
\mathcal{H}_{\mathrm{TB}}&=\sum_{\mathbf{k}}\sin k_x \big(c^\dag_{A\mathbf{k}}c_{A\mathbf{k}+\frac{\pi}{a}\hat{z}}+c^\dag_{A\mathbf{k}+\frac{\pi}{a}\hat{z}}c_{A\mathbf{k}}\nonumber\\
&\quad\quad\quad\quad\quad\quad-c^\dag_{B\mathbf{k}}c_{B\mathbf{k}+\frac{\pi}{a}\hat{z}}-c^\dag_{B\mathbf{k}+\frac{\pi}{a}\hat{z}}c_{B\mathbf{k}}\big)\quad,\nonumber\\
&=\sum_{\mathbf{k}}\Psi_{\mathbf{k}}^\dag\sin k_x\,\sigma^x s^z\Psi_{\mathbf{k}}\quad,
\label{eq:HTB}
\end{align}
where $\Psi_{\mathbf{k}}=(c_{A\mathbf{k}},c_{B\mathbf{k}},c_{A\mathbf{k}+\frac{\pi}{a}\hat{z}},c_{B\mathbf{k}+\frac{\pi}{a}\hat{z}})^T$ and $\sigma$ is the momentum degree of freedom. When we add this term to the simple Weyl semimetal Hamiltonian, we arrive at a Dirac semimetal, described by $\mathcal{H}_{\mathrm{DSM}}=\mathcal{H}_{\mathrm{WSM}}+\mathcal{H}_{\mathrm{TB}}$.
Since translational symmetry is broken by connecting states with momentum $k_z\sim k_z+\frac{\pi}{a}$, the Brillouin zone is reduced by half in the $\hat{z}$ direction, and our periodicity in $\hat{z}$ is $\tilde{a}=2a$ instead of $a$. The Hamiltonian becomes
\begin{align}
\mathcal{H}_{\mathrm{DSM}}=&\sum_{\delta\mathbf{k}}\Psi_{\mathbf{k}}^\dag \bigg[\sin k_x\, s^x+\sin k_y\, s^y +m_W(\mathbf{k})s^z\nonumber\\
&-\sin\left(\frac{ k_z\tilde a}{2}\right)\sigma^z s^z+\sin k_x\,\sigma^x s^z\bigg]\Psi_{\mathbf{k}}\quad,
\label{eq:motham}
\end{align}
which has a four-band degenerate single Dirac node at $\mathbf{k}=(0,0,\frac{\pi}{\tilde a})$. Note that the $s^z$ term destroys the doubly degenerate bands away from the node - it essentially acts as a time-reversal breaking term when $(A,B)=(\uparrow,\downarrow)$. One may notice that this Hamiltonian is not completely accurate since $k_z$ is now ill defined in terms of the periodicity of the $\sin(\frac{ k_z\tilde a}{2})$ term~\footnote{The odd periodicity actually comes about from a forbidden basis transformation using the real space lattice model - the lattice model will give the same dispersion but with a different proper (healthy) basis.}. To truly see a lattice realization of this state, we must look at the real space model, which will require some further modifications.

\subsubsection{Real space Hamiltonian}

Let us now attempt to write the previous Hamiltonian $\mathcal{H}_{\mathrm{WSM}}$ and $\mathcal{H}_{\mathrm{TB}}$ in real space. We will deal with a simple cubic lattice with a spin degree of freedom at each site ($A=\uparrow$, $B=\downarrow$). The real space representation of Eq.~\ref{eq:HWSM} in this lattice system is
\begin{align}
\mathcal{H}_{\mathrm{WSM}}=\frac{1}{2}\sum_{\mathbf{r}_i}\bigg[&-i
\Psi_{\mathbf{r}_i}^\dag
s^x
\Psi_{\mathbf{r}_i+a\hat{x}}
-i
\Psi_{\mathbf{r}_i}^\dag
s^y
\Psi_{\mathbf{r}_i+a\hat{y}}
\nonumber\\
&+2\Psi_{\mathbf{r}_i}^\dag s^z
\Psi_{\mathbf{r}_i}
-\Psi_{\mathbf{r}_i}^\dag s^z
\Psi_{\mathbf{r}_i+a\hat{x}}
\\
&-\Psi_{\mathbf{r}_i}^\dag
s^z
\Psi_{\mathbf{r}_i+a\hat{y}}+i
\Psi_{\mathbf{r}_i}^\dag
s^z
\Psi_{\mathbf{r}_i+a\hat{z}}\bigg]+h.c.,\nonumber
\end{align}
where $\Psi_{\mathbf{r}_i}=(c^\dag_{\uparrow\mathbf{r}_i}, c^\dag_{\downarrow\mathbf{r}_i})^{T}$.
As stated previously, this system is protected by both $U(1)$ and translational symmetry, and no further symmetries are required to guarantee the local stability of this gapless phase. Additionally, this lattice model also possesses other symmetries that will become important as we proceed: specifically, the $\pi$ rotation symmetry around the $\hat{z}$ axis, we label $C_{2z}$, which acts as
\begin{align*}
\Psi_{\mathbf{r}_i}
&\mapsto e^{i\frac{\pi}{2}s^z}\Psi_{2_{001}\mathbf{r}_i}
\quad,\\
\Psi_{\mathbf{k}}
&\mapsto e^{i\frac{\pi}{2}s^z}
\Psi_{2_{001}\mathbf{k}}\quad.
\end{align*}
This symmetry commutes with $\mathcal{H}_{\mathrm{WSM}}$, and the product of this $C_{2z}$ rotation followed by a $\hat{z}$ translation is naturally also a symmetry of the system, which we will call $C_{2z}^z$.
Now to arrive at our single node Dirac semimetal, we will independently break both the $\hat{z}$ translational symmetry and the $C_{2z}$ symmorphic rotation symmetry, but we retain the product $C_{2z}^z$ symmetry. This symmetry will end up being a non-symmorphic screw rotation symmetry in our new system that protects our single Dirac node.
To break the $\hat{z}$ translational symmetry and the $C_{2z}$ symmorphic rotation symmetry, we add the real space representation of the $\mathcal{H}_{\mathrm{TB}}$ (Eq.~\ref{eq:HTB}):
\begin{align}
\mathcal{H}_{\mathrm{TB}}=&\sum_{\mathbf{r}_i}\cos\left(\frac{\pi}{a}r_i^z\right)i\big[c^\dag_{\uparrow\mathbf{r}_i}c_{\uparrow\mathbf{r}_i+a\hat{x}}-c^\dag_{\uparrow\mathbf{r}_i+a\hat{x}}c_{\uparrow\mathbf{r}_i}\nonumber\\&-c^\dag_{\downarrow\mathbf{r}_i}c_{\downarrow\mathbf{r}_i+a\hat{x}}+c^\dag_{\downarrow\mathbf{r}_i+a\hat{x}}c_{\downarrow\mathbf{r}_i}\big]\quad.
\end{align}
This term breaks the original translation symmetry since it is periodic over $2a$ instead of $a$, and also does not commute with $C_{2z}$ due to the linear $k_x$ dependence. So now in the enlarged unit cell we have to account for a new sublattice degree of freedom for lattice sites at $\mathbf{r}_i$ and $\mathbf{r}_i+a\hat{x}$, which we will denote as $\mathbf{\tilde{r}}_i\alpha$ and $\mathbf{\tilde{r}}_i\beta$. In terms of Pauli matrices this degree of freedom will be denoted by $\tau$, and we will use the basis $\Psi_{\mathbf{\tilde{r}}_i}=(c_{\uparrow\mathbf{\tilde{r}}_i\alpha},c_{\downarrow\mathbf{\tilde{r}}_i\alpha},c_{\uparrow\mathbf{\tilde{r}}_i\beta},c_{\downarrow\mathbf{\tilde{r}}_i\beta})^T$, for which the total Hamiltonian is
\begin{align}
\mathcal{H}_{\mathrm{DSM}}
=&\frac{1}{2}\sum_{\mathbf{\tilde{r}}_i}\bigg[-i
\Psi^\dag_{\mathbf{\tilde{r}}_i}
s^x\Psi_{\mathbf{\tilde{r}}_i+a\hat{x}}
-i
\Psi^\dag_{\mathbf{\tilde{r}}_i}
s^y\Psi_{\mathbf{\tilde{r}}_i+a\hat{y}}\nonumber\\
&+2\Psi^\dag_{\mathbf{\tilde{r}}_i}
s^z\Psi_{\mathbf{\tilde{r}}_i}-\Psi^\dag_{\mathbf{\tilde{r}}_i}
s^z\Psi_{\mathbf{\tilde{r}}_i+a\hat{x}}-\Psi^\dag_{\mathbf{\tilde{r}}_i}
s^z\Psi_{\mathbf{\tilde{r}}_i+a\hat{y}}
\nonumber\\
&-\frac{1}{2}\Psi^\dag_{\mathbf{\tilde{r}}_i}
\sigma^y s^z\Psi_{\mathbf{\tilde{r}}_i}
+\frac{1}{2}\Psi^\dag_{\mathbf{\tilde{r}}_i}
\sigma^y s^z\Psi_{\mathbf{\tilde{r}}_i+\tilde{a}\hat{z}}\\
&+\frac{i}{2}
\Psi^\dag_{\mathbf{\tilde{r}}_i}
\sigma^x s^z\Psi_{\mathbf{\tilde{r}}_i+\tilde{a}\hat{z}}
+i\Psi^\dag_{\mathbf{\tilde{r}}_i}
\sigma^z s^z\Psi_{\mathbf{\tilde{r}}_i+a\hat{x}}\bigg]+h.c.,\nonumber
\end{align}
\footnote{Note that it requires some manipulation to show
\begin{align*}
&\sum_{\mathbf{r}_i}
\begin{pmatrix}
c^\dag_{\uparrow\mathbf{r}_i} & c^\dag_{\downarrow\mathbf{r}_i}
\end{pmatrix}
\begin{pmatrix}
1 & 0\\
0 & -1
\end{pmatrix}
\begin{pmatrix}
c_{\uparrow\mathbf{r}_i+a\hat{z}} \\ c_{\downarrow\mathbf{r}_i+a\hat{z}}
\end{pmatrix}\\
&=\sum_{\mathbf{\tilde{r}}_i}\left[\frac{1}{2}\Psi^\dag_{\mathbf{\tilde{r}}_i}
\sigma^x s^z\Psi_{\mathbf{\tilde{r}}_i}
+\frac{1}{2}\Psi^\dag_{\mathbf{\tilde{r}}_i}
\sigma^x s^z\Psi_{\mathbf{\tilde{r}}_i+\tilde{a}\hat{z}}
-\frac{i}{2}
\Psi^\dag_{\mathbf{\tilde{r}}_i}
\sigma^y s^z\Psi_{\mathbf{\tilde{r}}_i+\tilde{a}\hat{z}}\right]
\end{align*}} which in momentum space becomes
\begin{align}
&\mathcal{H}_{\mathrm{DSM}}=\sum_{\mathbf{k}}\Psi^\dag_{\mathbf{\tilde{k}}}\bigg[\sin k_x\,
s^x
+\sin k_y\,s^y+m_W(\mathbf{k})s^z\\
&+\frac{1}{2}(\cos \, k_z \tilde{a}-1)\sigma^y s^z
+\frac{1}{2}\sin k_z \tilde a \,\sigma^x
s^z
+\sin k_x 
\sigma^z s^z\bigg]\Psi_{\mathbf{\tilde{k}}}.\nonumber
\end{align}
This Hamiltonian, which has the same dispersion as Eq.~\ref{eq:motham}, now describes a single symmetry-protected Dirac node at $\mathbf{k}=\mathbf{0}$, while respecting Bloch's theorem.

The symmetry-protected nature of this model comes about from the fact that in the extended unit cell we retain the $C_{2z}^z$ non-symmorphic symmetry, which acts as
\begin{align*}
\Psi_{\mathbf{\tilde{r}}_i}&\mapsto  \begin{pmatrix}
0 & 1\\
0 & 0
\end{pmatrix}_\sigma e^{i\frac{\pi}{2}s^z}\Psi_{2_{001}\mathbf{\tilde{r}}_i}+\begin{pmatrix}
0 & 0\\
1 & 0
\end{pmatrix}_\sigma e^{i\frac{\pi}{2}s^z}\Psi_{2_{001}\mathbf{\tilde{r}}_i+\tilde{a}\hat{z}}
\quad, \\
\Psi_{\mathbf{\tilde{k}}}
&\mapsto 
\begin{pmatrix}
0 & 1\\
e^{ik_z\tilde{a}} & 0
\end{pmatrix}_\sigma e^{i\frac{\pi}{2}s^z}\Psi_{2_{001}\mathbf{\tilde{k}}}\quad.
\end{align*}
Along with $U(1)_V$ charge symmetry, the presence of the non-symmorphic symmetry guarantees that the perturbative mass terms, which could open a gap, are forbidden and gives the system a local stability. Of course a non-perturbative term may still open a gap such as via pair annihilation of the constituent Weyl nodes. To see this, let us have a look at the low energy expansion around the $k_x=k_y=0$ axis if we add a symmetry preserving mass $\delta s^z$, we arrive at the Weyl semimetal dispersion
\begin{align*}
E^\pm_{a}&=\pm\sqrt{k_x^2+k_y^2+\left[\delta+\sqrt{\sin\left(\frac{k_z\tilde{a}}{2}\right)^2+k_x^2}\right]^2}\quad,\\
E^\pm_{b}&=\pm\sqrt{k_x^2+k_y^2+\left[\delta-\sqrt{\sin\left(\frac{k_z\tilde{a}}{2}\right)^2+k_x^2}\right]^2}\quad,
\end{align*}
with Weyl nodes at $\mathbf{k}_W^\pm=(0,0,\pm 2\sin^{-1}(\delta)/\tilde{a}$). At $\delta=1$ the two Weyl nodes meet at $\mathbf{k}=\frac{\pi}{\tilde{a}}$ and for $\delta>1$ the nodes annihilate and open up a gap. This is consistent with the idea of a tunable anomaly that has been realized via the non-symmorphic symmetry. Let us now take a closer look at the symmetries and anomalies involved.

\subsubsection{Spacial symmetries \& anomalies protecting system}

Let us first re-examine the $\hat{z}$ translational symmetry by lattice constant $a$ of the WSM which we will denote as $T_a$. The momentum space action of this original translation symmetry acts in four band DSM model momentum space as
\begin{align}
T_a(k_z)=\begin{pmatrix}
0 & 1\\
e^{ik_z\tilde{a}} & 0
\end{pmatrix}_\sigma\quad,
\end{align}
which means that it has the following commutation relations with the second quantized Hamiltonians
\begin{align}
[\mathcal{H}_{\mathrm{WSM}},T_a]=0\quad,\nonumber\\
[\mathcal{H}_{\mathrm{TB}},T_a]\neq0\quad,\nonumber\\
[\mathcal{H}_{\mathrm{DSM}},T_a]\neq0\quad.\nonumber
\end{align}
Additionally, as previously discussed, the simple Weyl semimetal system has a natural $\pi$ rotation around the $\hat{z}$ axis centered on the lattice sites symmetry ($C_{2z}$), which in momentum space acts as $C_{2z}(\mathbf{k})=i s^z$.
We see that this commutes with $\mathcal{H}_{\mathrm{WSM}}$. However, we see that $C_{2z}$ does not commute with $\mathcal{H}_{\mathrm{TB}}$, so in general we have
\begin{align}
[\mathcal{H}_{\mathrm{WSM}},C_{2z}]=0\quad,\nonumber\\
[\mathcal{H}_{\mathrm{TB}},C_{2z}]\neq0\quad,\nonumber\\
[\mathcal{H}_{\mathrm{DSM}},C_{2z}]\neq0\quad.\nonumber
\end{align}
We may consider the non-symmorphic counterpart of a $\pi$ rotation around the $\hat{z}$ axis with a full (half) translation in the $\hat{z}$ direction for the WSM (DSM) cases, $C_{2z}^z$. In momentum space, this takes the form
\begin{align}
C^z_{2z}(\mathbf{k})
=i\begin{pmatrix}
0 & 1\\
e^{ik_z\tilde{a}} & 0
\end{pmatrix}_\sigma s^z\quad,
\end{align}
which has the following commutation relations
\begin{align}
[\mathcal{H}_{\mathrm{WSM}},C^z_{2z}]=0\quad,\nonumber\\
[\mathcal{H}_{\mathrm{TB}},C^z_{2z}]=0\quad,\nonumber\\
[\mathcal{H}_{\mathrm{DSM}},C^z_{2z}]=0\quad.\nonumber
\end{align}
These relations mean that the symmetry is present in the new model. In fact, its presence guarantees the gaplessness of the model, since it prevents any mass terms from gapping the spectrum, as long as we also have $U(1)_V$ charge conservation symmetry, which naturally remains in our DSM model.

\subsubsection{Chiral anomaly}
\label{sec:nonsymchiralanomaly}

The folding of the Brillouin zone in half changes the original Weyl semimetal 4+1d chiral anomaly in such a manner
\begin{align}
\frac{\pi}{2a}\int z\wedge \frac{dA}{2\pi}\wedge \frac{dA}{2\pi} \rightarrow \frac{\pi}{2a}\int z_c\wedge \frac{dA}{2\pi}\wedge \frac{dA}{2\pi}\quad,
\end{align}
where $z$ stands for the translation gauge field and $z_c$ stands for the new non-symmorphic $\hat{z}$ screw rotation gauge field. Essentially $\frac{1}{a}\oint z_c$ counts the amount of half layers that have been traversed. There is no issue when $\frac{1}{a}\oint z_c\in 2\mathbb{Z}$ which gives you twice the number of proper layers in the $\hat{z}$ direction.  However, when $\frac{1}{a}\oint z_c\in 2\mathbb{Z}+1$ we end up with an extra half layer of the system - this issue can be ruled out since our Hamiltonian is explicitly constructed to be an even number of layers. In order for $z_c$ to be a well-defined gauge field, we must also be able to describe the defect line for the screw rotation symmetry. In this case, the line is along the $\hat{z}$ direction and takes the form of a spiral defect, where one can think of it as a disclination line in $\hat{z}$ with a dislocation line wrapping around it to form a spiral defect.

We see that the new anomaly term essentially gives the same anomalous quantum Hall effect as in the original WSM system, except now per every half-layer of the new system~\footnote{To establish connections with the new translational gauge field $\tilde{z}$ of the DSM, handwavingly one can think of $\int_{0}^{2a} z_c\sim \int_0^{\tilde{a}}\tilde {z}$. In this case the anomaly term may then appear as 
\begin{align}
\frac{1}{8\pi^2}\frac{2\pi}{\tilde{a}}\int  \tilde{z}\wedge dA\wedge dA\quad,
\end{align}
from which one might conclude that no anomaly exists. However, this is misleading because the fundamental gauge field is actually $z_c$, not $\tilde{z}$.}.
Let us now properly derive the associated chiral anomaly by starting with the low energy expansion around the $\hat{z}$ axis:
\begin{align}
\mathcal{H}_{\mathrm{axis}}&=\sum_{\mathbf{k}}\Psi^\dag_{\mathbf{k}}\bigg[k_x
s^x
+k_y s^y
-\frac{1}{2}\left(1+\cos \, k_z \tilde{a}\right)\sigma^x s^z\nonumber\\
&-\frac{1}{2}\sin k_z \tilde a\,\sigma^y s^z
+k_x
\sigma^z s^z\bigg]\Psi_{\mathbf{k}}\quad.
\end{align}
We may diagonalise the squared Hamiltonian to take the form
\begin{align}
H_{axis}^2=&2\sqrt{2}e B_z\left(a^\dag a+\frac{1\pm1}{2}\right)
+\cos\left(\frac{k_z\tilde{a}}{2}\right)^2\nonumber
\quad,
\end{align}
where $\xi_l(k_z)=\sqrt{2}$, $\xi_s(k_z)=1$, and we have $a^\dag a$ eigenstates $|n\rangle$ for $n\geq0$.
We see that there is a four-fold degeneracy, on top of a momentum degeneracy (e.g., $k_y$), for all energy levels $n$ except for the lowest energy Landau level when $n=0$, for which we only have the negative energies, giving us $E(k_z)_-^2=\cos\left(\frac{k_z\tilde{a}}{2}\right)^2$. Since the Hamiltonian commutes with the $C_{2z}^z$ operator, we may choose a simultaneous eigenbasis for which we have
\begin{align}
|u^-_1\rangle&=\frac{1}{2\sqrt{-2+\sqrt{2}}}\begin{pmatrix}
(-1+\sqrt{2})e^{-\frac{ik_z\tilde{a}}{2}} \\
e^{-\frac{ik_z\tilde{a}}{2}} \\
1-\sqrt{2}\\
1
\end{pmatrix}|n\rangle\quad, \\
|u_2^-\rangle&=\frac{1}{2\sqrt{-2+\sqrt{2}}}\begin{pmatrix}
-(-1+\sqrt{2})e^{-\frac{ik_z\tilde{a}}{2}} \\
-e^{-\frac{ik_z\tilde{a}}{2}} \\
1-\sqrt{2}\\
1
\end{pmatrix}|n\rangle \,\,,
\end{align}
where $u^-_1$ and $u^-_2$ have $C_{2z}^z$ eigenvalues $-e^{\frac{i k_z\tilde{a}}{2}}$ and $e^{\frac{i k_z\tilde{a}}{2}}$, respectively. Notice that these eigenvectors wrap twice around the Brillouin zone, which is an essential feature that allows the realization of a single Dirac node. Now, it is easy to show that
\begin{align}
\langle u^-_1|H_{axis}|u^-_1\rangle=&\cos\left(\frac{k_z\tilde{a}}{2}\right)=-\langle u^-_2|H_{axis}|u^-_2\rangle ,
\end{align}
so we see that the lowest Landau levels have energies $\pm \cos\left(\frac{k_z\tilde{a}}{2}\right)$ with $C_{2z}^z$ eigenvalues $\mp e^{\frac{i k_z\tilde{a}}{2}}$. Let us note that the $C_{2z}^z$ eigenvalues wrap into each other at the edge of the Brillouin zone which is a key feature that allows for a consistent theory. In fact, for a pure point group symmetry protecting a single node, we would run into an inconsistency at the edge of the Brillouin zone. The low energy theory is described by the action
\begin{align}
S=\int dt dz\left(i\psi_{r}^\dag\partial_-\psi_{r}+i\psi_{l}^\dag\partial_+\psi_{l}\right)\quad,
\end{align}
where $\psi_r$ ($\psi_l$) is the right (left) mover.
Once again, we can apply the known results of the $1d$ chiral anomaly in which we push a left mover to a right mover when an electric flux is applied. These left and right movers carry a $C_{2z}^z=e^{i\pi Q_{z\alpha}}$ charge $Q_{z\alpha}=k_z\tilde{a}/2\pi+\delta_{\alpha r}$, where $\alpha\in\{l,r\}$, so the total non-symmorphic $C_{2z}^z$ charge at $k_z=-\pi/\tilde{a}$ is given by
\begin{align}
Q_c&=\left(-\frac{1}{2}+1\right)\psi_R^\dag \psi_R-\frac{1}{2}\psi_L^\dag \psi_L\quad,\nonumber\\
&=\frac{1}{2}\left(\psi_R^\dag \psi_R-\psi_L^\dag \psi_L\right)\quad.\nonumber
\end{align}
For the smallest quantum of flux, i.e., change in particle number by 2, we have
\begin{align*}
\Delta Q_c=1\quad,
\end{align*}
which, when we recall that $Q_c$ is only conserved modulo $2\mathbb{Z}$, means that the associated chiral anomaly is non-trivial.
Since momentum is related by $P=2\pi Q_c/\tilde{a}$, we have
\begin{align*}
\Delta P=\frac{2\pi}{\tilde{a}}\quad,
\end{align*}
which means that the momentum change is trivial, since $2\pi/\tilde{a}\sim0$ in the BZ, meaning that the chiral anomaly associated with new translation gauge field $\tilde{z}$ is trivial. A change of integer value means we have the integer quantum Hall effect per layer. Overall, this type of behaviour is described by the $3+1$d topological term
\begin{align}
S=\frac{1}{4\pi^2}\frac{\pi}{2a}\int z_c\wedge A\wedge dA\quad.
\end{align}
Naturally, this Hall conductivity breaks time-reversal symmetry. Since the existence of the Hall conductivity does not depend on whether we have time-reversal or not, i.e., it just depends on the low energy expansion and $C_{2z}^z$ symmetry, it follows that this type of system cannot exist with time-reversal.

When such a system is symmetrically gapped via the vortex condensation method, we find that odd-fold vortices possess non-gappable fermionic modes in its core thus making them un-condensable, and two-fold vortices break the non-symmorphic $C_{2z}^z$ symmetry upon an insertion of the spiral defect, so we are forced to condense 4-fold vortices to restore our $U(1)_V$ symmetry. We essentially end up with the same $\mathbb{Z}_4$ topological order and fractional quantum Hall effect as when the magnetic WSM is gapped.

\section{Distorted Dirac spinel model}
\label{app:distorteddirac}
\subsection{Protecting symmetries and the chiral anomaly}

The distorted Dirac spinel model is a lattice DSM model with potential material realizations being several Bismuth compounds (e.g., BiZnSiO$_4$). Initially, the model was constructed by elongating two bonds per each lattice site of the diamond lattice in the Fu-Kane-Mele model. This elongation changes the diamond lattice's space group 227 to the distorted spinel's space group 74. The FKM model possesses three distinct symmetry-enforced Dirac nodes on the $X^{x/y/z}$ points of the BZ. By elongating the relevant bonds to arrive at the distorted spinel case we break certain symmetries such that two Dirac nodes remain on the $X^x\rightarrow T$ and $X^y\rightarrow T'$ points but the Dirac node is gapped on $X^z\rightarrow X$, leaving only two of the three initial nodes. There has been some previous confusion with regards to the number of symmetry-enforced Dirac nodes in the spinel model: here, we show through the tightbinding model, symmetries, and anomaly arguments that there exist two distinct nodes at the high-symmetry $T$ and $T'$ points.

\subsubsection{Symmetries of the model}

The space group that governs the distorted Dirac spinel is space group 74 (Imma). The spacial symmetries of the lattice are as follows in Seitz notation, where the lattice $\mathbf{0}$ reference point is shown in Figure~\ref{fig:distortedlattice}(a)
\begin{enumerate}
\item $\mathds{1}=\{\mathds{1}|\mathbf{0}\}$ [identity transformation] 
\item $C_{2z}^y=\{2_{001}|(0,\frac{b}{2},0)\}$ [$\pi$ rotation around $\hat{z}$ with half-translation in $y$]
\item $C_{2y}^y=\{2_{010}|(0,\frac{b}{2},0)\}$ [$\pi$ rotation around $y$ with half-translation in $y$ (screw rotation)]
\item $C_{2x}=\{2_{100}|\mathbf{0}\}$ [$\pi$ rotation around $x$]
\item $P=\{-\mathds{1}|\mathbf{0}\}$ [inversion] 
\item $M_z^y=\{m_{001}|(0,\frac{b}{2},0)\}$ [reflection of $\hat{z}$ with half-translation in $y$ (glide symmetry)]
\item $M_y=\{m_{010}|(0,\frac{b}{2},0)\}$ [reflection of $y$ with half-translation in $y$]
\item $M_x=\{m_{100}|\mathbf{0}\}$ [reflection of $x$]
\end{enumerate}
alongside the translational symmetries $T_{\mathbf{t}_1}=\{\mathds{1}|\mathbf{t}_1\}$, $T_{\mathbf{t}_2}=\{\mathds{1}|\mathbf{t}_2\}$, and $T_{\mathbf{t}_3}=\{\mathds{1}|\mathbf{t}_3\}$, which generate the complete translation group. Additionally, our system also obeys time-reversal symmetry $\Theta=is^y\mathcal{K}$, where $\mathcal{K}$ is the complex conjugation operator.
The relevant spacial symmetries at the high symmetry points are those that take the high-symmetry point to an equivalent, i.e., related by a reciprocal lattice, $\mathbf{k}$ point. These symmetries form the \textit{little group}. The list of the little groups of each relevant high-symmetry point is as follows
\begin{align*}
X &: \{\mathds{1},C_{2z}^y,C_{2y}^y,C_{2x},P,M_z^y,M_y^y,M_x,\Theta\}\\
T &: \{\mathds{1},C_{2z}^y,P,M_z^y,\Theta\}\\
T' &: \{\mathds{1},C_{2z}^y,P,M_z^y,\Theta\}
\end{align*}
The key to FDIR representations lies in the occurrence of projective representations that are necessitated by a non-trivial action of a non-symmorphic symmetry followed by an inversion transformation. We will not go into detail regarding FDIR representation except to state that $X$ does not possess such a representation, while $T$ and $T'$ have a minimal set of symmetry generators $\mathcal{H}=\{C_{2z}^y,P,\Theta, U(1),T_{\mathbf{t}_3}\}$ that symmetry-enforce an FDIR thus resulting in a four-fold degeneracy, i.e., a Dirac node. Here we have added $U(1)$ charge and $T_{\mathbf{t}_3}$ symmetries such that the nodes may not be gapped via a superconducting-type term or a charge density wave, respectively. Although the necessity for $U(1)$ is self-explanatory, let us elaborate on the need for $T_{\mathbf{t}_3}$ rather than a pure $\hat{z}$ translational symmetry. Take, for example, a perturbation of the form $c_{\alpha, k_z}^\dag c_{\beta,k_z+2\pi}$, which can gap the spectrum: this perturbation can not be prevented with a pure unit translation along the $\hat{z}$ direction. Instead, we need a symmetry involving a half-translation in the $\hat{z}$ direction such as $T_{\mathbf{t}_3}$ to prevent a gap.

Let us now derive the forms of the relevant symmetries using our tight binding model, for which an example will be presented for the $C_{2z}^y$ symmetry As a quick refresher, $C_{2z}^y$ has the following action on the annihilation operators in real and momentum space
\begin{align*}
&c_{\mathbf{r}_i,A,s}\mapsto e^{i \frac{\pi}{2}s^z}c_{2_{001}\mathbf{r}_i+b\hat{y},A,s}, &c_{\mathbf{r}_i,B,s}&\mapsto e^{i \frac{\pi}{2}s^z}c_{2_{001}\mathbf{r}_i,B,s},\\ 
&c_{\mathbf{k},A,s}\mapsto e^{i \frac{\pi}{2}s^z}e^{-ik_yb}c_{2_{001}\mathbf{k},A,s}, &c_{\mathbf{k},B,s}&\mapsto e^{i \frac{\pi}{2}s^z}c_{2_{001}\mathbf{k},B,s},
\end{align*}
so when transforming the Bloch Hamiltonian, the symmetry acts with the operator
\begin{align}
C_{2z}^y(\mathbf{k})= i \begin{pmatrix}
e^{-ik_yb} & 0\\
0 & 1
\end{pmatrix}_\sigma s^z \quad,
\end{align}
where the subscript indicates which subspace the matrix lives in (in this case, corresponding to the $A,B$ degree of freedom)
By construction $C_{2z}^y(\mathbf{k})$ is a symmetry of our system, so the Hamiltonians in Eqs.~\ref{eq:HX}, \ref{eq:HT}, and \ref{eq:HTp} must, and indeed do, obey
\begin{align*}
\left[C_{2z\alpha}^y\right]^{-1}H_\alpha(\mathbf{q})\,C_{2z\alpha}^y&=H_\alpha(2_{001}\mathbf{q})\quad,
\end{align*}
where $\alpha\in\{X,T,T'\}$ and $\tilde{C}_{2z\alpha}^y$ is the zeroth order expansion of $C_{2z}^y(\mathbf{k})$ around the $\alpha$ high-symmetry points. Similarly, one finds
$P=\sigma^x$ in momentum space.
When we combine the symmetry elements $\{C_{2z}^y,P,\Theta,U(1),T_{\mathbf{t}_3}\}$, we see that indeed the individual nodes at $T$ and $T'$ are fixed at $\mathbf{q}=0$ since all free-fermion masses or splitting terms are prohibited.

Now to study the relevant anomalies, let us instead find the symmetry elements that \textit{symmetry-protect}, rather than enforce, the gaplessness of $H_T$ and $H_{T'}$. This condition is satisfied by a smaller sets $\mathcal{G}_1=\langle\{C_{2z}^y,U(1),T_{\mathbf{t}_3}\}\rangle\subseteq\mathcal{H}$ or $\mathcal{G}_2=\langle\{M_{z}^y,U(1),T_{\mathbf{t}_3}\}\rangle\subseteq\mathcal{H}$, where we have now eliminated time-reversal and parity symmetry, which allows for perturbations that alter the Dirac semimetal into gapless Weyl, nodalline, or nodalsphere semimetal states. Importantly, the $C_{2z}^y$ symmetry remains in order to prevent any mass terms, i.e., any term that anticommutes with $\sigma^zs^z$, from perturbatively opening a gap.

As discussed in the main text, such a system must possess cancellation of the anomalies that protect the individual Dirac nodes so that there is no contradiction with the anomaly matching condition and gaugeability. We will now explore this feature in the next section.

\subsection{Chiral anomalies}
\label{sec:chiralanomaly}

\subsubsection{Symmorphic $C_2$ chiral anomaly}

\begin{figure}[htb]
    \centering
    \vspace{0.5cm}
        \includegraphics[width=0.8\columnwidth]{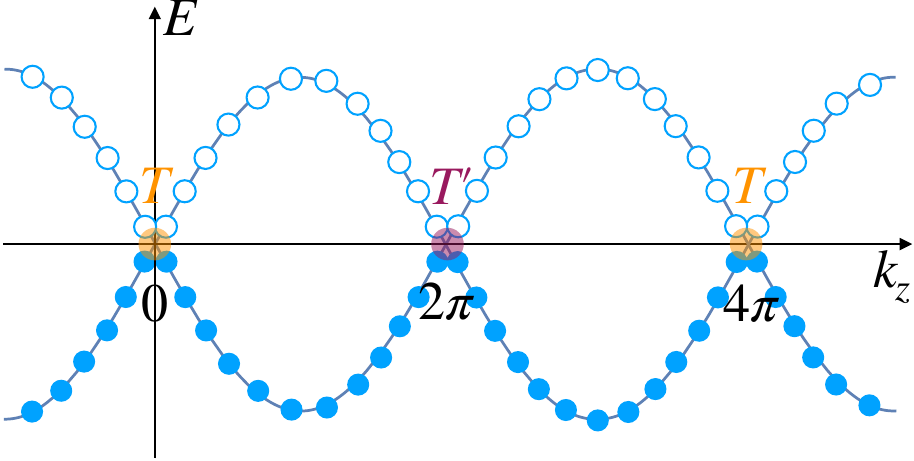}
    \caption{\label{fig:distortedspineLL}Lowest Landau levels for the distorted Dirac spinel model. Note that along the $k_z$ axis we have a $4\pi/c$ periodicity of the Brillouin zone rather than the usual $2\pi/c$ (as long as the $T_{\mathbf{t}_3}$ symmetry is preserved).}
\end{figure}

To consider the $\{C_{2z}^y,U(1)\}$ chiral anomaly let us work along the $k_y=k_x=\pi$ axis which connects the two Dirac nodes at $T$ and $T'\equiv T'+\mathbf{g}_3=\left(\pi,\pi,2\pi\right)$ (see Figure~\ref{fig:distortedlattice}(b) for visualization). An expansion around this axis gives the Hamiltonian
\begin{align}
H_{axis}=&-\left(t_sq_y+t_l\cos\left(\frac{k_z}{2}\right)q_x\right)\sigma^y-t_l\sin\left(\frac{k_z}{2}\right)q_x\sigma^x\nonumber\\
&+ \left(4\gamma \lambda_{sl}\cos\left(\frac{k_z}{2}\right)\,q_y-(1-4\gamma)\lambda_{ll}\,q_x\right)\sigma^z s^y\nonumber\\
&-\left(4\gamma \lambda_{ss}\,q_y-(1-4\gamma)\lambda_{sl}\cos\left(\frac{k_z}{2}\right)\,q_x\right)\sigma^z s^x\nonumber\\
&-2\lambda_{sl}\,\sin\left(\frac{k_z}{2}\right)\,\sigma^z s^z, 
\end{align}
which in response to a magnetic field $B_z$ in the $\hat{z}$ direction, we may replace $q_{\alpha}\rightarrow \pi_{\alpha}=-i\partial_\alpha-A_\alpha$, where $\nabla\times\mathbf{A}=B_z \hat{z}$, $[\pi_x,\pi_y]=-ieB_z$, and $\alpha\in\{x,y\}$. To determine the chiral anomaly, we work with the squared Hamiltonian, which we can diagonalize to take a form
\begin{align}
H_{axis}^2=&2eB_z\xi_l(k_z)\xi_s(k_z)\left(a^\dag a+\frac{1}{2}\right)\pm eB_z\xi_l(k_z)\xi_s(k_z)\nonumber\\
&+\left(2\lambda_{sl}\right)^2\sin\left(\frac{k_z}{2}\right)^2\quad,
\label{eq:Haxisdiag}
\end{align}
where the annihilation operator is given by
\begin{align*}
a=\frac{1}{\sqrt{2eB\xi_l(k_z)\xi_s(k_z)}}(\xi_l(k_z)\tilde{\pi}_x-i\xi_s(k_z)\tilde{\pi}_y)\quad,
\end{align*}
with $[a,a^\dag]=1$, and
\begin{align*}
\xi_q(k_z)&=\sqrt{t_q^2+(\delta_{lq}-4\gamma)^2\left[\lambda_{qq}^2+\lambda_{sl}^2\cos\left(\frac{k_z}{2}\right)^2\right]}\quad.
\end{align*}
Here $q\in\{l,s\}$, $\delta$ is the Kronecker delta, and $\tilde{\pi}_\alpha$ is an $SO(2)$ rotated version of $\pi_\alpha$, which possess the relations $[\tilde{\pi}_x,\tilde{\pi}_y]=-ieB_z$.
The eigenstates are given by
\begin{align}
|\chi_{n}^\pm\rangle=\mathcal{N}_\chi
\begin{pmatrix}
0\\
\frac{\pm\sqrt{(\xi_a-\xi_f)^2+(\xi_b+\xi_e)^2+(\xi_c-\xi_d)^2}-\xi_a+\xi_f}{\xi_b+\xi_e+i (\xi_c-\xi_d)}\\
1\\
0
\end{pmatrix}
|n\rangle\,, \\
|\eta_{n}^\pm\rangle=\mathcal{N}_\eta
\begin{pmatrix}
\frac{\pm\sqrt{(\xi_a+\xi_f)^2+(\xi_b-\xi_e)^2+(\xi_c+\xi_d)^2}+\xi_a+\xi_f}{\xi_b- \xi_e+i (\xi_c+\xi_d)}\\
0\\
0\\
1
\end{pmatrix}
|n\rangle\,.
\end{align}
where $\xi_p$, $p\in\{a,b,c,d,e,f\}$, are coefficients that depend on $k$, $\mathcal{N}_\chi$ and $\mathcal{N}_\eta$ are the respective normalization factors, and $\left(a^\dag\right)^n|0\rangle=|n\rangle$. Their exact forms are unimportant for the concept derivation presented here.
These Landau energy levels are quadruply degenerate (in addition to a degeneracy in $k_y$ which keeps $C_{2z}^y$ well defined) except for the $0th$ Landau level where it has a double degeneracy with degenerate eigenstates. We also see that the $\chi$ and $\eta$ eigenstates are eigenstates of the zeroth-order ($k_y=\pi$) $C^y_{2z}\approx -i\sigma^zs^z$ symmetry with eigenvalue $i$ and $-i$ respectively. For the lowest energy eigenstates, we have $k_z$ dispersions
\begin{align}
\langle\chi_0^-|H_{axis}|\chi_0^-\rangle=2\lambda_{sl}\sin\left(\frac{k_z}{2}\right)=-\langle\eta_0^-|H_{axis}|\eta_0^-\rangle\quad,
\end{align}
which essentially describes two sets of $1d$ left and right moving chiral modes, situated at $k_z=0$ and $k_z=2\pi$ respectively (recall that the BZ is $4\pi$ periodic in $k_z$). We have depicted the lowest Landau level dispersions in Figure~\ref{fig:distortedspineLL}. Near the nodes we can capture the physics using the continuum action
\begin{align}
S=\int dt dz\sum_{\xi\in\{T,T'\}}i\left(\left[\psi^{\xi}_{r}\right]^\dag\partial_-\psi^{\xi}_{r}+\big[\psi^{\xi}_{l}\big]^\dag\partial_+\psi^{\xi}_{l}\right),
\end{align}
where $\partial_\pm=\partial_t\pm\partial_z$, and $\psi^{\xi}_{r/l}$ at the right and left-movers at the high-symmetry point $\xi$.
Now using the $1$d chiral anomaly (where the charge operator is the $C^y_{2z}=e^{i\pi Q_{cz}}$ rotation charge $Q_{cz}=-\sigma^zs^z/2$) it is simple to show that the overall chiral anomaly disappears since we have two nodes, i.e., $\Delta Q_{cz}=0$ even when we have an electric field in the $\hat{z}$ direction. Specifically, the node at $T$ has charge
\begin{align}
Q^T_{cz}=\frac{1}{2}\left([\psi^T_{r}]^\dag\psi_{r}^T-[\psi^T_{l}]^\dag\psi_{l}^T\right)\quad,
\end{align}
where $r$ ($l$) indicate the 1d right (left) movers. So the change in charge, when we thread a flux that takes a left mover to a right mover is
\begin{align}
\Delta Q^{T}_{cz}=1\,(\mathrm{mod}\, 2)\quad,
\end{align}
where the $\mathrm{mod}\, 2$ comes from $\mathbb{Z}_2$ nature of any $\pi$ rotation point-group.
Now if we also consider the $\tilde{C}^y_{2z}$ charge of the $T'$ node, we have
\begin{align}
Q^{T'}_{cz}=-\frac{1}{2}\left([\psi^{T'}_r]^\dag\psi^{T'}_r-[\psi_l^{T'}]^\dag\psi^{T'}_l\right)=-Q^T_{cz}\quad,
\end{align}
which the opposite of the charge of $X^y$, and thus the change, when the smallest amount of flux is inserted, is
\begin{align}
\Delta Q^{T'}_{cz}=-1 \,(\mathrm{mod}\,2)=-\Delta Q^T_{cz}\quad.
\end{align}
Thus the change of the total charge is
\begin{align}
\Delta Q_{cz}=\Delta Q^{T}_{cz}+\Delta Q^{T'}_{cz}=0\quad,
\end{align}
which means that there is no total anomaly associated with this charge. However on the individual nodes, there is a non-conservation of $C_{2z}$ charge which is the chiral anomaly that protects the local stability of the $T$ and $T'$ nodes.

\subsubsection{Translation chiral anomaly}

In addition to the $C_{2z}$ type of anomaly, we can also consider the chiral anomaly that comes from the translation symmetry $T_{\mathbf{t}_3}$ and $U(1)$. To observe this anomaly cancellation we need to now thread a magnetic field that preserves the $T_{\mathbf{t}_3}$ symmetry, and perform the same study as before. We do this by choosing a basis and gauge where $A_\mu$ depends spatially purely on the $\hat{\mathbf{x}}_{\perp}=(1/\sqrt{2},-1/\sqrt{2},0)$ axis, thereby preserving the $T_{\mathbf{t}_3}$ symmetry, and the $k_z$ quantum number.

\begin{figure}[htb]
    \centering
    \vspace{0.5cm}
           \includegraphics[width=0.45\columnwidth]{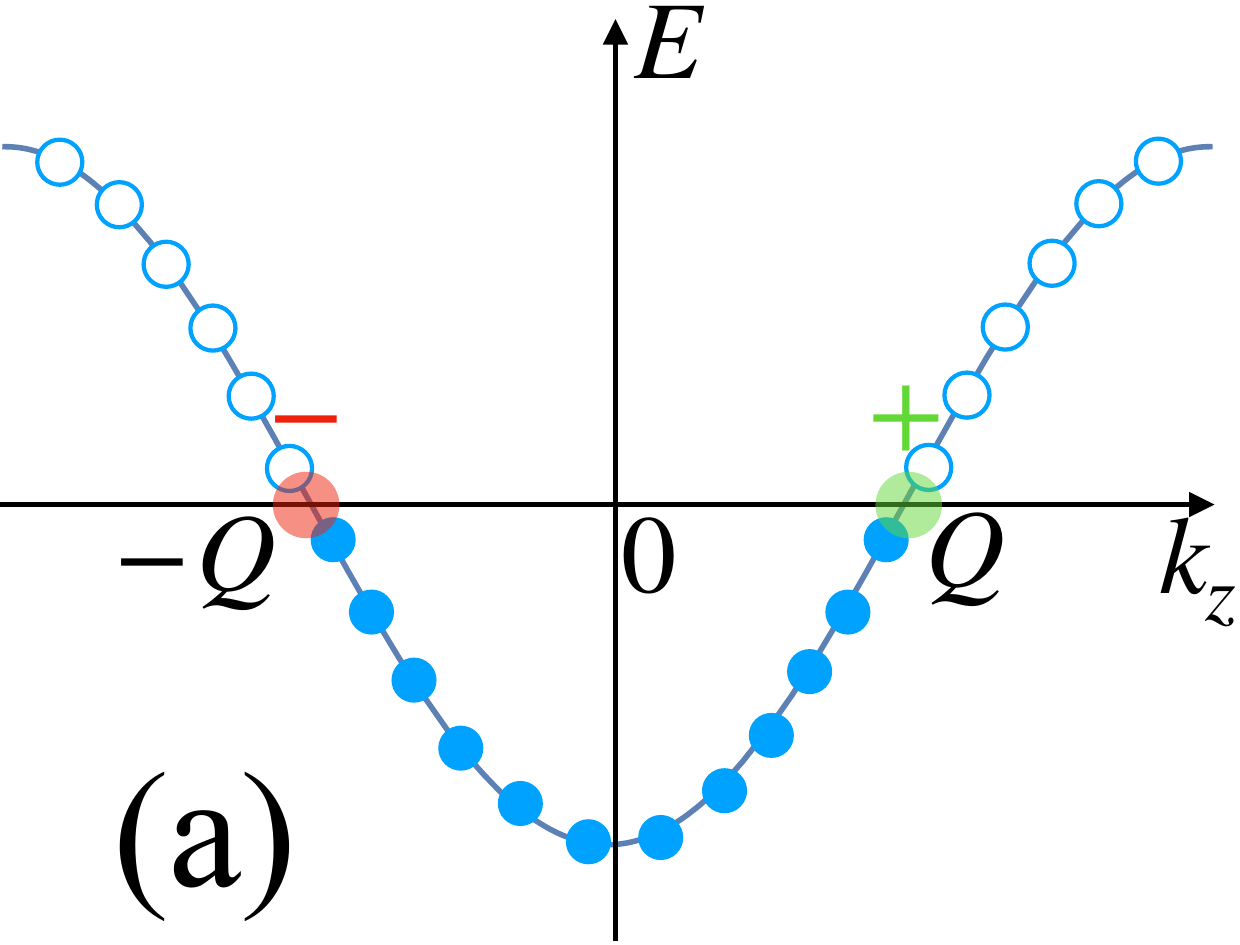}
           \includegraphics[width=0.45\columnwidth]{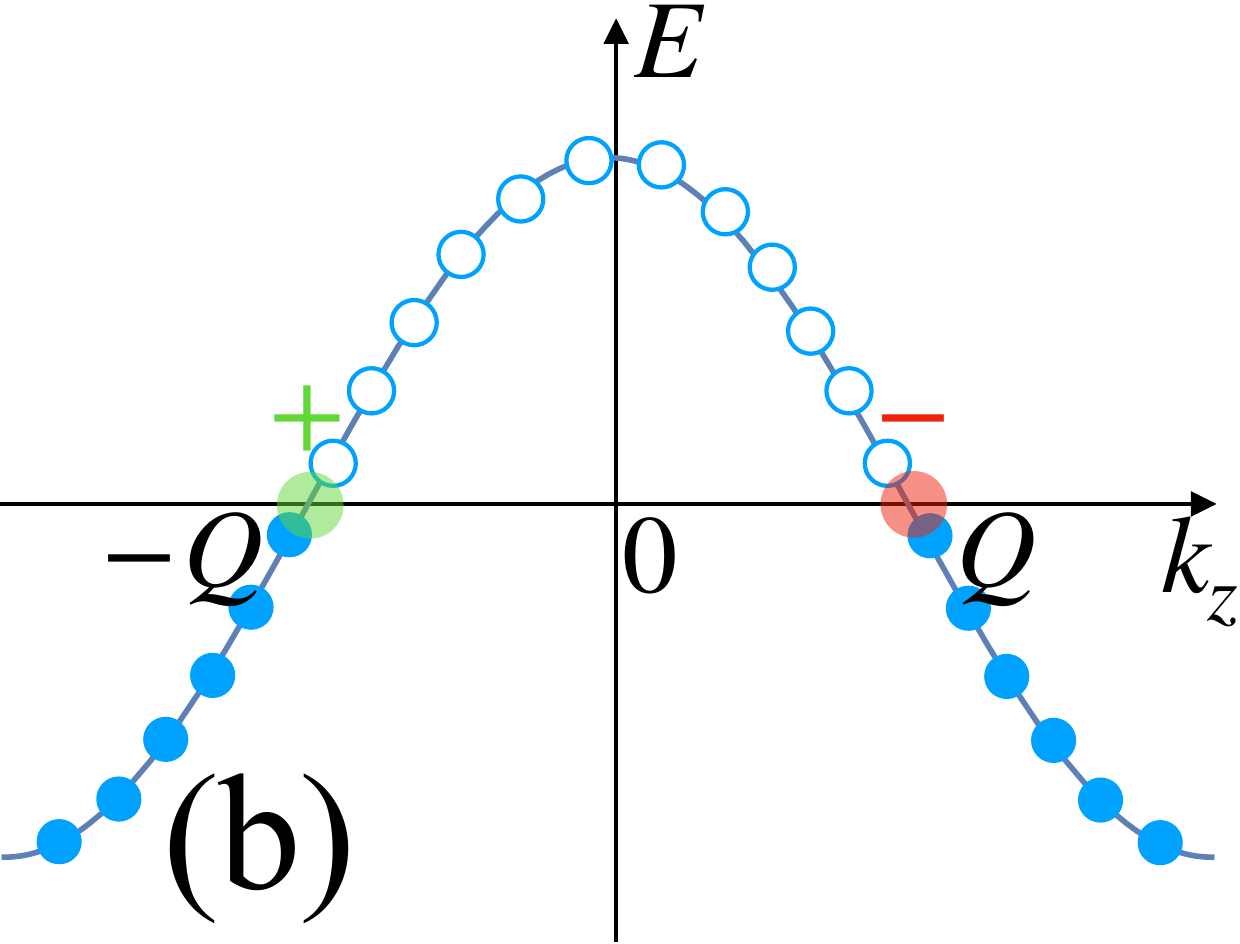}
    \caption{\label{fig:WSMLL}Lowest Landau level for simple magnetic WSM with positive ($+$, green) and negative ($-$, red) nodes separated by $2Q$. We depict (a) $\pm$ chirality nodes at $\pm Q$, and (b) $\mp$ chirality nodes at $\pm Q$. The overlap between these two systems and their respective Landau levels can be thought to give the full lowest Landau levels depiction for the distorted spinel model, as shown in Figure~\ref{fig:distortedspineLL}.}
\end{figure}

The translation symmetry in momentum space takes the form
$$T_{\mathbf{t}_3}(k_{||},k_z)=e^{\frac{i}{\sqrt{2}}\left(k_{||}+\frac{k_z}{\sqrt{2}}\right)}\quad,$$
where $k_{||}=k_x/\sqrt{2}+k_y/\sqrt{2}$ is well defined, and thus the symmetry remains well-defined. The physics here is identical to two overlapping copies of a magnetic WSM with a positive and negative chirality node. In this case the two sets of overlapping Weyl nodes, forming Dirac nodes at $T$ and $T'$, are separated by $k_z=2\pi$ at $k_{||}=\sqrt{2}\pi$. The charge of the $T_{\mathbf{t}_3}=e^{i\pi Q_{\mathbf{t}_3}}$ translational symmetry is simply $Q_{\mathbf{t}_3}=\left(\sqrt{2}k_{||}+k_z\right)/2\pi$. Using the results from the previous section, the low-energy model with $C_{2z}$ charge of $\frac{\pi}{2}$ is given by
\begin{align}
Q_{\mathbf{t}_3}^\chi=[\psi^{T}_r]^\dag\psi^T_r+2[\psi^{T'}_l]^\dag\psi^{T'}_l\quad \mathrm{for}\,\chi_0^-\,\mathrm{eigenstate}\\
Q_{\mathbf{t}_3}^\eta=[\psi^T_l]^\dag\psi^T_l+2[\psi^{T'}_r]^\dag\psi^{T'}_r\quad \mathrm{for}\,\eta_0^-\,\mathrm{eigenstate}
\end{align}
where we have written the charge in terms of the $\chi$ and $\eta$ eigenstates, corresponding to the $0th$ Landau level of two overlapping magnetic oppositely-charged WSMs. See Figure~\ref{fig:WSMLL} for intuition.

Similarly to the rotation case, leads to a change of charge of 
\begin{align*}
\Delta Q^\chi_{\mathbf{t}_3}&=-1\quad,\\
\Delta Q^\eta_{\mathbf{t}_3}&=1\quad,
\end{align*}
when the smallest flux is applied. So the total charge changes by
\begin{align}
\Delta Q_{\mathbf{t}_3}=\Delta Q^\chi_{\mathbf{t}_3}+\Delta Q^\eta_{\mathbf{t}_3}=0\quad,
\end{align}
which, when we recall that the charge is equivalent to momentum so this is just the trivial change of momentum. We see that the physics here is identical to the magnetic WSM case with a positive and negative chirality node, except now we have two sets of overlapping Weyl nodes separated by $k_z=2\pi$, thus cancelling the anomalous Hall conductivities. Of course since this is a $\mathbb{Z}$ gauge field we may have a non-zero chiral anomaly effect as we add say a time-reversal breaking $m s^z$ term~\footnote{Note that this actually reduces the DMS to a time-reversal broken nodal-line semi-metal.}, which causes a non-trivial chiral anomaly with 
\begin{align}
\Delta Q_{\mathbf{t}_3}=\Delta Q^\chi_{\mathbf{t}_3}+\Delta Q^\eta_{\mathbf{t}_3}=\frac{2m}{\pi}\quad,
\end{align}
leading to a Hall effect of
\begin{align}
\sigma_{xy}=\frac{2m}{\pi}\quad.
\end{align}

So we see that the $C_2$ chiral anomaly guarantees the individual low energy stability of the $T$ and $T'$ Dirac nodes, and the $T_{\mathbf{t}_3}$ chiral anomaly ensures the stability of the two separate overlapping simple magnetic Weyl semimetals. However, the total chiral anomaly for the Dirac semimetal is zero when the time-reversal symmetry is preserved.

\section{Fu-Kane-Mele model}
\label{app:FKM}

\subsection{Protecting symmetries and the chiral anomaly}

Now that we have explored the spinel case in detail, the properties of the FKM model follow rather nicely. Crucially, the FKM model possess two extra rotation $C_2$ symmetries which give rise to its three nodes at $X^{\hat{r}}$, which are each protected similarly by a rotation and translational symmetry. The physics in the FKM model consists of three copies of the same chiral anomalies found in the distorted spinel model. We will start with a quick review of the model and relevant symmetries.

\subsubsection{Symmetries of the model}

Here we will skip over the symmetry-enforced nature of the $X^{\hat{r}}$ points and move straight to studying the symmetries that protect their gaplessness. The relevant spacial symmetries are given as follows in Seitz notation:
\begin{enumerate}
\item $C_{2z}=\{2_{001}|0\}$ [$\pi$ rotation around $\hat{z}$],
\item $C_{2y}=\{2_{010}|0\}$ [$\pi$ rotation around $\hat{y}$],
\item $C_{2x}=\{2_{100}|0\}$ [$\pi$ rotation around $\hat{x}$],
\item $P=\{-\mathds{1}|\mathbf{0}\}$ [inversion] 
\item $M_{z}=\{m_{001}|\frac{a}{4}(1,1,1)\}$ [mirror of $\hat{z}$ axis],
\item $M_{x}=\{m_{100}|\frac{a}{4}(1,1,1)\}$ [mirror of $\hat{x}$ axis],
\item $M_{z}=\{m_{010}|\frac{a}{4}(1,1,1)\}$ [mirror of $\hat{y}$ axis],
\end{enumerate}
and time reversal symmetry $\Theta$, and the translation symmetries $T_{\mathbf{t}_1}=\{\mathds{1},\mathbf{t}_1\}$, $T_{\mathbf{t}_2}=\{\mathds{1},\mathbf{t}_2\}$, and $T_{\mathbf{t}_3}=\{\mathds{1},\mathbf{t}_3\}$. The translation symmetries act as the usual exponential $e^{i\mathbf{k}\cdot\mathbf{t}_\alpha}$. The parity and time-reversal operators take the forms $P=\sigma^x$ and $\Theta=is^y\mathcal{K}$. Importantly we have non-symmorphic protecting symmetries with the following momentum space forms
\begin{align*}
M_z\Theta&=i \begin{pmatrix}
0 & e^{i\mathbf{k}\cdot\mathbf{t}_3} \\
1 & 0 
\end{pmatrix}_{\sigma}s^x\mathcal{K}\quad,\\
M_y\Theta&=-  \begin{pmatrix}
0 & e^{i\mathbf{k}\cdot\mathbf{t}_2} \\
1 & 0
\end{pmatrix}_\sigma\mathcal{K} \quad,\\
M_x\Theta&=-i  \begin{pmatrix}
0 & e^{i\mathbf{k}\cdot\mathbf{t}_1}\\
1 & 0
\end{pmatrix}_\sigma s^z\mathcal{K} \quad.
\end{align*}

\begin{figure}[htb]
    \centering
    \vspace{0.5cm}
        \includegraphics[width=0.5\columnwidth]{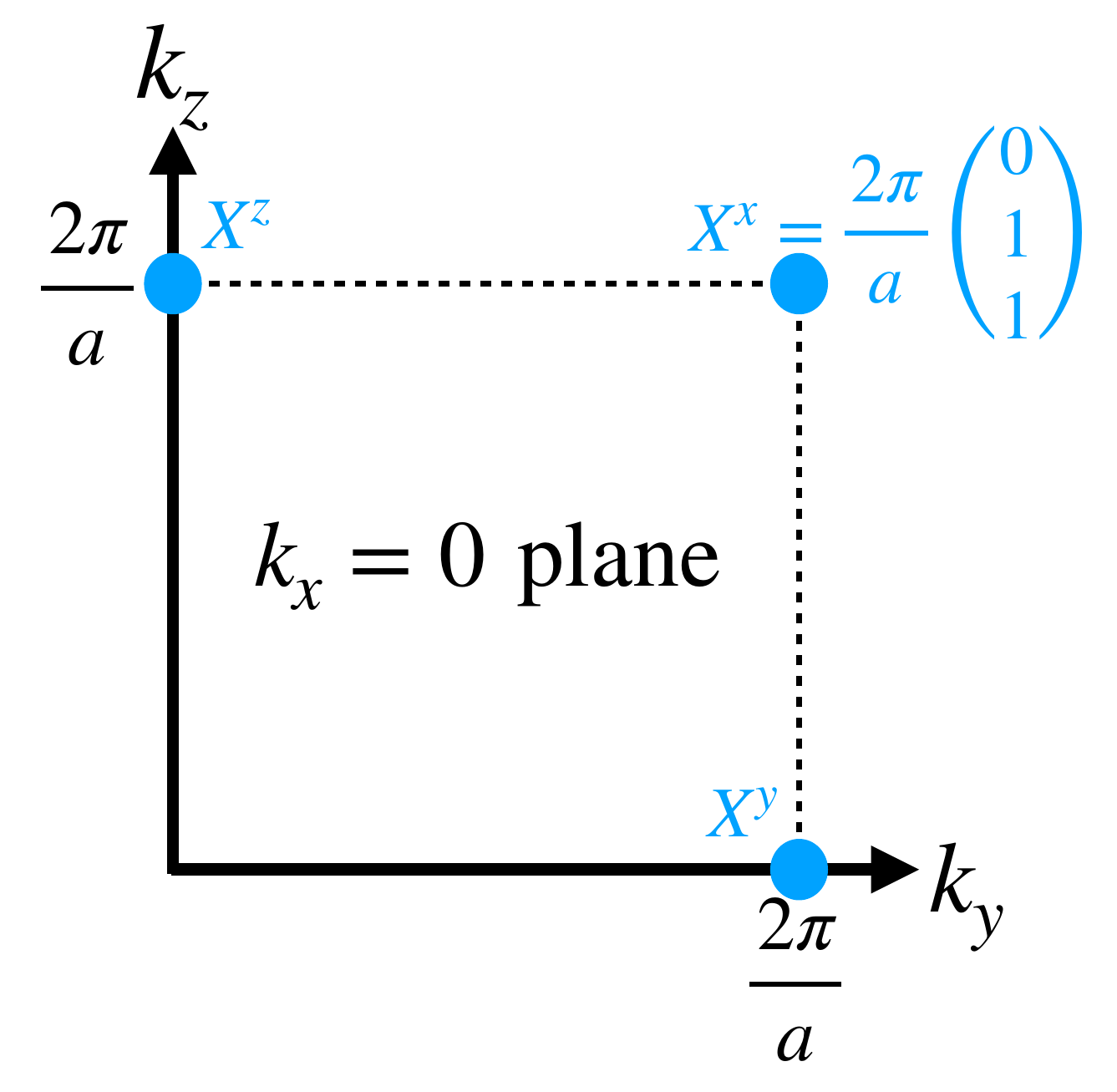}
    \caption{\label{fig:FKMlattice}We show a depiction of the relevant high-symmetry points all in the $k_x=0$ plane. Note that along the $k_x=0,\,k_y=2\pi/a$ axis we have a $4\pi/a$ periodicity of the Brillouin zone rather than the usual $2\pi/a$.}
\end{figure}

Now, let us see which low-energy Dirac nodes are protected by which symmetries. We see that the $\pi$ rotations change as momentum varies so not all the above symmetries  contribute to the stability of individual node's gaplessness. Recall that gaplessness does not imply retaining the Dirac dispersion, i.e., we allow for Weyl semimetal phases and nodalline semimetal phases. Gaplessness is also only locally stable: stable to perturbations but not to non-perturbative effects. This sort of gaplessness is protected as follows
\begin{enumerate}
\item
Gaplessness of $H_{X^x}$ is protected if we have either $\{C_{2y},U(1),T_{\mathbf{t}_1}\}$ or $\{C_{2z},U(1),T_{\mathbf{t}_1}\}$ or $\{M_y\Theta,U(1),T_{\mathbf{t}_1}\}$ or $\{M_z\Theta,U(1),T_{\mathbf{t}_1}\}$
\item
Gaplessness of $H_{X^y}$ is protected if we have either $\{C_{2x},U(1),T_{\mathbf{t}_2}\}$ or $\{C_{2z},U(1),T_{\mathbf{t}_2}\}$ or $\{M_x\Theta,U(1),T_{\mathbf{t}_1}\}$ or $\{M_z\Theta,U(1),T_{\mathbf{t}_1}\}$
\item
Gaplessness of $H_{X^z}$ is protected if we have either $\{C_{2x},U(1),T_{\mathbf{t}_3}\}$ or $\{C_{2y},U(1),T_{\mathbf{t}_3}\}$ or $\{M_x\Theta,U(1),T_{\mathbf{t}_1}\}$ or $\{M_y\Theta,U(1),T_{\mathbf{t}_1}\}$
\end{enumerate}
To understand this in more depth, let us take a pair of nodes, say, the $X^x$ and $X^y$ nodes. These nodes can be most conveniently explored along an axis in the $\hat{z}$ direction, centered at $k_x=0,\,k_y=2\pi/a$ (see Figure~\ref{fig:FKMlattice} for reference)~\footnote{Note that we can have $X^x=\frac{2\pi\hat{x}}{a}\equiv \frac{2\pi\hat{x}}{a}+\mathbf{g}_1=\frac{2\pi}{a}(0,1,1)$.}. It is immediately visible that the gaplessness is mandated by $\{C_{2z},U(1),T_{\mathbf{t}_1},T_{\mathbf{t}_2}\}=\{C_{2z},U(1),T_{\mathbf{t}_1}\}\cup\{C_{2z},U(1),T_{\mathbf{t}_2}\}$. Near the axis, we have the zeroth order expansion of our rotation operator as
\begin{align*}
C_{2z}=-is^z\sigma^z \quad,
\end{align*}
which prevents any mass terms such as $ms^y\sigma^z$ and $m\sigma^x$ from occurring along the axis. Thus a gapped state can only be achieved by pair-annihilation of nodes. So far we have not considered the third node at $X^z$ that lies on another $\hat{z}$ invariant axis, centered at $k_x=k_y=0$. Along this axis, an expansion of $C_{2z}$ is
\begin{align*}
C_{2z}=is^z\quad,
\end{align*}
which does not prevent a mass term such as $m s^z\sigma^z$ from gapping out the $H_{X^z}$ spectrum. So, crucially, this third node's gaplessness is not symmetry-protected by $C_{2z}$. This means that from the perspective of anomalies, we only need to consider the first two nodes that lie along the non-trivial rotation axis, which we will do now.

\subsection{Chiral anomaly}

We will demonstrate the chiral anomaly for a single set of symmetries $\{C_{2z},U(1),T_{\mathbf{t}_1},T_{\mathbf{t}_2}\}$, where the other sets can be calculated using the same process. This calculation follows the distorted Dirac spinel calculation, so here we will skip over details.
To derive the chiral anomaly of $U(1)$ and $C_2$ rotation symmetry, let us expand the Hamiltonian around the $\hat{z}$ axis centered at $k_x=0,\,k_y=2\pi/a$:
\begin{align}
&H_{axis}=-8\lambda_{\mathrm{SO}}a\sin\left(\frac{k_z}{2a}\right)s^z\sigma^z-\frac{t a}{2} \, (k_x+k_y)\sigma^y\nonumber\\
&+\frac{t a}{2} \,(k_x-k_y)\left(\cos\left(\frac{k_z}{2a}\right)\sigma^y-\sin\left(\frac{k_z}{2a}\right)\sigma^x\right)\\
&+4\lambda_{\mathrm{SO}}ak_x\cos\left(\frac{k_z}{4a}\right)^2s^x\sigma^z-4\lambda_{\mathrm{SO}}ak_y\sin\left(\frac{k_z}{4a}\right)^2s^y\sigma^z.\nonumber
\end{align}
Following the same chiral anomaly calculations as in Sec.~\ref{sec:chiralanomaly}, where we insert a magnetic field in $\hat{z}$ and diagonalize to find a squared Hamiltonian in the form of Eq.~\ref{eq:Haxisdiag}, we arrive at the same results except with
\begin{align*}
&\xi_q(k_z)=\nonumber\\
&2a\sqrt{\left(\frac{t}{2}\right)^2\sin\left(\frac{k_z}{4a}+\delta_{qs}\frac{\pi}{2}\right)^2+\lambda_\mathrm{SO}^2\sin\left(\frac{k_z}{4a}+\delta_{ql}\frac{\pi}{2}\right)^4}.
\end{align*}
The low-energy physics is described by
\begin{align}
S=\int dt dz\sum_{\xi}\left(i\left[\psi^{\xi}_{r}\right]^\dag\partial_-\psi^{\xi}_{r}+i\big[\psi^{\xi}_{l}\big]^\dag\partial_+\psi^{\xi}_{l}\right)\quad,
\end{align}
where $\xi\in\{X^x,X^y\}$, and $\psi_r^\xi$ ($\psi_l^\xi$) is the right (left) moving chiral field at $\xi$.
Now using the $1$d chiral anomaly we can again show that the overall chiral anomalies disappears since we have two nodes. The charge of $C_{2z}=e^{i\pi Q_{cz}}$ is $Q_{cz}=-\sigma^zs^z/2$, i.e., $\Delta Q_{cz}=0$ even  when we thread an electric flux that takes a left mover to a right mover is
\begin{align}
\Delta Q^{X^y}_{cz}=1=-\Delta Q^{X^x}_{cz}\implies \Delta Q_{cz}=0\quad,
\end{align}
which means that there is no total anomaly associated with this charge. On the individual nodes, there is a non-conservation of the $C_{2z}$ charge which is the chiral anomaly that protects the local stability of the $X^{x/y}$ nodes. If we were to consider the physics of the third node at $X^z$, we would find that $\Delta Q^z=0$, since both left and right moving chiral states possess the same $C_{2z}$ charge. In other words, this means that the $C_{2z}$ symmetry, and associated chiral anomaly, do not contribute to the local stability of the $X^z$ node.

In addition to the $C_{2z}$ type of anomaly, we also need to consider the chiral anomaly that comes from the translation symmetries $T_{\mathbf{t}_1}=e^{i\pi Q_{\mathbf{t}_1}}$ and $T_{\mathbf{t}_2}=e^{i\pi Q_{\mathbf{t}_2}}$, where the charges are $Q_{\mathbf{t}_1}=\left(k_y+k_z\right)/2\pi$ and $Q_{\mathbf{t}_2}=\left(k_x+k_z\right)/2\pi$~\footnote{To ensure that the charges remain well-defined, we must choose a gauge $A_\mu$ for $B_z=\partial_y A_x-\partial_xA_y$ such that either $k_y$, thus $Q_{\mathbf{t}_1}$, or $k_x$, thus $Q_{\mathbf{t}_2}$ remain well-defined.}. As in the distorted spinel case, we find that both charges vanish, i.e., $Q_{\mathbf{t}_2}=0$ and $\Delta Q_{\mathbf{t}_2}=0$, for a minimum quantum of electric flux. When a time-reversal breaking mass $m\sigma^z$ is inserted, the system suffers from a non-zero chiral anomaly and gains a quantum Hall conductivity of
\begin{align}
\sigma_{xy}=\frac{2m}{\pi}\quad,
\end{align}
which is compatible with our statement that the only non-zero anomalies can come from tuneable $\mathbb{Z}$ anomalies. Similarly to the two above cases, we can analyze the remaining symmetry groups and conclude that the rotation charge chiral anomalies always cancel, while the translation associated chiral anomalies vanish when time-reversal is preserved. In fact, all these anomalies are just three copies of those seen in the distorted spinel section.

Now that we have fully understood the two cases where the Dirac nodes are protected by a symmorphic $C_2$ symmetries, along with a translation symmetries, let us move on to nodes that are protected by non-symmorphic $C_2$ symmetries. Firstly we shall build a system that has a single symmetry-protected Dirac node. We will then see that the associated chiral anomaly does not vanish, unlike the distorted spinel or FKM cases.

\section{Time-reversal symmetry ($\Theta^2=(-1)^{\hat{F}}$) and its effect on filling}
\label{sec:pottingsoil}

Here we will explain in the free-fermion picture why a $\Theta^2=(-1)^{\hat{F}}$ (where $F$ is the fermion parity) system requires an electron filling of $2\mathbb{Z}$ per unit cell to be trivial. To understand why this is the case, let us examine the effects of flux threading $\Phi$ on the energy spectrum of the system.

Take a 1+1d system with length $L$ and let us first observe what happens when $L$ is odd. In this case, when the filling is an odd integer per unit cell, then we must have at an odd number of Kramer's pair at zero energy to maintain half-filling. Each energy in the spectrum represents a Bloch state with a specific kinetic momentum ranging from $(-\pi,\pi]$ in steps of $\frac{2\pi\mathbb{Z}}{L}$ (for $T^L=1$), and due to Kramers degeneracy this energy is doubly degenerate with associated momentum $\pm k$. After an insertion of $2\pi$ flux the energy spectrum returns to itself, but the filled energy states change their kinetic momentum by $\frac{2\pi}{L}$. At the fluxes $\Phi=\pi\mathbb{Z}$ there should always be a Kramer's degeneracy since $\Phi$ is only well-defined mod $2\pi$. By design, such a system is necessarily gapless and long-range entangled.

The argument is slightly more complicated when $L$ is even: in this case at $\Phi\in\{0,\pi\}$ no Kramers pairs need to be pinned at $E=0$ unlike the odd length situation. However, due to the odd-integer filling condition, $L/2\,(\mathrm{mod}{\,L})$ Kramers pairs are below the zero energy and $L/2\,(\mathrm{mod}{\,L})$ Kramers pairs are above the zero energy. Since Kramers pairs consist of Bloch states of opposite momentum. Of these Kramers pairs two are special: the Kramers pairs at the time-reversal invariant momenta $k=0$ and $k=\pi$. Without loss of generality let us assume that the $k=0$ Kramers pair is below the Fermi level (the argument also works if above the Fermi level). If the system is gapped and a $2\pi$ flux is thread through the system the $k=0$ pair both flow to the two distinct $k=2\pi/L$ state which must also be below the gap. We may do this argument reiteratively to show that all momentum states between $k\in\{0,\frac{2\pi}{L},...,2\pi-\frac{2\pi}{L}\}$ are doubly present below the gap, which represents $2L$ number of occupied electron states or, equivalently, $L$ occupied Kramers pairs. This is in contradiction with a gapped odd electron filling. The resolution is that at some momentum one or both states have to flow above the Fermi surface once a $2\pi$ flux is thread. Similarly to the odd length situation, this implies that the system is gapless.

Now we may observe what happens when we add more spacial dimensions: take a 2+1d system with size $L_x\times L_y$. Again, we examine why an odd integer filling per unit cell results in a gapless state. For a flux thread through the $x$ loop there are two $k_y\in\{0,\pi\}$ lines in the Brillouin zone that transform to themselves under time reversal symmetry. The same logic that we applied to the 1+1d systems holds for these BZ lines thus implying their gaplessness. One may wonder whether there could be a fully filled/empty $k_y\in\{0,\pi\}$ line: this is only possible if there exists a Fermi surface which can be verified by a flux-threading argument in the $y$ loop. If we purely desire gapless points, rather than surfaces, then there must be at least four such points: one situated at each TRIM point $(k_x,k_y)=(0,0),(\pi,0),(0,\pi),(\pi,\pi)$. 

Now if we fold up the BZ in these cases and turn the translation symmetry to a non-symmorphic symmetry, we see that we can get two twofold degenerate nodes per axis so at least two fourfold degenerate nodes in the whole BZ. This also implies that in non-symmorphic symmetry-protected 2+1d Dirac nodes (four-fold degenerate point) must occur in pairs. Similar logic follows in higher dimensions.

\subsection{Fine-tuned Dirac points}

A model of a time-reversal ($\Theta_{\rm UV}^2=(-1)^{\hat{F}}$), $U(1)_{V,{\rm UV}}$, and translation symmetric fine-tuned single Dirac node is
\begin{align}
    H_{\rm FT}=\sum_\mathbf{k}c_{\mathbf{k}}^\dag \bigg[&\sin k_x \sigma_z s_x + \sin k_y \sigma_z s_y +\sin k_z \sigma_z s_z \nonumber\\
    +(3-&\cos k_x -\cos k_y -\cos k_z)\sigma_x\bigg] c_{\mathbf{k}}\,.
\end{align}
This model possesses a single fine-tuned Dirac node at the $\Gamma\equiv \mathbf{0}$ point. This situation is fully compatible with $U(1)_{V,{\rm UV}}$ flux threading when there is no non-symmorphic symmetry, since two electrons per unit cell is fully compatible with a $\Theta^2=(-1)^{\hat{F}}$ band insulator. This can be understood via the filling response term
\begin{align}
    S_{\rm FT}=2\int x\wedge y\wedge z\wedge A\quad,
\end{align}
which is fully gauge invariant even in the presence of time-reversal symmetry with $\Theta_{\rm UV}^2=(-1)^{\hat{F}}$.

However if the system possessed a non-symmorphic symmetry, say in the $\hat{z}$ direction, then we would also have the term
\begin{align}
    S_{\rm NS}=\int x\wedge y\wedge \tilde{z}\wedge A\quad,
\end{align}
where $\tilde{z}$ is the non-symmorphic gauge field. Although this term still describes two electron per unit cell, it also details the presence of a single electron per half unit cell which is not consistent with time-reversal symmetric $\Theta_{\rm UV}^2=(-1)^{\hat{F}}$ band insulator. This can be seen via the flux threading argument.

\end{document}